\documentclass[%
floatfix,
reprint,
superscriptaddress,
amsmath,amssymb,
aps,
prx,
]{revtex4-2}

\usepackage{graphicx}% Include figure files
\usepackage{bm, color}% bold math and color
\usepackage{physics}
\usepackage{amsmath}
\usepackage{amssymb}
\usepackage{cases}

\usepackage{chngcntr} % Change figure index.

\usepackage[colorlinks=true, citecolor=cyan, linkcolor=blue, urlcolor=black]{hyperref}
\usepackage[utf8]{inputenc}

\usepackage{xparse}  % For advanced command definitions
\usepackage{xcolor}  % For colored text
\begin{document}
\title{Anomalous diffusion in coupled viscoelastic media: A fractional Langevin equation approach}
\author{Chan Lim}
\affiliation{Department of Physics, Pohang University of Science and Technology, Pohang 37673, Republic of Korea}
\author{Jae-Hyung Jeon}
\email{jeonjh@postech.ac.kr}
\affiliation{Department of Physics, Pohang University of Science and Technology, Pohang 37673, Republic of Korea}
\affiliation{Institute for Theoretical Science, Pohang University of Science and Technology, Pohang 37673, Republic of Korea}
\affiliation{Asia Pacific Center for Theoretical Physics, Pohang 37673, Republic of Korea}
\begin{abstract}
Anomalous diffusion often arises in complex environments where viscoelastic or crowded conditions influence particle motion. In many biological and soft-matter systems, distinct components of the medium exhibit unique viscoelastic responses, resulting in time-dependent changes in the observed diffusion exponents. Here, we develop a theoretical model of two particles, each embedded in a distinct viscoelastic medium, and coupled via a harmonic potential. By formulating and solving a system of coupled fractional Langevin equations (FLEs) with memory exponents $0<\alpha<\beta\leq1$, we uncover rich transient anomalous diffusion phenomena arising from the interplay of memory kernels and bilinear coupling. Notably, we identify ``recovery dynamics'', where a subdiffusive particle ($\alpha$) transiently accelerates and eventually regains its intrinsic long-time mobility. This recovery emerges only when memory exponents differ ($\alpha<\beta$), whereas identical exponents ($\alpha=\beta$) suppress recovery. 
{ Our FLE models are further validated in comparison with Langevin dynamics simulations of polymer-based real systems.}
Our theoretical predictions offer insight into experimentally observed transient anomalous diffusions, such as polymer–protein complexes and cross-linked cytoskeletal networks, highlighting the critical role of memory heterogeneity and mechanical interactions in biological anomalous diffusion.
\end{abstract}

\maketitle

\section{Introduction}
The motion of particles in biological and soft-matter systems often exhibits anomalous diffusion, in which the mean square displacement (MSD) of the particle follows a power-law:  
$\langle \Delta x^2 (t) \rangle \sim t^\nu$, where $\nu$ is anomalous exponent in the range of $0<\nu\leq 2$ \cite{metzlerAnomalousDiffusionModels2014,baggioliAnomalousDiffusionNoethers2021}. In particular, it is known that subdiffusive motions ($0<\nu<1$) are observed in many soft-matter systems, mainly due to the viscoelasticity and molecular crowding of the embedding media, which gives rise to memory effects in the diffusing dynamics of the embedded particles. A physics-based mathematical framework to describe such anomalous diffusion phenomena is the generalized Langevin equation (GLE) \cite{fordStatisticalMechanicsAssemblies1965,moriTransportCollectiveMotion1965,zwanzigNonlinearGeneralizedLangevin1973}, which extends the Langevin equation by incorporating a time-dependent frictional kernel. The GLE takes the form:
\begin{equation}
\label{eq:GLE}
    m\ddot{x}(t) = - \int_0^t K(t-t') \dot{x}(t') \, dt' + \xi(t),
\end{equation}
where $K(t)$ is the memory kernel that governs the frictional response, and $\xi(t)$ is the thermal noise satisfying the fluctuation-dissipation theorem (FDT) of the second kind \cite{kuboFluctuationdissipationTheorem1966}:
\begin{equation}
    \langle \xi(t) \xi(t') \rangle = k_B T K(|t-t'|)
\end{equation}
where $k_B$ is the Boltzmann constant and $T$ is the absolute temperature. When $K(t)=2\gamma\delta(t)$, the GLE reduces to the ordinary Langevin equation described with the friction term $-\gamma \dot{x}$. In viscoelastic systems, friction has memory such that the velocities and positions in the past influence the diffusion dynamics at present.  

An important class of the GLE is the so-called \emph{fractional Langevin equation} (FLE) \cite{lutzFractionalLangevinEquation2001,goychukViscoelasticSubdiffusionAnomalous2009, goychukViscoelasticSubdiffusionGeneralized2012}, where the memory kernel decays over time in a power-law manner, as follows;
\begin{equation}
\label{eq:memory_kernel}
    K_{\mu}(t) = \frac{\gamma_{\mu}}{\Gamma(1-\mu)} t^{-\mu}, \quad 0<\mu<1.
\end{equation}
Here $\gamma_\mu$ is a generalized friction coefficient in the unit of [$\mathrm{kg} \cdot \mathrm{s^{\mu-{2}}}$], $\mu$ is a memory exponent, and $\Gamma(z)$ is the Gamma function. Such power-law forms of the memory kernel are often observed in numerous biological and soft-matter systems, e.g.,  proteins~\cite{kouGeneralizedLangevinEquation2004, minObservationPowerLawMemory2005}, intracellular fluids~\cite{fabryScalingMicrorheologyLiving2001,wilhelmOutofEquilibriumMicrorheologyLiving2008,guoProbingStochasticMotorDriven2014}, chromosomes~\cite{weberBacterialChromosomalLoci2010,lampoPhysicalModelingChromosome2015}, polymer networks~\cite{masonRheologyFactinSolutions2000,gislerScalingMicrorheologySemidilute1999,schnurrDeterminingMicroscopicViscoelasticity1997}, and complex fluids~\cite{dasguptaMicrorheologyPolyethyleneOxide2002}. Plugging the kernel \eqref{eq:memory_kernel} into Eq.~\eqref{eq:GLE} leads to the FLE (neglecting the inertia term): 
\begin{equation}
\label{eq:overdamped_FLE}
    \int_0^t K_{\mu}(t-t') \dot{x}(t') \, dt' =\gamma_\mu \frac{{\rm d}^{\mu}}{{\rm d}t^{\mu}}x(t) = \xi_\mu(t),
\end{equation}
where $\frac{{\rm d}^{\mu}}{{\rm d}t^{\mu}}x(t)$ is the Caputo fractional derivative of order $\mu$~ \cite{caputoLinearModelsDissipation1967,deoliveiraReviewDefinitionsFractional2014}.
The MSD of the FLE \eqref{eq:overdamped_FLE} exhibits the scaling law
\begin{equation}
\label{eq:MSD_k=0}
    \langle \Delta x^2(t) \rangle = \frac{2k_BT}{\gamma_\mu\Gamma(1+\mu)} t^{\mu}, \quad 0<\mu<1,
\end{equation}
describing the subdiffusive motion observed in biological and soft-matter systems \cite{masonOpticalMeasurementsFrequencyDependent1995,goychukAnomalousEscapeGoverned2007}. In the limit of $\mu \to 1$, the power-law memory kernel reduces to a delta function, and the FLE becomes the ordinary Langevin equation, recovering the Brownian motion.

The FLE framework has been widely applied to model various bio-complex systems. Theoretically, it can be derived from the general elastic models, including those incorporating hydrodynamic interactions \cite{taloniGeneralizedElasticModel2010}. A broad range of exponents $\mu$ emerges when considering the dynamics of a single monomer or a local segment as part of polymeric or membrane systems. For instance, polymer models predict specific values of $\mu$: an entangled polymer has $\mu = 1/4$ \cite{panjaAnomalousPolymerDynamics2010}; a flexible polymer exhibits $\mu = 1/2$ \cite{lizanaFoundationFractionalLangevin2010,vandebroekGeneralizedLangevinEquation2017,maesLangevinGeneralizedLangevin2013,shinkaiGeneralizedLangevinDynamics2024}, while a semiflexible polymer has $\mu = 3/4$ \cite{hanNonequilibriumDiffusionActive2023,durangGeneralizedLangevinEquation2024} and a self-avoiding (flexible) chain reveals $\mu \approx 0.54$ \cite{panjaAnomalousPolymerDynamics2010,sakaueMemoryEffectFluctuating2013,kapplerCyclizationRelaxationDynamics2019}; The Zimm model considering hydrodynamic interactions has $\mu = 2/3$ \cite{panjaGeneralizedLangevinEquation2010,taloniGeneralizedElasticModel2010}. Interestingly, the transverse undulation dynamics in lipid membranes has $\mu = 2/3$ \cite{taloniGeneralizedElasticModel2010}.

Apart from these model systems, the power-law memory kernel is revealed via the generalized Stokes--Einstein relation \cite{masonOpticalMeasurementsFrequencyDependent1995}, using the microrheology experiments for viscoelastic gels and biological systems \cite{masonRheologyFactinSolutions2000,rigatoHighfrequencyMicrorheologyReveals2017,fabryScalingMicrorheologyLiving2001,wilhelmOutofEquilibriumMicrorheologyLiving2008,guoProbingStochasticMotorDriven2014}. Furthermore, molecular dynamics simulations have reported viscoelastic subdiffusion in lipid bilayers \cite{jeonAnomalousDiffusionPhospholipids2012,yamamotoDynamicInteractionsMembrane2017} and protein condensates \cite{watanabeDiffusionIntrinsicallyDisordered2024}, demonstrating the relevance of FLE models in describing real-world systems.

On the other hand, anomalous diffusions in biological and soft-matter systems are, in many cases, \emph{transient} such that the anomalous exponent $\nu(t)$ changes over time. Biological examples showing such transient anomalous diffusions includes the chromatin motion in biological nuclei~\cite{germierRealTimeImagingSingle2017, zidovskaMicronscaleCoherenceInterphase2013,bronsteinTransientAnomalousDiffusion2009,bronshteinLossLaminFunction2015, ashwinOrganizationFastSlow2019,khannaChromosomeDynamicsSolgel2019,leviChromatinDynamicsInterphase2005}, cytoplasmic macromolecular transport~\cite{hoffmanConsensusMechanicsCultured2006,ottenLocalMotionAnalysis2012,annibaleSingleCellVisualization2015,makIntegratedAnalysisIntracellular2017,rogersIntracellularMicrorheologyMotile2008,leeRealTimeTrackingVesicles2024}, membrane-protein diffusion on biological membranes~\cite{muraseUltrafineMembraneCompartments2004, suzukiRapidHopDiffusion2005}, tracer diffusion in synthetic gels~\cite{wulsteinTopologydependentAnomalousDynamics2019, garamellaAnomalousHeterogeneousDNA2020,andersonSubtleChangesCrosslinking2021, leeMyosindrivenActinmicrotubuleNetworks2021, sheungMotordrivenAdvectionCompetes2022} or cultured cells~\cite{krajinaMicrorheologyRevealsSimultaneous2021}, and the migration of a cell in extracellular matrices~\cite{bakerExtracellularMatrixStiffness2009, higginsIntracellularMechanicsTBX32023}.

As a possible mechanism, such transient anomalous diffusions originate from the presence of multiple viscoelastic components within complex biological systems and their coupled dynamics. When isolated in free space, each viscoelastic component would exhibit anomalous dynamics described by FLE~\eqref{eq:overdamped_FLE} with a time-independent monoscaling exponent ($\mu$). When several viscoelastic components comprise a complex system through mechanical couplings, the system can achieve temporally heterogeneous anomalous dynamics in which the anomalous exponent $\nu$ changes over time, depending on which viscoelastic component dominates or how strong the coupling between components is. 

\begin{figure}
\includegraphics[width=1\columnwidth]{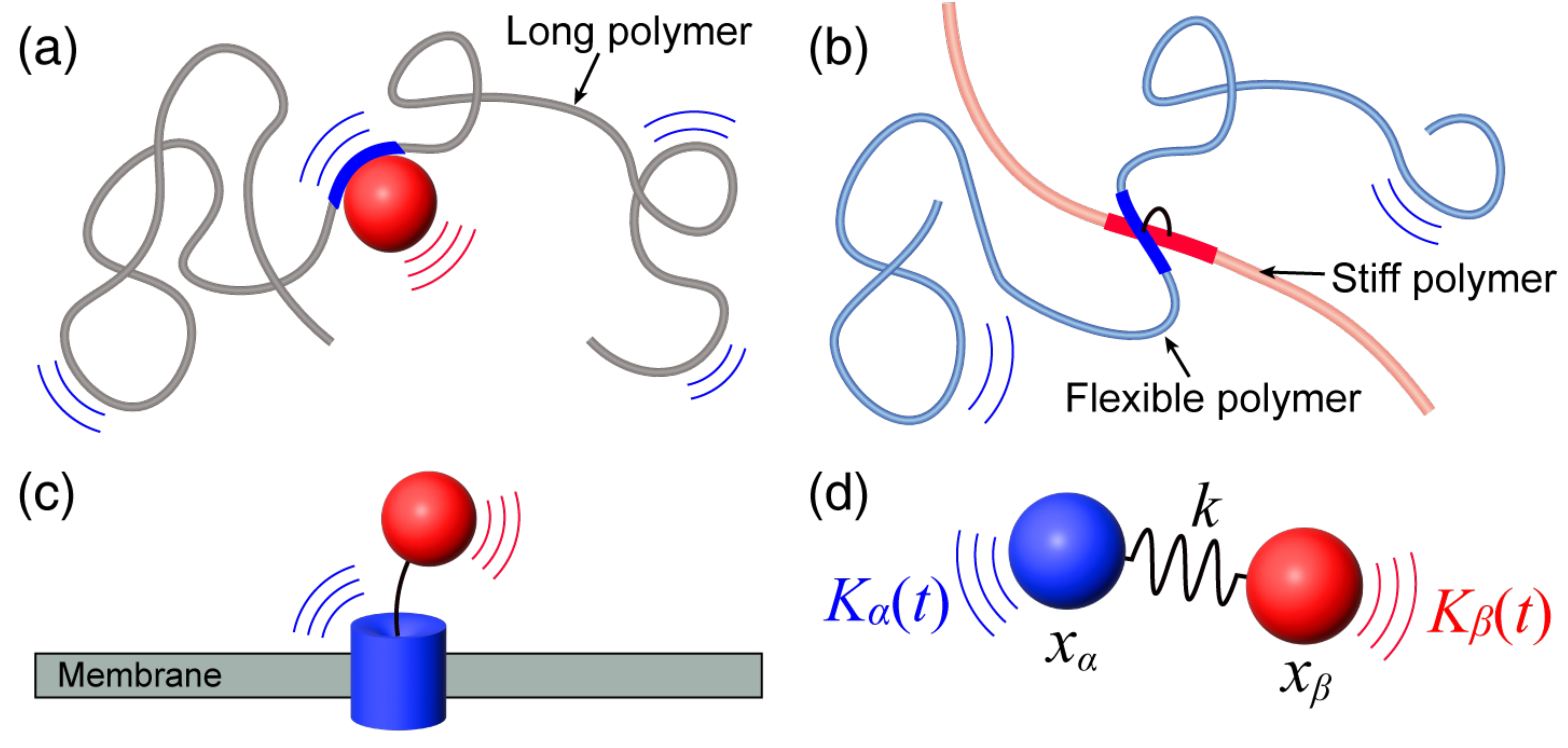}
\caption{\label{fig1}
Schematic illustrations of biological and soft-matter systems modeled by two coupled FLEs. (a) A polymer chain (e.g., chromatin) tethered to a macromolecular complex or protein condensate. (b) A composite cytoskeletal network comprising flexible polymers cross-linked with stiff polymers. (c) A membrane protein embedded within a lipid bilayer, interacting with both trans- and extracellular domains. (d) Our minimal bipartite model of these systems. Two particles, each governed by its own viscoelastic memory kernel characterized by distinct memory exponents ($\alpha$ and $\beta$), are mechanically coupled by a harmonic potential.
}
\end{figure}

There are multiple biological examples that can be understood with our model of coupled FLE systems (or multi-component viscoelastic systems). One such example is the chromosome dynamics where a chromatin polymer is associated with DNA-binding proteins, e.g., transcription factors \cite{spitzTranscriptionFactorsEnhancer2012}, chromatin remodelers \cite{clapierMechanismsActionRegulation2017}, and large protein condensates that function as membrane-less organelles within the nucleus \cite{suLiquidliquidPhaseSeparation2021, hiroseGuideMembranelessOrganelles2023}. The schematic illustration for this example is depicted in Fig.~\ref{fig1}(a). Here, the system consists of two components: the chromatin (i.e., a polymer chain) and a macromolecular complex (red). The chromatin is a viscoelastic material whose dynamics is modeled with FLE with the momery exponent $\alpha=1/2$ (flexible chains) or $3/4$ (semiflexible chains). 
The macromolecular complex represents the other viscoelastic counterpart, governed by another FLE with the memory exponent $\beta$,  which is unity if the embedding medium is a viscous fluid or $<1$ in a crowded (viscoelastic) medium. This macromolecular complex is bound to the chromatin, making the two FLE systems coupled.

Another interesting example is the cytoskeleton dynamics, as schematically illustrated in Fig.~\ref{fig1}(b). For instance, a flexible polymer ($\alpha=1/2$) cross-links with a stiff filament ($\beta=3/4$). Then, the cross-linked polymer network exhibits the coupled monomer dynamics of these two distinct polymers. As an example, a microrheology experiment showed that when a composite gel is made with two distinct types of polymers with markedly different persistence lengths ($l_p$), the viscoelastic memory of the gel has two distinct power-law regimes at short and long timescales~\cite{rickettsCoEntangledActinMicrotubuleComposites2018,wulsteinTopologydependentAnomalousDynamics2019,andersonSubtleChangesCrosslinking2021,garamellaAnomalousHeterogeneousDNA2020}. In the cell, the cytoskeleton is mainly composed of actin filaments ($l_p \approx 10\,\mathrm{\mu m}$) and microtubules ($l_p \approx 1\,\mathrm{mm}$) \cite{gittesFlexuralRigidityMicrotubules1993,kikumotoFlexuralRigidityIndividual2006,huberCytoskeletalCrosstalkWhen2015}. These filaments are sterically entangled and frequently crosslinked by accessory proteins, facilitating essential cellular processes such as proliferation, differentiation, and migration. 

An additional relevant example is the diffusion of membrane proteins on a plasma membrane. Membrane proteins, embedded in a lipid bilayer through its transmembrane domain (blue), typically possess a large cytoplasmic part (red) or form a complex (red) with macromolecules in the cytoplasm [Fig.~\ref{fig1}(c)]. 
The transmembrane domain lies in a viscoelastic medium (lipid bilayer, $\alpha$), which is connected to the extracellular part in contact with a cytoplasmic viscoelastic fluid ($\beta$).  Such a mechanical coupling between the two distinct viscoelastic dynamics can result in a complex diffusion dynamics of membrane proteins. Furthermore, additional interactions with the cytoskeletal filaments can amplify the coupling effects, reinforcing the emergence of complex viscoelastic responses. 

In this work, we investigate the dynamics of coupled viscoelastic systems using a minimal bipartite model based on two coupled FLEs [see Fig.~\ref{fig1}(d)]. In this model, two particles are connected via a harmonic spring, and each undergoes motion governed by its own FLE with a distinct memory kernel (with the indexes $\alpha$ or $\beta$). We solve the model analytically and show that the system exhibits rich transient anomalous diffusion characterized by multiple scaling regimes. A particularly surprising consequence is the appearance of recovery dynamics, where the slower particle (with the anomalous exponent $\alpha$) temporarily accelerates and catches up with the faster one (with the exponent $\beta$), resulting in an effective scaling exponent $\nu(>\alpha)$ during an intermediate time window. The results highlight how coupling between components can lead to emergent anomalous behaviors that differ significantly from those of the individual subsystems.

The remainder of this paper is structured as follows. In Sec.~\ref{sec:Model}, we begin by introducing the coupled FLE model and defining relevant timescales based on system parameters. In Sec.~\ref{sec:Theory}, we then solve the coupled equations and analyze the resulting effective memory kernel. In Sec.~\ref{sec:MSD}, we compute the MSDs of the two particles and their relative displacement. { In Sec.~\ref{sec:simulation}, we validate our theoretical FLE models with Langevin dynamics simulations of two physical examples.} Finally, we summarize the key findings and discuss their physical implications.

\section{Coupled fractional Langevin equations}
\label{sec:Model}

Motivated by the coupled viscoelastic systems explained above, we consider a bipartite system shown in Fig.~\ref{fig1}(d) where each particle performs a viscoelastic anomalous diffusion with a distinct (power-law decaying) memory kernel [Eq.~\eqref{eq:memory_kernel}] and their motion is coupled via a harmonic spring. This coupled FLEs are described by the following equation of motion:
\begin{subequations}
\label{eq:coupled_FLE}
\begin{eqnarray}
\label{eq:coupled_FLE1}
    \int_0^t K_\mathrm{\alpha}(t-t')\dot{x}_\alpha(t')dt' = -k\big(x_\alpha(t)-x_\beta(t)\big)+\xi_\alpha(t)\\
\label{eq:coupled_FLE2}
    \int_0^t K_\mathrm{\beta}(t-t')\dot{x}_\beta(t')dt' = -k\big(x_\beta(t)-x_\alpha(t)\big)+\xi_\beta(t).
\end{eqnarray}   
\end{subequations}
In this expression, $x_\mu(t)$ is the position of the particle $\mu$ at time $t$, where $\mu=\alpha,\beta$, and $0<\alpha<\beta\leq1$ is assumed without loss of generality. Here, $K_\mu(t)$ denotes the power-law memory kernel given by Eq.~\eqref{eq:memory_kernel}, and $\xi_\mu(t)$ is the zero-mean fractional Gaussian noise satisfying FDT: $\langle \xi_\mu(t)\xi_{\mu'}(t')\rangle=k_B T \delta_{\mu\mu'} K_\mu(|t-t'|)$. The two particles interact with each other through the harmonic potential of an interaction strength $k$. {This setup can be considered as a generalization of two harmonically coupled Brownian particles in the context of viscoelastic media characterized with long-time memories~\cite{schulzTwoHarmonicallyCoupled2001}.} In our model, the initial condition of the relative displacement $x_\alpha(t)-x_\beta(t)$ is assumed to follow a Gibbs--Boltzmann distribution, $x_\alpha(0)-x_\beta(0)\sim\mathcal{N}(0,k_B T/k)$.

Importantly, our FLE model~\eqref{eq:coupled_FLE} can be directly derived from coupled polymer Langevin dynamics, ensuring a clear physical foundation. Consider a bipartite polymer system composed of two  sub-systems, $A$ and $B$, each represented by beads with coordinates $\mathbf r_i^{Z}(t)\in\mathbb R^d$ ($Z\in\{A,B\}$, $i=1,\dots,M_Z$).  The total potential energy of the system is
\begin{equation}\label{eq:poly_E_tot}
U_{\text{tot}}
= U^{A}\!\bigl(\{\mathbf r^{A}_i\}\bigr)
+ U^{B}\!\bigl(\{\mathbf r^{B}_i\}\bigr)
+ \tfrac{1}{2} k (\mathbf r^{A}_{a}-\mathbf r^{B}_{b})^2,
\end{equation}
where the bead with index $a$ in $A$ and the bead with index $b$ in $B$ are coupled via the harmonic potential in $U_{\text{tot}}$. 
Microscopically, this bipartite polymer system is described by the overdamped Langevin equation
\begin{equation}\label{eq:poly_EOM}
\gamma^{Z}\,\dot{\mathbf r}^{Z}_i(t)
= -\,\frac{\partial U_{\text{tot}}}{\partial \mathbf r^{Z}_i}
+ \boldsymbol{\eta}^{Z}_i(t),
\end{equation}
where $\gamma^{Z}$ is the bead friction and $\boldsymbol{\eta}^{Z}_i$ is Gaussian white noises with zero mean and covariance $
\big\langle \eta^{Z}_{i,u}(t)\,\eta^{Z'}_{j,v}(t')\big\rangle
= 2\gamma^{Z} k_BT\delta_{ZZ'}\delta_{ij}\delta_{uv}\delta(t-t')
$, with $u,v$ denoting Cartesian components. By integrating out all coordinate variables except for the tracers $\mathbf r^{A}_{a}(t)$ and $\mathbf r^{B}_{b}(t)$ and projecting along one Cartesian direction~\cite{panjaGeneralizedLangevinEquation2010,panjaAnomalousPolymerDynamics2010,taloniGeneralizedElasticModel2010,lizanaFoundationFractionalLangevin2010,maesLangevinGeneralizedLangevin2013,durangGeneralizedLangevinEquation2024,hanNonequilibriumDiffusionActive2023,vandebroekGeneralizedLangevinEquation2017,sakaueMemoryEffectFluctuating2013,shinkaiGeneralizedLangevinDynamics2024}, we obtain the one-dimensional coupled FLEs in the form of Eq.~\eqref{eq:coupled_FLE}. As an example, in Appendix~\ref{sec:FLE_derivation}, we have derived the FLE for the macromolecule--polymer coupled system depicted in Fig.~\ref{fig1}(a). The obtained FLE \eqref{eq:coupled_FLE_flex_bead} indeed follows the form in Eq.~\eqref{eq:coupled_FLE} with a power-law memory kernel of $\alpha=1/2$ and a delta-correlated kernel (or $\beta=1$). Refer to Appendix~\ref{sec:FLE_derivation} for a complete description of the mathematical formalism.

% We note that our coupled FLE \eqref{eq:coupled_FLE} provides a versatile framework for describing a wide range of coupled systems. It can be derived explicitly from polymer models by coarse-graining over all internal degrees of freedom while retaining only the tracer coordinates of interest, $x_\alpha$ and $x_\beta$ (see Appendix \ref{sec:FLE_derivation} for an explicit example). In this sense, our formulation generalizes earlier studies that derived single-particle FLEs from polymers \cite{panjaGeneralizedLangevinEquation2010,panjaAnomalousPolymerDynamics2010,taloniGeneralizedElasticModel2010,lizanaFoundationFractionalLangevin2010,maesLangevinGeneralizedLangevin2013,durangGeneralizedLangevinEquation2024,hanNonequilibriumDiffusionActive2023,vandebroekGeneralizedLangevinEquation2017,sakaueMemoryEffectFluctuating2013,shinkaiGeneralizedLangevinDynamics2024}, extending them to coupled viscoelastic systems. At the same time, it preserves a level of simplicity that enables transparent analytical treatment while remaining physically grounded.

Before solving the coupled FLEs \eqref{eq:coupled_FLE}, we define three important characteristic timescales through a comparative analysis of the memory kernels and the interaction part. In the Laplace domain (with the transformation $\tilde{f}(s) = \int_0^\infty e^{-st}f(t)ds$), the memory kernels become $\tilde{K}_\alpha(s)=\gamma_\alpha s^{-1+\alpha}$ and $\tilde{K}_\beta(s)=\gamma_\beta s^{-1+\beta}$, while the spring constant $k$ is to be $k s^{-1}$. In the Laplace space, any two curves among them intersect each other, which allows us to define the following characteristic times: 
\begin{equation}
\label{eq:taus}
    \tau_\alpha = \left(\frac{\gamma_\alpha}{k}\right)^{1/\alpha},\;
    \tau_\beta=\left(\frac{\gamma_\beta}{k}\right)^{1/\beta},\;
    \tau_c = \left(\frac{\gamma_\beta}{\gamma_\alpha}\right)^{1/(\beta-\alpha)}.
\end{equation}
The physical meaning of these timescales are as follows: The $\tau_\alpha$ and $\tau_\beta$ mark the times at which the particle $\alpha$ (or $\beta$) diffusing in free space ($k=0$) reaches the distance of $\sqrt{2k_B T/k}$ (i.e., the average confined distance in the harmonic potential). The $\tau_c$ is the time when the free-space diffusion for particles $\alpha$ and $\beta$ have the same diffusion distance, that is,
\begin{equation}\label{eq:taus_dynamics}
    \begin{aligned}
        \tau_\alpha:& \quad \langle \Delta x_\alpha^2(\tau_\alpha)\rangle _{k=0} \approx 2k_B T/k,\\
        \tau_\beta:& \quad \langle \Delta x_\beta^2(\tau_\beta)\rangle _{k=0} \approx 2k_B T/k,\\
        \tau_c:& \quad \langle \Delta x_\alpha^2(\tau_c)\rangle _{k=0} \approx\langle\Delta x_\beta^2(\tau_c)\rangle _{k=0},\\
    \end{aligned}
\end{equation}
where $\langle \Delta x_\mu^2(t)\rangle_{k=0}=\frac{2k_B T t^\mu}{\gamma_\mu \Gamma(1+\mu)}$.

As a special case, there exists a critical interaction strength
\begin{equation}
\label{eq:k_c}
    k_c = (\gamma^\beta_\alpha/\gamma^\alpha_\beta)^{1/(\beta-\alpha)}
\end{equation}
at which the three terms ($\tilde{K}_\alpha(s)$, $\tilde{K}_\beta(s)$, and $k s^{-1}$) intersect at one point in the Laplace domain. It is noteworthy that the value of $k/k_c$ determines the ordering of the three characteristic times: (1) the weak interaction regime ($k < k_c$), it is $\tau_\alpha > \tau_\beta > \tau_c$. (2) the strong interaction regime ($k>k_c$), $\tau_\alpha < \tau_\beta <\tau_c$ is obtained.

In the next section, we explicitly solve the coupled FLEs \eqref{eq:coupled_FLE} for $x_\alpha$ and $x_\beta$ as well as for the relative displacement
\begin{equation}\label{eq:rab}
    r_{\alpha\beta}(t)\equiv x_\alpha(t) -x_\beta(t).
\end{equation}
Because the two particles have distinct memory kernels, the normal mode decomposition is not applicable. Instead, we explicitly derive the generalized Langevin equation for each system, identifying the corresponding memory kernel and the effective noise term to find the analytic solution. 

\section{The effective decoupled Langevin equations}
\label{sec:Theory}

In this section, we introduce the GLE framework to derive effective equations for the relative displacement and individual particle dynamics, identifying the effective memory kernels and noise terms that incorporate the effects of the coupled dynamics.

\subsection{GLE for the relative displacement \texorpdfstring{$r_{\alpha\beta}$}{r ab}}
Rewriting the coupled FLEs~\eqref{eq:coupled_FLE} in the Laplace domain, we obtain
\begin{subequations}
\label{eq:coupled_FLE_laplace}
\begin{eqnarray}
\label{eq:coupled_FLE_laplace_a}
\tilde{\dot{x}}_\alpha(s)
&= -\frac{k}{\tilde K_\alpha(s)}(\tilde x_\alpha(s)-\tilde x_\beta(s)) + \frac{1}{\tilde K_\alpha(s)} \tilde{\xi}_{\alpha} (s)\\
\label{eq:coupled_FLE_laplace_b}
\tilde{\dot{x}}_\beta(s)
&= -\frac{k}{\tilde K_\beta(s)}(\tilde x_\beta(s)-\tilde x_\alpha(s)) + \frac{1}{\tilde K_\beta(s)} \tilde{\xi}_{\beta} (s).
\end{eqnarray}
\end{subequations}
Subtracting Eq.~\eqref{eq:coupled_FLE_laplace_b} from Eq.~\eqref{eq:coupled_FLE_laplace_a}, we readily obtain the GLE for $r_{\alpha\beta}$ as
\begin{equation}
\tilde{\dot{r}}_{\alpha\beta}=-k \frac{\tilde K_\alpha(s)+\tilde K_\beta(s)}{\tilde K_\alpha(s)\tilde K_\beta(s)}
\tilde{r}_{\alpha\beta}+\frac{\tilde\xi_\alpha(s)}{\tilde K_\alpha(s)}-\frac{\tilde\xi_\beta(s)}{\tilde K_\beta(s)}.
\end{equation}
From this expression, the corresponding GLE in the time domain reads 
\begin{equation}\label{eq:GLE_r_ab}
    \int_0^t K^{\mathrm{(eff)}}_{\alpha\beta}(t-t') \dot{r}_{\alpha\beta}(t')
= -k r_{\alpha\beta}(t)+\eta _{\alpha\beta}(t).
\end{equation}
Thus, the relative motion can be understood such that a thermal particle governed by a fictional memory kernel
\begin{equation}\label{eq:kernel_rab}
K^{\mathrm{(eff)}}_{\alpha\beta}(t) = \gamma_{\alpha} t^{-\alpha} E_{\alpha,\beta}(-(t/\tau_{c})^{\beta-\alpha})
\end{equation}
diffuses in a confining harmonic potential of spring constant $k$. 
In this expression, $E_{a,b}(z)$ denotes the generalized Mittag-Leffler function  defined through its Laplace transform~\cite{hauboldMittagLefflerFunctionsTheir2011,mainardiWhyMittagLefflerFunction2020}:
\begin{equation}
    \mathcal{L}\qty[t^{b-1} E_{a,b}(\pm ct^a)](s) = \frac{s^{-b}}{1\mp cs^{-a}}
\end{equation}
for $\Re{a}>0$ and $\Re{b}>0$. 
The $\eta _{\alpha\beta}(t)$ is the noise term written in the Laplace domain as
\begin{equation}\label{eq:noise_rab}
    \tilde\eta_{\alpha\beta}(s) = \frac{\tilde K_\beta(s) \tilde{\xi}_{\alpha} (s)
-\tilde K_\alpha(s) \tilde{\xi}_{\beta} (s)}{\tilde K_\alpha(s)+\tilde K_\beta(s)},
\end{equation}
which satisfies the FDT of the second kind:
\begin{equation}\label{eq:FDT_rab}
\langle \eta_{\alpha\beta}(t)\eta_{\alpha\beta}(t')\rangle = k_B T K^{\mathrm{(eff)}}_{\alpha\beta}(|t-t'|)
\end{equation}
(see Appendix~\ref{sec:noise_correlation} for details).

Using the asymptotic expressions for the Mittag-Leffler function around $z=0$ and $z\to\infty$:
\begin{equation}
    E_{a,b}(z) = \sum_{n=0}^\infty \frac{z^n}{\Gamma( b+an)}=\sum_{n=1}^\infty \frac{z^{-n}}{\Gamma(b-an)},
\end{equation}
we find that the memory kernel exhibits two distinct power-law characteristics: 
\begin{equation}
\begin{aligned}
    K^{\mathrm{(eff)}}_{\alpha\beta}(t)\simeq\begin{cases}
K_{\alpha}(t)=\frac{\gamma_\alpha}{\Gamma(1-\alpha)}t^{-\alpha}, &  t\ll \tau_c,\\
K_{\beta}(t)=\frac{\gamma_\beta}{\Gamma(1-\beta)}t^{-\beta}, & t\gg\tau_c.
\end{cases}
\end{aligned}
\end{equation}
Here, $\tau_c$ is the cross-over time introduced in Eq.~\eqref{eq:taus}. This indicates that the relative displacement follows the motion of particle $\alpha$ for times shorter than $\tau_c$ and then smoothly crosses over to that of particle $\beta$ for times longer than $\tau_c$. 

\subsection{GLE of particles \texorpdfstring{$\alpha$ and $\beta$}{a and b}\label{sec:GLE_a_b}}

We now solve the coupled FLE \eqref{eq:coupled_FLE} and derive the equation of motion for $x_\alpha$ and $x_\beta$. Taking the Laplace transform of Eq.~\eqref{eq:coupled_FLE2} and using the identity $\tilde{x}(s)=\left(\tilde{\dot{x}}(s)+x(0)\right)/s$, we isolate $\tilde K_\beta (s)\tilde{\dot{x}} _\beta(s)$ to obtain
\begin{multline}
\label{eq:CFLE_dev1}
    \tilde K_\beta (s) \tilde{\dot{x}}_\beta(s) = -k(x_\beta(0)-x_\alpha(0)) \frac{\tilde K_\beta(s)}{s\tilde K_\beta(s) + k} \\
    + \frac{\tilde K_\beta(s)}{s\tilde K_\beta(s) + k} k \tilde{\dot{x}}_\alpha(s) + \frac{s \tilde K_\beta(s)}{s\tilde K_\beta(s) + k} \tilde \xi_\beta(s).
\end{multline}
Next, the sum of Eqs.~\eqref{eq:coupled_FLE1} and \eqref{eq:coupled_FLE2} yields
\begin{equation}
\label{eq:CFLE_dev2}
    \tilde K_\alpha(s) \tilde{\dot{x}}_\alpha(s) = - \tilde K_\beta (s) \tilde{\dot{x}}_\beta(s) + \tilde \xi _\alpha (s) + \tilde \xi _\beta (s).
\end{equation}
We plug the expression for $\tilde K_\beta (s) \tilde{\dot{x}}_\beta(s)$ from Eq.~\eqref{eq:CFLE_dev1} into Eq.~\eqref{eq:CFLE_dev2}, obtaining the effective GLE for $x_\alpha$:
\begin{multline}
\label{eq:CFLE_dev3}
    \int_0^t \left[K_\alpha(t') +\Phi_\beta(t')\right] \dot x_\alpha(t-t') dt' =  \\
    -kr_{\alpha\beta}(0) E_{\beta,1}(-(\tfrac{t}{\tau_\beta})^\beta)+
    \xi_\alpha(t)+\eta_\beta(t).
\end{multline}
Here, $\Phi_\beta(t)$ and $\eta_\beta(t)$ are the memory and noise terms resulting from the motion of particle $\beta$, which read in the Laplace domain
\begin{equation}\label{eq:transmitted_laplace}
    \tilde \Phi_\beta(s)=\frac{k \tilde K_\beta(s)}{s\tilde K_\beta(s)+k}, \quad
    \tilde \eta_\beta(s) = \frac{s\tilde \xi_\beta(s)}{s\tilde K_\beta(s)+k}.
\end{equation}
The memory kernel $\Phi_\beta(t)$ is identified to 
\begin{equation}
%\label{eq:Mittag-Leffler}
\begin{aligned}
    \Phi_\beta(t)&=k E_{\beta,1}\qty(-\qty(t/\tau_\beta)^\beta), 
\end{aligned}
\end{equation}
which interpolates between two limiting behaviors: 
\begin{equation}
\label{eq:Mittag-Leffler}
\begin{aligned}
\Phi_\beta(t)\simeq 
\begin{cases}
        k, & t\ll\tau_\beta,\\
        K_\beta(t)=\tfrac{\gamma_\beta}{\Gamma(1-\beta)}t^{-\beta}, & t\gg\tau_\beta.
\end{cases}
\end{aligned}
\end{equation}
Hence, the additional memory term $\Phi_\beta(t)$ captures the confinement effect in the potential for times shorter than $\tau_\beta$ and becomes the memory kernel $K_\beta(t)$ for particle $\beta$.
It is noted that Eq.~\eqref{eq:CFLE_dev3} is seen as an effective GLE for $x_\alpha$ defined with the effective memory kernel
\begin{equation}
\label{eq:memory_eff}
    K_\alpha^\mathrm{(eff)}(t) = \frac{\gamma_\alpha}{\Gamma(1-\alpha)} t^{-\alpha} + kE_{\beta,1}\qty(-(t/\tau_\beta)^\beta)
\end{equation}
and the effective noise 
\begin{equation}\label{eq:noise_eff_laplace}
     \xi_\alpha(t)+\eta_\beta(t)=\xi _\alpha^\mathrm{(eff)} (t).
\end{equation}
It can be shown that the effective GLE is a thermal equilibrium model such that at longer times ($t\gg\tau_\beta$) $\xi_\alpha^\mathrm{(eff)}$ satisfies the FDT of the second kind (Appendix~\ref{sec:noise_correlation}),
\begin{equation}\label{eq:FDT_aeff}
    \langle\xi_\alpha^\mathrm{(eff)}(t)\xi_\alpha^\mathrm{(eff)}(t')\rangle \simeq k_B T K_\alpha^\mathrm{(eff)}(|t-t'|).
\end{equation}
Interestingly, the effective GLE \eqref{eq:CFLE_dev3} for $x_\alpha$ has the extra term $kr_{\alpha\beta}(0) E_{\beta,1}\big(-(t/\tau_\beta)^\beta\big)$ attributed to the relaxation in the confining harmonic potential from the initial relative distance between the two particles. The effect of this initial condition decays out in a power-law manner and, in general, survives for long times. This term can be ignored if the system started from $t=-\infty$ or the system started at $t=0$ with the equilibrium (stationary) condition: $\langle r_{\alpha\beta}^2 (0)\rangle=\frac{k_BT}{k}$ and $\langle r_{\alpha\beta} (0)x_\alpha(0)\rangle=0$. See Appendix~\ref{sec:MSD_exact} for details. 
In this study, we focus on the coupled FLE dynamics under these equilibrium conditions, so the GLE~\eqref{eq:CFLE_dev3} is rewritten in the following form: 
\begin{equation}
\label{eq:GLE_eff} 
    \int_0^t K_\alpha^\mathrm{(eff)}(t-t') \dot{x}_\alpha(t')dt' = \xi_\alpha^\mathrm{(eff)}(t).
\end{equation}
We repeat the derivation for the effective GLE for $x_\beta$. These expressions are obtained by symmetrically exchanging all $\alpha$- and $\beta$- dependent terms in the steps above, resulting in
\begin{equation}\label{eq:kernel_laplace_b_exact}
\tilde K^\mathrm{(eff)}_\beta(s) = \tilde K_\beta(s) + \tilde \Phi_\alpha(s)    
\end{equation}
\begin{equation}
\tilde \xi^\mathrm{(eff)}_\beta(s) = \tilde \xi_\beta(s) + \tilde \eta_\alpha(s)
\end{equation}
where $\tilde \Phi_\alpha(s)$ and $\tilde \eta_\alpha(s)$ are given by
\begin{equation}\label{eq:transmitted_laplace2}
    \tilde \Phi_\alpha(s)=\frac{k \tilde K_\alpha(s)}{s\tilde K_\alpha(s)+k}, \quad
    \tilde \eta_\alpha(s) = \frac{s\tilde \xi_\alpha(s)}{s\tilde K_\alpha(s)+k}.
\end{equation}
Although the dynamics of particles $\alpha$ and $\beta$ are separated using their respective GLEs, their motions remain correlated because the effective noise terms $\xi_\alpha^\mathrm{(eff)}$ and $\xi_\beta^\mathrm{(eff)}$ themselves are correlated. Consequently, their relative displacement $r_{\alpha\beta}$ exhibits the confined dynamics in the harmonic potential, as described by Eq.~\eqref{eq:GLE_r_ab}.

\begin{figure}
\includegraphics[width=0.9\columnwidth]{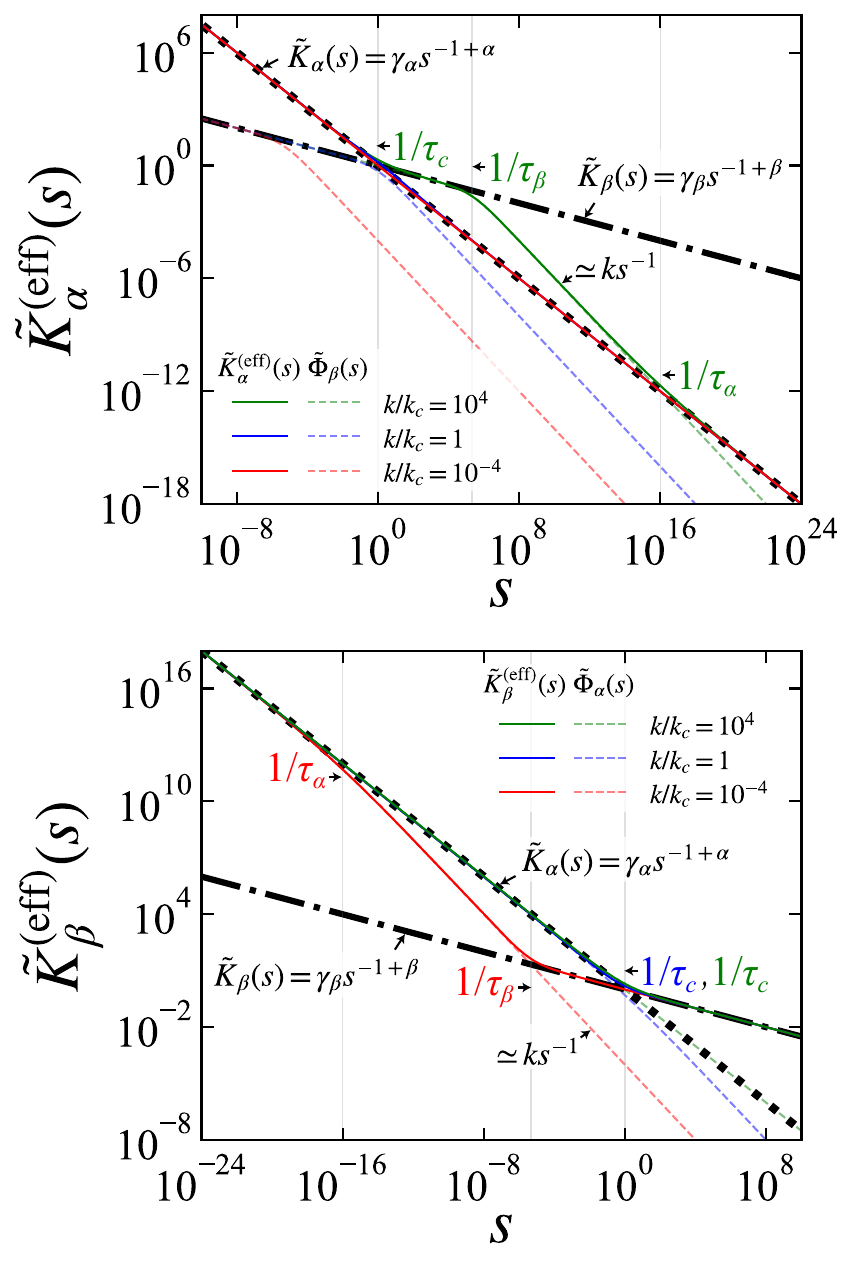}
\caption{\label{fig2}
Effective memory kernels in the Laplace domain: $\tilde K_\alpha^\mathrm{(eff)}(s)$ [Eq.~\eqref{eq:memory_eff}, upper panel] and $\tilde K_\beta^\mathrm{(eff)}(s)$ [Eq.~\eqref{eq:kernel_laplace_b_exact}, lower panel]. In both panels, the dotted and dotted-dashed curves depict $\tilde K_\alpha(s)$ and $\tilde K_\beta(s)$, respectively, while the dashed lines represent $\tilde \Phi_\alpha(s)$ and $\tilde \Phi_\beta(s)$ from Eqs.~\eqref{eq:transmitted_laplace} and \eqref{eq:transmitted_laplace2}. 
Gray vertical lines indicate $1/\tau_\alpha$, $1/\tau_\beta$, and $1/\tau_c$ shown in Eqs.~\eqref{eq:kernel_laplace_a} and \eqref{eq:kernel_laplace_b}. We set $\alpha=0.25$, $\beta=0.75$, $\gamma_\alpha=\gamma_\beta=1$, and varied $k$, so $\tau_c$ remains the same for all cases.
}
\end{figure}

In Fig.~\ref{fig2}, we analyze the asymptotic behavior of the memory kernels in the Laplace domain. In the strong interaction regime ($k>k_c$) with the ordering of $\tau_\alpha<\tau_\beta<\tau_c$, we find that the kernel $\tilde K_\alpha^\mathrm{(eff)}(s)$ scales as:
\begin{subnumcases}{
\label{eq:kernel_laplace_a}
\tilde K_\alpha^\mathrm{(eff)}(s)\simeq}
    \gamma_{\alpha} s^{-1+\alpha}, &\quad 
    $s\ll1/\tau_c$,\label{eq:kernel_laplace_a_1}\\
    \gamma_\beta s^{-1+\beta}, &\quad  $1/\tau_c\ll s \ll1/\tau_\beta$, \label{eq:kernel_laplace_a_2}\\
    ks^{-1}, &\quad  $1/\tau_\beta\ll s \ll1/\tau_\alpha$,\label{eq:kernel_laplace_a_3}\\
    \gamma_{\alpha} s^{-1+\alpha}, &\quad 
    $s\gg 1/\tau_\alpha $.\label{eq:kernel_laplace_a_4}
\end{subnumcases}
See the theoretical curve of $\tilde K_\alpha^\mathrm{(eff)}(s)$ for $k/k_c=10^{4}$ in Fig.~\ref{fig2}. The two scaling relations \eqref{eq:kernel_laplace_a_2} and \eqref{eq:kernel_laplace_a_3} emerge only in the strong interaction regime. 
In this case, the memory term $\tilde \Phi_\beta(s)$ dominates over $\tilde K_\alpha(s)$, producing the intermediate asymptotic regimes.
For the opposite conditions ($k\lesssim k_c$), simply $\tilde K_\alpha^\mathrm{(eff)}(s)\simeq\tilde K_\alpha(s)$ across all time scales (see $k/k_c=10^{-4}$ and $k/k_c=1$).

In contrast to $\tilde K_\alpha^\mathrm{(eff)}(s)$, the effective kernel $\tilde K_\beta^\mathrm{(eff)}(s)$ exhibits the full complicated scaling relations in the weak interaction regime ($k<k_c$), which reads:
\begin{subnumcases}{
\label{eq:kernel_laplace_b}
\tilde K_\beta^\mathrm{(eff)}(s)\simeq}
    \gamma_{\alpha} s^{-1+\alpha}, &\quad 
    $s\ll1/\tau_\alpha$,\label{eq:kernel_laplace_b_1}\\
    ks^{-1}, &\quad  $1/\tau_\alpha\ll s \ll1/\tau_\beta$, \label{eq:kernel_laplace_b_2}\\
    \gamma_\beta s^{-1+\beta}, &\quad  $s \gg1/\tau_\beta$\label{eq:kernel_laplace_b_3}.
\end{subnumcases}
The intermediate scaling relation~\eqref{eq:kernel_laplace_b_2} appears only when $\tau_c<\tau_\beta<\tau_\alpha$ (i.e., the weak interaction regime). Also note that the long-time (small $s\ll1/\tau_c$, $s\ll 1/\tau_\alpha$) behavior of the kernel $K_\beta^\mathrm{(eff)}(t)$ follows that of $K_\alpha(t)$, manifesting the impact of the slower particle $\alpha$. In the strong interaction regime, the scaling of $\tilde K_\beta^\mathrm{(eff)}(s)$ is simpler such that $\tilde K_\beta^\mathrm{(eff)}(s)$ is $\tilde K_\alpha(s)$ for $s<1/\tau_c$ and $\tilde K_\beta(s)$ for $s>1/\tau_c$.

\section{Mean square displacements}
\label{sec:MSD}
In the previous section, we show that the dynamics of the relative displacement can be understood as that of a non-Markovian particle in a harmonic potential, described by the GLE~\eqref{eq:GLE_r_ab} and that the individual particles diffuse as if in the free space governed by the effective memory kernel and noise--described in GLE~\eqref{eq:GLE_eff}--that incorporate the effect of the harmonic interaction between the two particles. 
The mean square displacements (MSDs) of these particles can be evaluated using the expressions of the MSDs for a GLE with a given memory kernel $K(t)$ and noise $\xi(t)$ subject to a confining harmonic potential or in free space.   
For the GLE in the presence of a harmonic potential $\frac{1}{2}k x^2$, the corresponding MSD is given by
\begin{equation}\label{eq:MSD_general_confined}
\begin{aligned}
\langle \Delta x^2(t)\rangle &= \langle  \left[x(t+t_0)-x(t_0) \right]^2\rangle\\
    &= 2k_B T \mathcal{L}^{-1} \left\{ \tfrac{1}{s\left(s\tilde K(s)+k\right)} \right\} (t)
\end{aligned}
\end{equation}
at the stationary state (see the derivation in Appendix~\ref{sec:GLE_MSD}). From the MSD, we define its (instantaneous) anomalous exponent $\nu(t)$ as the slope of $\langle \Delta x^2(t)\rangle $ in the log-log coordinate:
\begin{equation}\label{eq:anomaly_exponent_confined}
    \nu(t) = \frac{d \log{\expval{\Delta x^2(t)}}}{d \log{t}} = \frac{t \mathcal{L}^{-1} \qty{\tfrac{1}{s\tilde K(s)+k}}(t)}{\mathcal{L}^{-1} \qty{\tfrac{1}{s\left(s\tilde K(s)+k\right)}}(t)}.
\end{equation}
For a GLE in free space, the MSD of a particle is obtained as
\begin{equation}
\label{eq:MSD_general}
    \langle\Delta x^2(t) \rangle = 2k_B T \mathcal{L}^{-1} \qty{\frac{1}{s^2 \tilde K(s)}}(t)
\end{equation}
and the respective anomalous exponent is identified to  
\begin{equation}
\label{eq:anomaly_exponent}
    \nu(t) = \frac{t \mathcal{L}^{-1} \qty{\frac{1}{s \tilde K(s)}}(t)}{\mathcal{L}^{-1} \qty{\frac{1}{s^2 \tilde K(s)}}(t)}.
\end{equation}
The derivation of these results is explained in Appendix~\ref{sec:GLE_MSD}. 
We numerically compute the inverse Laplace transform for these analytic expressions using the Gaver--Stehfest algorithm~\cite{stehfestAlgorithm368Numerical1970}.

\subsection{The MSD for the relative displacement}
\begin{figure}
\includegraphics[width=0.9\columnwidth]{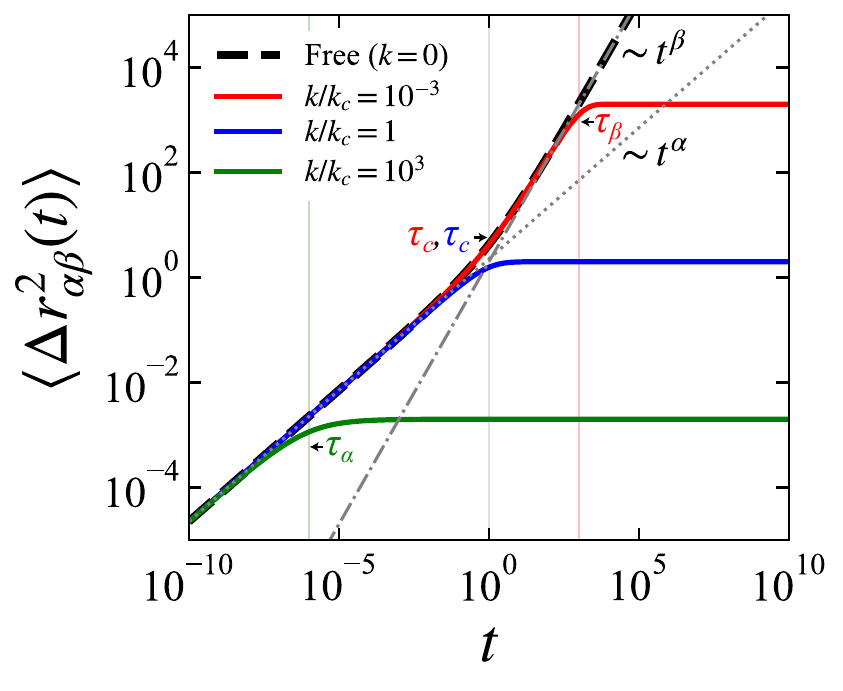}
\caption{\label{fig3}
The MSD of the relative displacement $ r_{\alpha\beta} $. Solid lines depict the MSDs computed from Eqs.~\eqref{eq:kernel_rab} and \eqref{eq:MSD_general_confined} for different values of $ k $, while the dashed line represents the uncoupled case ($ k = 0 $) [Eq.~\eqref{eq:MSD_rab_free}]. The dotted and dot-dashed lines correspond to the free subdiffusion [Eq.~\eqref{eq:MSD_k=0}] governed by $ K_\alpha(t) $ and $ K_\beta(t) $, respectively. Gray vertical lines indicate $\tau_\alpha$, $\tau_\beta$, and $\tau_c$ shown in Eqs.~\eqref{eq:MSD_rab_asymp_weak} and \eqref{eq:MSD_rab_asymp_str}. We used the parameters, $ \alpha=0.25 $, $ \beta=0.75 $, and $ \gamma_\alpha=\gamma_\beta=1 $, with $k$ varied such that $ \tau_c $ remains unchanged for all cases.
}
\end{figure}

Employing Eq.~\eqref{eq:MSD_general_confined} with plugging into $K(t)=K^\mathrm{(eff)}_{\alpha\beta}(t)$, we obtain the expression for the MSD of the relative displacement $r_{\alpha\beta}(t)$. As a baseline, we first examine the limiting case that the confining force becomes zero (i.e., $k\to0$). In this case, the MSD exhibits the time-dependence as
\begin{equation}
\label{eq:MSD_rab_free}
\begin{aligned}
\langle \Delta r_{\alpha\beta}^2(t)\rangle_{k=0}
&= \frac{2 k_B T}{\gamma_\alpha \,\Gamma(1+\alpha)}\,t^\alpha
 + \frac{2 k_B T}{\gamma_\beta \,\Gamma(1+\beta)}\,t^\beta \\[4pt]
&\simeq
\begin{cases}
   \displaystyle \frac{2 k_B T}{\gamma_\alpha \,\Gamma(1+\alpha)}\,t^\alpha,
   & t \ll \tau_c,\\
   \displaystyle \frac{2 k_B T}{\gamma_\beta \,\Gamma(1+\beta)}\,t^\beta,
   & t \gg \tau_c.
\end{cases}
\end{aligned}
\end{equation}
As expected, the MSD is dominated by that of the slower (stronger subdiffusive) particle ($\alpha$) for times shorter than the cross-over time $\tau_c$ and that of the faster one ($\beta$) beyond $\tau_c$.  See the numerical plot for $k=0$ in Fig.~\ref{fig3}.

When a harmonic potential is present, the relative displacement becomes confined at long times. 
In the weak interaction regime ($k < k_c$), the relative displacement displays the MSD of the following form (e.g., the case of $k/k_c=10^{-3}$ in Fig.~\ref{fig3}): 
\begin{subnumcases}{\label{eq:MSD_rab_asymp_weak}
\langle \Delta r_{\alpha\beta}^2(t)\rangle \simeq}
   \displaystyle \frac{2 k_B T}{\gamma_\alpha \,\Gamma(1+\alpha)}\,t^\alpha,
   & $t \ll \tau_c$, \label{eq:MSD_rab_asymp_weak1}\\[4pt]
   \displaystyle \frac{2 k_B T}{\gamma_\beta \,\Gamma(1+\beta)}\,t^\beta,
   & $\tau_c \ll t \ll \tau_\beta$, \label{eq:MSD_rab_asymp_weak2}\\[4pt]
   \displaystyle \frac{2 k_B T}{k},
   & $t \gg \tau_\beta$. \label{eq:MSD_rab_asymp_weak3}
\end{subnumcases}
For $t\ll \tau_\beta$,  the confining harmonic potential effects negligibly. Thus, the MSD has the same forms as the above MSD~\eqref{eq:MSD_rab_free} for $k=0$. For $t\gg \tau_\beta$, however, the confining effect becomes significant, and the relative displacement ends up with thermal saturation in the harmonic potential.  

In the critical and the strong interaction regimes (where $\tau_\alpha \lesssim \tau_\beta \lesssim \tau_c$), the MSD is simpler. In this regime, the harmonic force is strong such that the MSD of the relative displacement grows as $\sim t^\alpha$ at short times and then becomes saturated for $t>\tau_\alpha$. Thus, the MSD displays a two-stage behavior as follows:
\begin{subnumcases}{
\label{eq:MSD_rab_asymp_str}
\langle \Delta r_{\alpha\beta}^2(t)\rangle \simeq}
   \displaystyle \frac{2 k_B T}{\gamma_\alpha \,\Gamma(1+\alpha)}\,t^\alpha,
   & $t \ll \tau_\alpha$, \label{eq:MSD_rab_asymp_str1}\\[4pt]
   \displaystyle \frac{2 k_B T}{k},
   & $t \gg \tau_\alpha$. 
%\label{eq:MSD_rab_asymp_str2}
\end{subnumcases}
The numerical plots for $k/k_c=1$ and $k/k_c=10^3$ confirm these relations [Fig.~\ref{fig3}].

\subsection{The MSD for particles \texorpdfstring{$\alpha$ and $\beta$}{alpha and beta}\label{sec:MSD_of_each_particle}}
Having analyzed the relative displacement, we now investigate the MSDs of the individual particles $x_\alpha(t)$ and $x_\beta(t)$. By substituting the effective memory kernels $K(t)=K_\mu^\mathrm{(eff)}(t)$ from Eqs.~\eqref{eq:memory_eff} \& \eqref{eq:kernel_laplace_b_exact} into Eq.~\eqref{eq:MSD_general}, we can evaluate the MSDs for each particle.

The asymptotic form of each particle’s MSD depends on whether it reaches the confinement-induced plateau $2k_BT/k$ before or after its counterpart. As mentioned in Eq.~\eqref{eq:taus_dynamics}, $\langle \Delta x_\mu^2(t)\rangle$ reaches $2k_B T/k$ at $t\approx \tau_\mu$ ($\mu=\alpha$, $\beta$), and the order of timescales $\tau_\alpha$ and $\tau_\beta$ [Eq.~\eqref{eq:taus}] depend on $k/k_c$. We now present our results in three parts: the weak ($k < k_c$), critical ($k=k_c$), and strong ($k > k_c$) interactions.

\begin{figure}
\includegraphics[width=0.9\columnwidth]{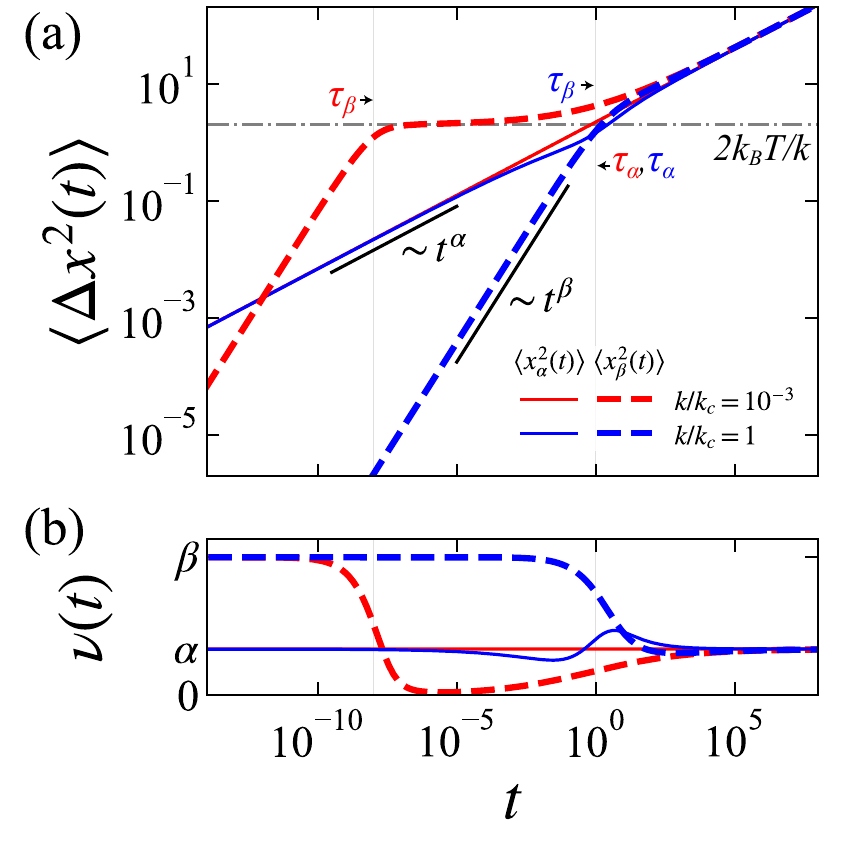}
\caption{\label{fig4}
The MSDs and the anomalous exponents in the weak interaction regime ($k<k_c$).
(a) MSDs \eqref{eq:MSD_general} for $x_\alpha(t)$ (solid) and $x_\beta(t)$ (dashed) as a function of time $t$ for two distinct values of $k/k_c$. 
(b) The time-dependent anomalous exponent \eqref{eq:anomaly_exponent} derived from the MSDs. Gray vertical lines indicate $\tau_\alpha$ and $\tau_\beta$. 
We set $\alpha=0.25$, $\beta=0.75$, $k=1$, $\gamma_\alpha=1$, and $k_B T=1$ and varied $\gamma_\beta$ while fixing $\tau_\alpha$ for all cases.
}
\end{figure}

In the weak-interaction regime ($k < k_c$ \& $\tau_\beta < \tau_\alpha$), the MSD of particle $\beta$ reaches the thermal plateau $2k_B T/k$ at $t\approx\tau_\beta$, while particle $\alpha$ is (sub)diffusing with negligible confinement effect. The particle $\beta$ diffuses in the following way: For $t\ll \tau_\beta$, it moves as if in free space with the MSD of $\langle \Delta x_\beta^2(t)\rangle\approx \langle \Delta x_\beta^2(t)\rangle_{k=0}$. Beyond this timescale, the particle's movement is affected by the harmonic potential, and eventually it reaches the confined state with $\langle \Delta x_\beta^2(t)\rangle\approx 2 k_B T/k$ for $\tau_\beta \ll t \ll \tau_\alpha$. For $t \gg \tau_\alpha$, it follows the motion of the slower particle ($\alpha$). We note that for $t\ll\tau_\alpha$, the MSD  of particle $\beta$ is identical with that of a single FLE particle confined to a harmonic potential~\cite{jeonInequivalenceTimeEnsemble2012}:
\begin{equation}
\label{eq:FLE_conf}
    \langle\Delta x_\beta^2(t)\rangle
    \simeq
    2\,k_B T\,\frac{t^\beta}{\gamma_\beta}\,
    E_{\beta,\,1+\beta}
    \Bigl(-\tfrac{k}{\gamma_\beta} t^\beta\Bigr).
\end{equation}
For particle $\alpha$, the confining effect is negligible at short times ($t\ll\tau_\alpha$), and its MSD follows that of free subdiffusion, $\langle \Delta x^2_\alpha(t)\rangle\simeq\langle\Delta x^2_\alpha(t)\rangle_{k=0}\propto t^\alpha$. 
After then, the MSD of particle $\alpha$ approaches $2k_B T/k$ without confinement, and the two particles move together through the harmonic coupling. Hence, their MSDs share the same asymptotic form ($t\gg \tau_\alpha$): 
\begin{equation}
    \langle \Delta x_\alpha^2(t)\rangle\simeq\langle \Delta x_\beta^2(t)\rangle \simeq \frac{2k_B T}{\gamma_\alpha\Gamma(1+\alpha)} t^\alpha.
\end{equation}
In this regime, both effective kernels $K_\alpha^\mathrm{(eff)}(t)$ and $K_\beta^\mathrm{(eff)}(t)$ are approximated by $K_\alpha(t)$ [See Eqs.~\eqref{eq:kernel_laplace_a}~and~\eqref{eq:kernel_laplace_b}].

Summing up, particle $\alpha$ maintains the power-law scaling $\sim t^\alpha$ over the entire time domain, with a negligible intermediate cross-over. 
In contrast, particle $\beta$ undergoes a three-stage evolution ($k/k_c=10^{-3}$ in Fig.~\ref{fig4}):
\begin{subnumcases}{\label{eq:MSD_weak_beta_asymps}
\langle \Delta x_\beta^2(t)\rangle \;\simeq\;}
    \displaystyle \frac{2\,k_B T}{\gamma_\beta \,\Gamma(1+\beta)}\,t^\beta,
    & $t\ll\tau_\beta$,\label{eq:MSD_weak_beta_asymp1}\\[6pt]
    \displaystyle \frac{2\,k_B T}{k},
    & $\tau_\beta\ll t \ll\tau_\alpha$,\label{eq:MSD_weak_beta_asymp2}\\[6pt]
    \displaystyle \frac{2\,k_B T}{\gamma_\alpha \,\Gamma(1+\alpha)}\,t^\alpha,
    & $t \gg \tau_\alpha$.\label{eq:MSD_weak_beta_asymp3}
\end{subnumcases}
The cross-over from faster to slower subdiffusion highlights how harmonic interaction induces a delayed coherence in the coupled FLE system. 

At the critical interaction strength ($k=k_c$), the three characteristic timescales become identical, i.e., $\tau_\alpha=\tau_\beta=\tau_c$. In this case, both particles terminate their free subdiffusion regime simultaneously at $t\approx\tau_c$, and immediately enter the long-time regime with $\langle \Delta x_\alpha^2(t)\rangle\simeq\langle \Delta x_\beta^2(t)\rangle \simeq \frac{2k_B T}{\gamma_\alpha\Gamma(1+\alpha)} t^\alpha$. There is no intermediate plateau in $\langle \Delta x_\beta^2(t)\rangle$, unlike in the weak interaction case. The resulting asymptotic expressions are:
\begin{equation}
\label{eq:MSD_crit_alpha_asymp}
    \langle \Delta x_\alpha^2(t)\rangle
    \;\simeq\;
    \frac{2\,k_B T}{\gamma_\alpha \,\Gamma(1+\alpha)}\,t^\alpha,
\end{equation}
and
\begin{subnumcases}{\label{eq:MSD_crit_beta_asymps}
\langle \Delta x_\beta^2(t)\rangle \;\simeq\;}
    \displaystyle \frac{2\,k_B T}{\gamma_\beta \,\Gamma(1+\beta)}\,t^\beta,
    & $t\ll\tau_c$,\label{eq:MSD_crit_beta_asymp1}\\[6pt]
    \displaystyle \frac{2\,k_B T}{\gamma_\alpha \,\Gamma(1+\alpha)}\,t^\alpha,
    & $t \gg \tau_c$.\label{eq:MSD_crit_beta_asymp2}
\end{subnumcases}
These behaviors are illustrated in the numerical plot for $k/k_c=1$ in Fig.~\ref{fig4}.

\begin{figure}[t!]
\includegraphics[width=0.9\columnwidth]{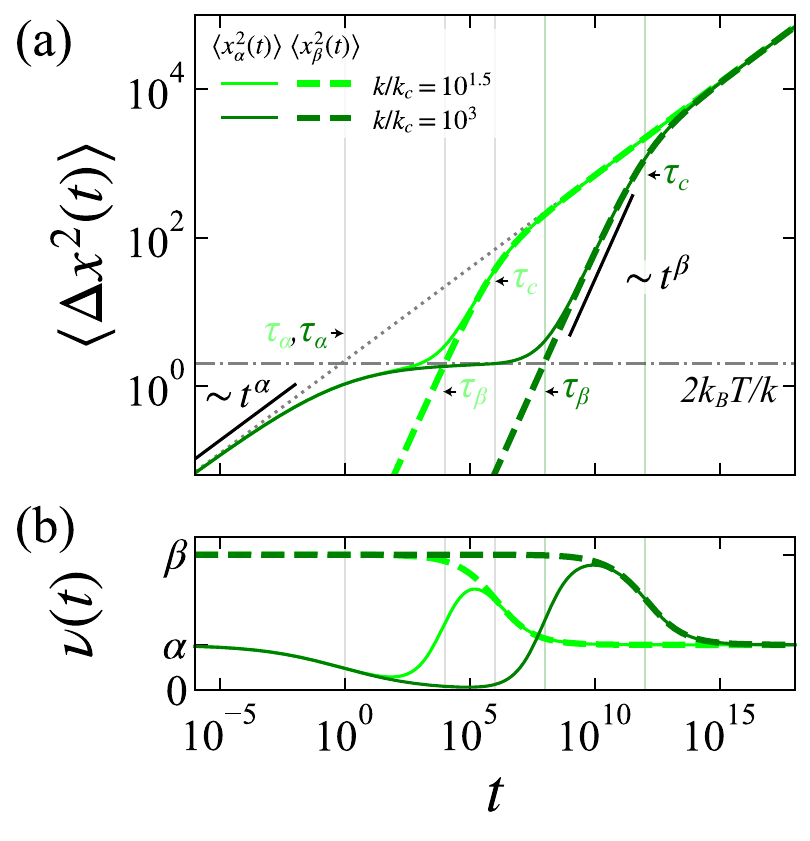}
\caption{\label{fig5}    
 The MSDs and the anomalous exponents in the strong interaction regime ($k>k_c$). 
(a) MSDs~\eqref{eq:MSD_general} for $x_\alpha(t)$ (solid) and $x_\beta(t)$ (dashed) as a function of time $t$ for different $k/k_c$. 
(b) The anomalous exponent \eqref{eq:anomaly_exponent}. Gray vertical lines represent $\tau_\alpha$, $\tau_\beta$, and $\tau_c$ shown in Eqs.~\eqref{eq:MSD_strong_alpha_asymps} and \eqref{eq:MSD_strong_beta_asymps}.
We set $\alpha=0.25$, $\beta=0.75$, $k=1$, $\gamma_\alpha=1$, $k_B T=1$, and varied $\gamma_\beta$ while fixing $\tau_\alpha$ for all cases.
}
\end{figure}

In the strong-interaction regime ($k>k_c$ \& $\tau_\alpha < \tau_\beta < \tau_c$), Here, the confinement becomes dominant early in the dynamics. The MSD of particle $\alpha$ exhibits a four-regime behavior (Fig.~\ref{fig5}):
\begin{subnumcases}{
\label{eq:MSD_strong_alpha_asymps}
\langle \Delta x_\alpha^2(t)\rangle \,\simeq}
    \frac{2\,k_B T}{\gamma_\alpha \,\Gamma(1+\alpha)}\,t^\alpha, 
    & $t\ll\tau_\alpha$,\label{eq:MSD_strong_alpha_asymp1}\\
    \frac{2\,k_B T}{k}, 
    & $\tau_\alpha\ll t \ll\tau_\beta$,\label{eq:MSD_strong_alpha_asymp2}\\
    \frac{2\,k_B T}{\gamma_\beta \,\Gamma(1+\beta)}\,t^\beta, 
    & $\tau_\beta \ll t \ll \tau_c$,\label{eq:MSD_strong_alpha_asymp3}\\
    \frac{2\,k_B T}{\gamma_\alpha \,\Gamma(1+\alpha)}\,t^\alpha, 
    & $t \gg \tau_c$,\label{eq:MSD_strong_alpha_asymp4}
\end{subnumcases}
and particle $\beta$ has
\begin{subnumcases}{
\label{eq:MSD_strong_beta_asymps}
\langle \Delta x_\beta^2(t)\rangle\;\simeq}
    \frac{2\,k_B T}{\gamma_\beta \,\Gamma(1+\beta)}\,t^\beta, 
    & $t\ll\tau_c$,\label{eq:MSD_strong_beta_asymps1}\\
    \frac{2\,k_B T}{\gamma_\alpha \,\Gamma(1+\alpha)}\,t^\alpha, 
    & $t \gg \tau_c$.\label{eq:MSD_strong_beta_asymps2}
\end{subnumcases}
Distinguished from the weak and critical regimes, the long-time MSD displays a two-step scaling relation for $t\gg \tau_\beta$. In this regime, the MSD of both particles can be obtained using a different approach, directly from the asymptotic in the Laplace domain:  
\begin{equation}
\label{eq:MSD_strong_alpha_rec}
\begin{aligned}
\langle \Delta x_\alpha^2(t)\rangle 
&\simeq \langle \Delta x_\beta^2(t)\rangle\\
&\simeq 2\,k_B T\,\frac{t^\beta}{\gamma_\beta}\,
    E_{\beta-\alpha,\,1+\beta}
    \Bigl(-\bigl(t/\tau_c\bigr)^{\beta-\alpha}\Bigr)\\[4pt]
&\simeq
\begin{cases}
   \displaystyle \frac{2 k_B T}{\gamma_\beta \,\Gamma(1+\beta)}\,t^\beta,
   & t \ll \tau_c,\\
   \displaystyle \frac{2 k_B T}{\gamma_\alpha \,\Gamma(1+\alpha)}\,t^\alpha,
   & t \gg \tau_c.
\end{cases}
\end{aligned}
\end{equation}

A striking consequence of Eq.~\eqref{eq:MSD_strong_alpha_rec} is that particle $\alpha$, despite being intrinsically slower, undergoes transient acceleration. Specifically, it exhibits an effective power-law scaling $\sim t^\beta$ ($\beta>\alpha$) during the intermediate time window $\tau_\beta \ll  \tau_c$, albeit its native subdiffusion is $\sim t^\alpha$. We refer to this post-confinement acceleration as \emph{recovery dynamics} because it enables $\langle \Delta x_\alpha^2(t)\rangle$ to approach $\langle\Delta x_\alpha^2(t)\rangle_{k=0}$ in the long-time limit. Note that this recovery dynamics only occur for $k>k_c$.

As a result of this recovery dynamics, particle $\alpha$ ultimately regains the same generalized diffusivity it exhibits in the uncoupled limit ($k=0$), regardless of the interaction strength $k$. Defining short- and long-time generalized diffusivities as 
\begin{equation}
D_{\alpha}^{(0)}
\;\equiv\;
\lim_{t\to0}
\frac{\langle \Delta x_\alpha^2(t)\rangle}{2\,t^\alpha}
\;\;\text{and}\;\;
D_{\alpha}^{(\infty)}
\;\equiv\;
\lim_{t\to\infty}
\frac{\langle \Delta x_\alpha^2(t)\rangle}{2\,t^\alpha},
\end{equation}
we find that they are equal to each other:
\begin{eqnarray}
\label{eq:D_alpha_recovery}
D_{\alpha}^{(0)}
\;=\;
D_{\alpha}^{(\infty)}
\;=\;
\frac{2\,k_B T}{\gamma_\alpha\,\Gamma(1+\alpha)}.
\end{eqnarray}
This indicates that particle $\alpha$ eventually recovers the same long-time behavior as in free space. (This long-time diffusivity is revisited in Sec.~\ref{sec:alpha_eq_beta}.) 

\begin{figure}
\includegraphics[width=0.9\columnwidth]{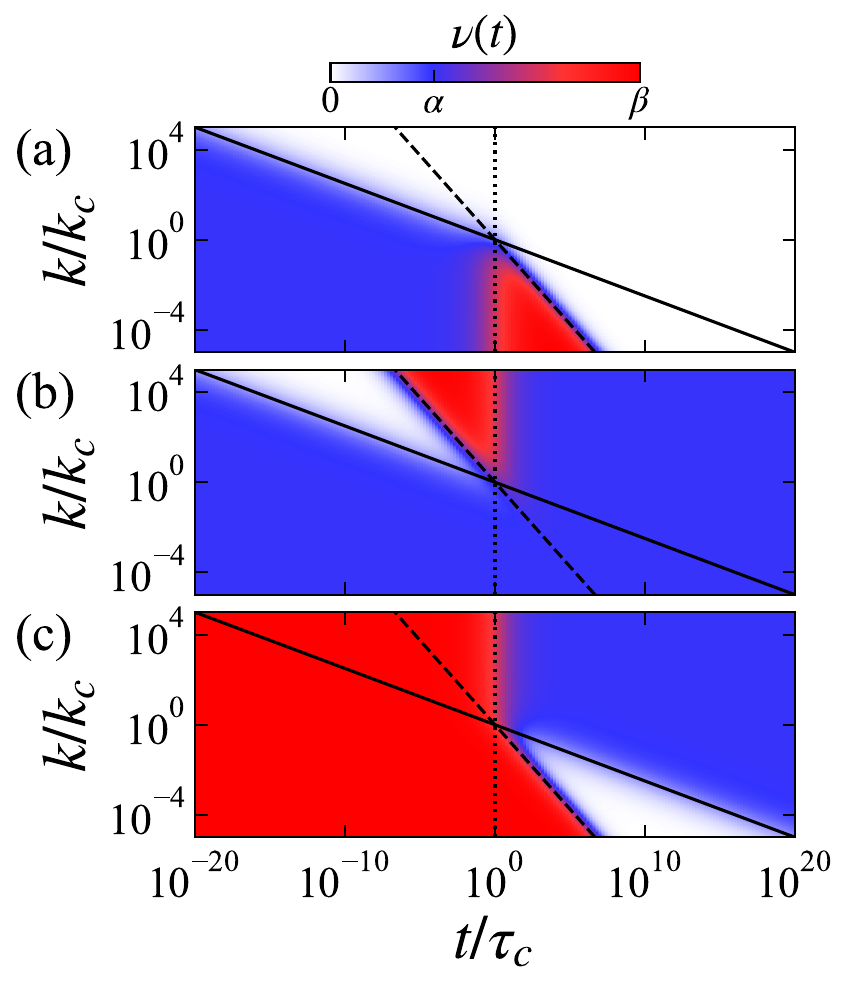}
\caption{\label{fig6}
Heatmap of the anomalous exponent $\nu(t)$ for the relative displacement (a) and the two individual particles (b) \& (c), plotted as functions of the dimensionless time $t/\tau_c$ and the interaction strength $k/k_c$. 
The color scale encodes the anomalous exponents: 0 (white), $\alpha$ (blue), and $\beta$ (red), 
while the solid and dashed diagonal lines mark $t = \tau_\alpha$ and $t = \tau_\beta$. The vertical dotted line indicates $t = \tau_c$. We used $\alpha=0.25$,  $\beta=0.75$ in the plot.
}
\end{figure}

To consolidate our MSD findings, we present in Fig.~\ref{fig6} a heatmap of the anomalous exponent $\nu(t)$ as a function of dimensionless time $t/\tau_c$ (horizontal axis) and interaction strength $k/k_c$ (vertical axis). Each horizontal slice of the heatmap represents $\nu(t)$ for a fixed value of $k/k_c$.

Panel~(a) shows $\nu(t)$ for the relative displacement $r_{\alpha\beta}$. Since $\langle \Delta r_{\alpha\beta}^2(t)\rangle$ eventually saturates at $\sqrt{2\,k_B T/k}$, the corresponding $\nu(t)$ always decays to zero at long times for any condition of $k/k_c$. Notably, a clear $t^\beta$ power-law region (highlighted in red) is present only when $k < k_c$. This fast dynamics emerges in this regime because $\langle \Delta r_{\alpha\beta}^2(t)\rangle\simeq\langle \Delta x_{\alpha}^2(t)\rangle+\langle \Delta x_{\beta}^2(t)\rangle$ is dominated by the MSD of particle $\beta$. 

Panel~(b) displays $\nu(t)$ for particle $\alpha$, revealing the hallmark features of \emph{recovery dynamics}. For $k > k_c$, particle $\alpha$ experiences a transient confinement in the intermediate regime, followed by the recovery regime where the particle moves with a larger anomalous exponent $\nu(t) \approx \beta > \alpha$. This transient acceleration reflects the post-confinement catch-up process, after which $x_\alpha$ smoothly regains its native subdiffusive dynamics ($t^\alpha$) and the long-time generalized diffusivity.

Panel~(c) shows $\nu(t)$ for particle $\beta$. In the strong-interaction regime, it exhibits subdiffusion of $\sim t^\beta$ at short times and later follow the diffusion ($\sim t^\alpha$) of the slower particle. When the interaction is weak ($k<k_c$), particle $\beta$ has the intermediate regime of transient confinement between the short-time ($\sim t^\beta$) and long-time ($\sim t^\alpha$) subdiffusion.

Altogether, the heatmaps illustrate a rich landscape of \emph{transient anomalous diffusion} dynamics for the coupled FLE system. The intricate effects from the viscoelastic memories ($\alpha$, $\beta$) and their coupling through the harmonic interaction lead to such complex dynamic behaviors, including the absence/emergence of transient confinement and an apparent speedup within a certain time window via the recovery process.

\subsection{Coupled dynamics with identical memory exponents\label{sec:alpha_eq_beta}}
Up to now, we have analyzed the coupled system with distinct viscoelastic memories ($\alpha\neq\beta$), where the intricate effects from the distinct viscoelastic memories and the mechanical coupling result in a hierarchy of the system's characteristic timescales and rich transient anomalous dynamics. Here, we consider the special case that the two memory exponents are equal to each other ($\alpha=\beta$).

Now let us consider two particles (particle 1 and 2) whose positions are $x_{\alpha,1}(t)$ and $x_{\alpha,2}(t)$, respectively. We assume that their memory kernels share the same power-law exponent $\alpha$ but differ in amplitude, i.e., 
\begin{equation}
    K_{\alpha,1}(t) = \frac{\gamma_{\alpha}}{\Gamma(1-\alpha)}\,t^{-\alpha}, 
    \quad
    K_{\alpha,2}(t) = \lambda K_{\alpha,1}(t), \quad \lambda\geq1.
\end{equation}
Because both particles have the same MSD scaling law, among the three timescales in Eq.~\eqref{eq:taus}, $\tau_c$ diverges in the limit of $\beta\to\alpha$. However, the other two characteristic timescales remain,
\begin{equation}
    \tau_{\alpha,1}=\left(\frac{\gamma_{\alpha}}{k}\right)^{1/\alpha}, \quad
    \tau_{\alpha,2}=\left(\lambda\frac{\gamma_{\alpha}}{k}\right)^{1/\alpha}
\end{equation}
which satisfy the fixed ordering in magnitude and the ratio $\tau_{\alpha,2}/\tau_{\alpha,1}=\lambda^{1/\alpha}>1$ remains the same independently of $k$. Hence, when $\alpha=\beta$, there is no critical strength $k_c$, and the qualitative shape of MSD is identical regardless of $k$.

For the relative displacement $r_{\alpha,12}(t)\equiv x_{\alpha,1}(t)-x_{\alpha,2}(t)$, the effective memory kernel for GLE \eqref{eq:GLE_r_ab} is given as a single power-law,
\begin{equation}
    K^\mathrm{(eff)}_{\alpha,12}(t)=\frac{\lambda}{1+\lambda}\frac{\gamma_\alpha}{\Gamma(1-\alpha)}t^{-\alpha},
\end{equation}
leading to
\begin{equation}
    \langle \Delta r_{\alpha,12}^2(t)\rangle \simeq
    \begin{cases}
        \frac{2k_B T}{\gamma_\alpha\Gamma(1+\alpha)} (1+\frac{1}{\lambda}) t^\alpha,  &  t\ll\tau_{\alpha,12},\\
        \frac{2 k_B T}{k}, & t\gg \tau_{\alpha,12}
    \end{cases}
\end{equation}
where $\tau_{\alpha,12}=[\lambda\gamma_\alpha/(1+\lambda)k]^{1/\alpha}$. The MSD crosses directly from the subdiffusion $\sim t^\alpha$ to the plateau $2k_B T/k$. No intermediate, $k$-dependent transient dynamics appears, which is in sharp contrast with the unequal-exponent case ($\alpha\neq\beta$).

For $x_{\alpha,1}(t)$ and $x_{\alpha,2}(t)$, their effective memory kernels for GLE \eqref{eq:GLE_eff} take the form:
\begin{equation}
\begin{aligned}
    K_{\alpha,1}^\mathrm{(eff)}(t) 
    &=     K_{\alpha,1}(t) \;+\; k\,E_{\alpha,1}\!\Bigl(-\tfrac{k}{\lambda \gamma_{\alpha}}\,t^\alpha\Bigr)\\
    &\simeq
    \begin{cases}
        K_{\alpha,1}(t), & t \ll \tau_{\alpha,1},\\[4pt]
        k, & \tau_{\alpha,1} \ll t \ll \tau_{\alpha,2},\\[4pt]
        (1+\lambda)K_{\alpha,1}(t), & t \gg \tau_{\alpha,2},
    \end{cases}
\end{aligned}
\end{equation}
and
\begin{equation}
\begin{aligned}
    K_{\alpha,2}^\mathrm{(eff)}(t) 
    &=     K_{\alpha,2}(t) \;+\; k\,E_{\alpha,1}\!\Bigl(-\tfrac{k}{ \gamma_{\alpha}}\,t^\alpha\Bigr)\\
    &\simeq
    \begin{cases}
        K_{\alpha,2}(t), & t \ll \tau_{\alpha,2},\\[4pt]
        (1+\lambda)K_{\alpha,1}(t), & t \gg \tau_{\alpha,2}.
    \end{cases}
\end{aligned}
\end{equation}
Because both the kernel $K_{\alpha,1}(t)$ and the Mittag-Leffler term $k\,E_{\alpha,1}\!(-k t^\alpha /\lambda\gamma_{\alpha})$ share the same power-law exponent $\alpha$, their long-time contributions are simply added to be $(1+\lambda)K_{\alpha,1}(t)$ for $t\gg\tau_{\alpha,2}$. In the unequal-exponent system ($\alpha\neq\beta)$, this does not happen: the slower-decaying kernel $K_\alpha(t)\sim t^{-\alpha}$ always dominates the long-time asymptotics, so $K_\beta(t)\sim t^{-\beta}$ becomes sub-leading and no factor like ($1+\lambda$) appears.

Inserting these kernels to the general expression of MSD~\eqref{eq:MSD_general}, we obtain
\begin{subnumcases}{
    \langle \Delta x^2_{\alpha,1}(t)\rangle\simeq\label{eq:MSD_a1_asymp}
    }
    \displaystyle \tfrac{2\,k_B T}{\gamma_{\alpha}\,\Gamma(1+\alpha)}\,t^\alpha, 
    & $t \ll \tau_{\alpha,1}$, \label{eq:MSD_a1_asymp_1}\\[4pt]
    \displaystyle \tfrac{2\,k_B T}{k}, 
    & $\tau_{\alpha,1} \ll t \ll \tau_{\alpha,2}$, \label{eq:MSD_a1_asymp_2}\\[4pt]
    \displaystyle \tfrac{2\,k_B T}{(1+\lambda)\gamma_\alpha\,\Gamma(1+\alpha)}\,t^\alpha, 
    & $t \gg \tau_{\alpha,2}$\label{eq:MSD_a1_asymp_3}
\end{subnumcases}
for particle 1 and
\begin{subnumcases}{
    \langle \Delta x^2_{\alpha,2}(t)\rangle\simeq\label{eq:MSD_a2_asymp}
    }
    \displaystyle \tfrac{2\,k_B T}{\lambda \gamma_{\alpha}\,\Gamma(1+\alpha)}\,t^\alpha, 
    & $t \ll \tau_{\alpha,2}$, \label{eq:MSD_a2_asymp_1}\\[4pt]
    \displaystyle \tfrac{2\,k_B T}{(1+\lambda)\gamma_\alpha\,\Gamma(1+\alpha)}\,t^\alpha, 
    & $t \gg \tau_{\alpha,2}$\label{eq:MSD_a2_asymp_2}
\end{subnumcases}
for particle 2. We note that the recovery dynamics and the regaining of the long-time diffusivity are absent in the condition of $\alpha=\beta$. Instead, at longer times, the individual particle eventually attains the same reduced generalized diffusivity:
\begin{equation}
    D_{\alpha,1}^{(\infty)}=D_{\alpha,2}^{(\infty)} =\frac{1}{1+\lambda} D_{\alpha,1}^{(0)},
\end{equation}
demonstrating that the mechanical coupling slows down the subdiffusive motion of both particles. This behavior contrasts sharply with the unequal-exponent case, where particle $\alpha$ eventually recovers its free-space generalized diffusivity [Eq.~\eqref{eq:D_alpha_recovery}].

\section{Comparison with polymer simulations}\label{sec:simulation}
\begin{figure}
\includegraphics[width=0.85\columnwidth]{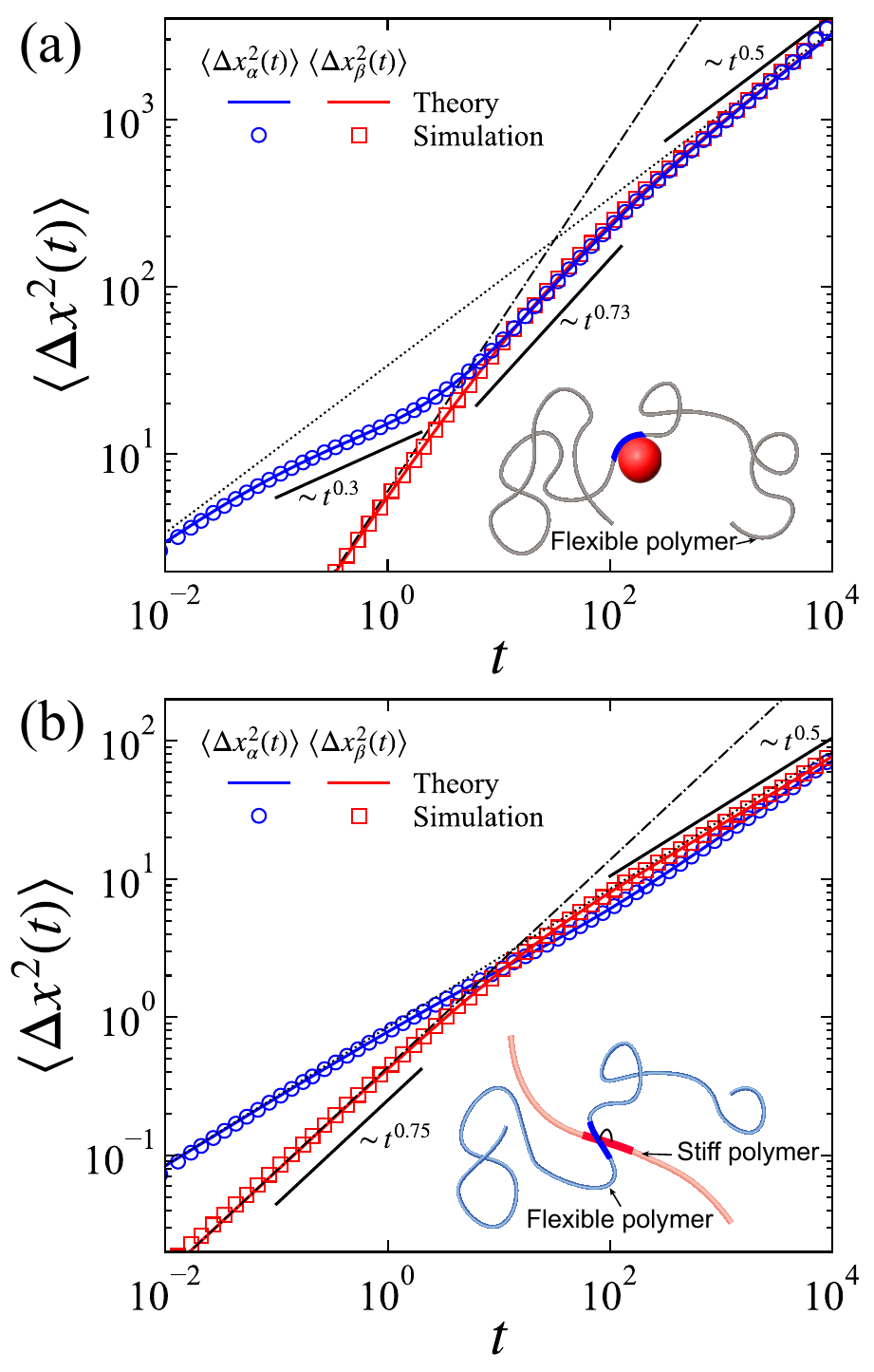}
\caption{\label{fig7}
{Comparison of our FLE models [Eq.~\eqref{eq:coupled_FLE}] with the Langevin dynamics simulations of the two polymer-based physical examples. Simulation parameters are provided in Appendix~\ref{sec:simulation_detail}. (a) A tracer macromolecule is bound to the central bead of a flexible polymer in a viscous medium. The blue symbol is the simulated MSD of the central bead while the red symbol represents the (simulated) MSD of the bound tracer. Solid lines are the theoretical MSD from the FLE model, with $\alpha=1/2$ and $\beta=1$. (b) Flexible (blue symbol) and stiff (red symbol) polymers whose center beads are crosslinked. Solid lines are the theoretical MSDs from the FLE model, with $\alpha=1/2$ and $\beta=3/4$. For both (a) and (b), dotted and dotted-dashed lines represent the MSDs of $\langle\Delta x_\mu^2(t)\rangle_{k=0}=\frac{2k_B T}{\gamma_\mu \Gamma(1+\mu)}t^\mu$ ($k=0$) for particles $\alpha$ and $\beta$, respectively. 
}
}
\end{figure}

In the preceding sections, we have analytically studied the dynamics of coupled FLEs~\eqref{eq:coupled_FLE} in terms of MSDs. To assess how well our FLE model captures coupled viscoelastic dynamics of real systems, in this section, we perform Langevin dynamics simulations of two polymer-based physical systems schematically introduced in Figs.~\ref{fig1}(a) \& \ref{fig1}(b) and examine whether the simulated polymer dynamics indeed agree with our FLE models.  Namely, we simulate the following two scenarios: (i) a flexible polymer whose central bead is bound to a Brownian tracer; and (ii) two polymers—one flexible and one stiff—whose central beads are crosslinked. Below we summarize the comparison of the simulation results with our FLE model. Further information about the simulation model and parameters is described in Appendix~\ref{sec:simulation_detail}.

\subsection{Flexible polymer bound to a Brownian particle}
We simulate a Rouse polymer with its central monomer tethered to a macromolecule (see Fig.~\ref{fig1}(a)), motivated by a chromatin locus tethered to a macromolecular complex or protein condensate \cite{rippeRNAPolymeraseII2025,alghoulCompartmentalizationDNADamage2023,larsonLiquidDropletFormation2017}. The central bead of the Rouse chain is labeled as particle~$\alpha$ and the tracer as particle~$\beta$. In our framework, this system is explained by our coupled FLE: particle $\alpha$ with the memory exponent $\alpha=1/2$ and particle $\beta$ of $\beta=1$. The corresponding MSDs are numerically obtained from Eq.~\eqref{eq:MSD_general} with the insertion of the corresponding memory kernels,  Eqs.~\eqref{eq:memory_eff}~\&~\eqref{eq:kernel_laplace_b_exact}.

Figure~\ref{fig7}(a) shows the comparison of MSDs obtained from the Langevin simulation (symbols) with the analytic counterpart (solid lines). After the kernel parameters are properly identified (Appendix~\ref{sec:simulation_detail}), our FLE model ($\alpha=1/2$, $\beta=1$) shows excellent agreement with the simulated dynamics of this macromolecule--polymer bipartite system over the entire time window of our interest. Under the given simulation condition, the system lies in the strong-coupling regime 
($k/k_c\simeq 7.5>1$), in which the MSD of $x_\alpha$ displays the four distinct scaling regimes described in Eq.~\eqref{eq:MSD_strong_alpha_asymps}, which is clearly visible in the figure.

An interesting note is that, despite the overall agreement between our FLE model and the simulation, the apparent MSD scalings can slightly differ from the asymptotic predictions, Eq.~\eqref{eq:MSD_strong_alpha_asymps}, obtained in the strong-coupling limit ($k/k_c\gg 1$). In particular, the deviation is noticeable in the two intermediate scalings; instead of the plateau ($\sim t^0$) the system exhibits subdiffusion of $\sim t^{0.3}$, and the Fickian recovery regime ($\sim t$) is replaced by an anomalous diffusion regime of $\sim t^{0.76}$. These results suggest that real coupled viscoelastic systems may display richer dynamical behavior than predicted by asymptotic theory.
The apparent exponents may be explained by the numerical solution of the full coupled FLE~\eqref{eq:MSD_general} at the chosen parameters.

\subsection{Crosslinked flexible and stiff polymers}
To test the generality of our framework, we further considered a second system where two polymers of differing stiffness are crosslinked. We simulate a flexible and a stiff polymers crosslinked at their central beads [Fig.~\ref{fig1}(b)], labeled as particles $\alpha$ and $\beta$, respectively.  In our theoretical framework, this system should be modeled by the FLE with the memory exponents $\alpha=1/2$ and $\beta=3/4$, respectively.  See Appendix~\ref{sec:simulation_detail} for further details on the simulated polymer system and its connection to the corresponding FLE model.  In Fig.~\ref{fig7}(b), we present the simulated MSDs of the two crosslinked beads in comparison with the FLE prediction. Here again, it is confirmed that the coupled FLE \eqref{eq:coupled_FLE} with the given memory exponents {captures the dynamics of this cross-linked polymer system with very good agreement}.  
These two examples provide strong support that the coupled FLE framework presented in our study can serve as a quantitative dynamic model that accurately describes the complex viscoelastic dynamics in soft-matter and biological systems via couplings between components with distinct memory exponents.

%====================================================================
\section{Concluding Remarks}
\label{sec:Conclusion}

In this work, we have introduced and analytically studied a minimal model of two viscoelastic systems coupled by a harmonic potential, each governed by a fractional Langevin equation (FLE) with distinct memory exponents ($\alpha$ and $\beta$). Our analytic and numerical studies revealed that the coupling between distinct viscoelastic environments can generate rich, time-dependent behaviors not present in uncoupled systems. As a generic outcome, the system exhibits multiple transient anomalous diffusion behaviors in a wide range of time domain.

First, we identified and characterized distinct scaling regimes resulting from the interplay between the viscoelastic memories and the harmonic interaction. It turns out that there are three relevant timescales [Eq.~\eqref{eq:taus}], $\tau_\alpha$, $\tau_\beta$, and $\tau_c$, characterizing the diffusion dynamics of individual particles ($x_\alpha$ and $x_\beta$) and their relative distance ($r_{\alpha \beta}$), and their dynamic behaviors differ whether the coupled FLE system is in the weak, critical, or strong interaction regime, depending on the strength of the harmonic interaction.  

Notably, we discovered the so-called \emph{recovery dynamics} phenomenon: when a slower particle (or system) with exponent $\alpha$ is coupled with a faster one with exponent $\beta$ (with $\alpha<\beta$), the slower particle suffers transient acceleration and eventually recovers its intrinsic long-time generalized diffusivity. This remarkable behavior disappears when both particles (or systems) share the same memory exponent, instead resulting in a persistent reduction of diffusivity.

A potentially important application of our model is to elucidate transient anomalous diffusion phenomena widely observed in various biological or soft-matter systems. For example, numerous \emph{in vivo} chromatin-tracking experiments have reported transient anomalous diffusions and time-dependent anomalous exponents~\cite{bronsteinTransientAnomalousDiffusion2009,bronshteinLossLaminFunction2015,germierRealTimeImagingSingle2017,zidovskaMicronscaleCoherenceInterphase2013,ashwinOrganizationFastSlow2019,leviChromatinDynamicsInterphase2005,khannaChromosomeDynamicsSolgel2019}. 
In human nuclei, Germier \textit{et al.} tracked a single gene locus and found a crossover from a slow subdiffusion, $\nu\approx0.3$ (0.4--4~s), to a faster subdiffusion, $\nu\approx0.8$ (4--20~s)~\cite{germierRealTimeImagingSingle2017}. Using the displacement-correlation spectroscopy technique over the time window of 0.2--20~s, Zidovska \textit{et al.} observed a similar two-step pattern: an initial slow subdiffusion with $\nu\approx0.2$ (0.2--2~s) that crosses over to a faster subdiffusion with $\nu\approx 0.7$ in interphase or nearly diffusive motion with $\nu\approx1$ in mitosis for 2--20~s~\cite{zidovskaMicronscaleCoherenceInterphase2013}. Meanwhile, at a much longer time window (2--2000~s), Khanna \textit{et al.} examined the relative genomic motion in B-cell nuclei and reported that its MSD has $\nu\approx0.7$ for 2--20~s, then decreasing to $\nu\approx0.5$ for 40--2000~s in B-cell nuclei~\cite{khannaChromosomeDynamicsSolgel2019}. These three eukaryotic-cell studies reveal a consistent picture: chromatin motion has a common subdiffusive exponent ($\nu\approx 0.7$--$0.8$) in the time window of 1--10 s, yet evolves into distinct regimes at shorter and longer times. 
Because a simple polymer model yields only a single power-law (e.g. $\nu=0.5$ for a flexible chain), this progression of multiple scaling regimes implies that additional processes—condensate formation, chromatin looping, and chromatin remodeling, confinement, and so on—must be at work.

Our theory provides a minimal quantitative model for the complex diffusion of a locus via the interaction with DNA--protein assemblies. It is viewed that the tracked locus in a nucleus is mechanically tethered to macromolecular (protein) complexes or phase-separated condensates (e.g. transcription factories~\cite{rippeRNAPolymeraseII2025}, DNA-repair hubs~\cite{alghoulCompartmentalizationDNADamage2023}, and HP1 droplets~\cite{ larsonLiquidDropletFormation2017}). In this picture (e.g., see Fig.~\ref{fig1}(a)), we treat the chromatin fibre as a viscoelastic medium with memory exponent $\alpha = 0.5$ (a flexible polymer), while the bound focus, if it diffused freely, would follow a larger anomalous exponent $\beta \approx 0.7$--1.0, consistent with inert tracer measurements in biological nuclei~\cite{wachsmuthAnomalousDiffusionFluorescent2000,bancaudMolecularCrowdingAffects2009,daddysmanRevisitingPointFRAP2013}.
Under the strong-interaction regime in our theory, the MSD of a chromatin locus has four distinct scaling regimes: (i) the short-time chromatin-dominated subdiffusion ($\nu=0.5$); (ii) the confined plateau ($\nu\sim 0$); (iii) the recovery dynamics ($\nu=0.7$--1); and (iv) the late-time return to the chromatin-dominated subdiffusion ($\nu=0.5$). For a certain finite value of $k/k_c$, the plateau in (ii) can tilt into a shallow subdiffusive slope (e.g. $\nu\approx 0.2$), as observed in Fig.~\ref{fig5}(b) with $k/k_c=10^{1.5}$. 
{We confirm that the four-stage dynamic pattern is indeed observed in our explicit bead–spring polymer simulation [Fig.~\ref{fig7}(a)]: with  $k/k_c\simeq 7.5$, the tilted plateau shows $\nu\approx 0.3$ and the recovery $\nu\approx 0.76$.}
{Within this framework,} the increase of the anomalous exponent from $\nu\approx 0.2$–0.3 to $\nu\approx 0.7$–0.8 reported by Germier et al.~\cite{germierRealTimeImagingSingle2017} and Zidovska et al.~\cite{zidovskaMicronscaleCoherenceInterphase2013} {seems to be consistent with} the transition from regime (ii) to (iii). Meanwhile, the decrease from $\nu\approx 0.7$ to $\nu\approx 0.5$ observed by Khanna et al.~\cite{khannaChromosomeDynamicsSolgel2019} likely reflects the transition from regime (iii) to (iv).

%%%%%%%%%%%%%%%%%%%%%%%%%%%%%%%%%%%%%%%%%%%%
A second illustrative example is the thermal transport of particles in cell-like composite gels comprising semiflexible actin filaments and stiff microtubules crosslinked by specialized proteins. Without crosslinkers (that mediate the interaction between filaments), tracer particles exhibit a single subdiffusive regime with $\nu \approx 0.75$, i.e., the thermal undulation dynamics of a semi-flexible filament ~\cite{andersonSubtleChangesCrosslinking2021}. However, with crosslinkers present, the MSD reveals two distinct scaling regimes, depending on the tracer and the type of crosslinkers: for microspheres, they reveal the short-time subdiffusion with $\nu \approx 0.5$--$0.75$ and the long-time subdiffusion with $\nu \approx 0.25$--$0.5$~\cite{andersonSubtleChangesCrosslinking2021}. For a fluorescent-labeled DNA, it exhibits the short-time subdiffusion with $\nu \approx 0.75$ and the long-time subdiffusion with $\nu \approx 0.5$~\cite{wulsteinTopologydependentAnomalousDynamics2019,garamellaAnomalousHeterogeneousDNA2020}. Within our framework, the microtubules can be modeled as a stiff filament with a thermal undulation of $\beta = 0.75$ while the actin behaves as a flexible chain having a thermal undulation of $\alpha = 0.5$ at the timescales of interest \cite{durangGeneralizedLangevinEquation2024}. In this scenario, the transition from faster to slower subdiffusion ($t^\beta \to t^\alpha$) can arise only when the components are mechanically coupled through the crosslinkers. Although these experiments do not directly measure fluctuations of the network itself, our model nonetheless provides a qualitative explanation for the observed two-regime behavior in the experiment that emerges only in the presence of the crosslinkers.

Our work bridges the gap between theoretical models and real biological systems, highlighting the critical role of memory heterogeneity and mechanical interactions in understanding transient, heterogeneous anomalous dynamics. Indeed, numerous biological and soft-matter systems operate far from equilibrium, thereby violating FDT. Active processes—e.g., transcription, loop extrusion, or motor protein activity—can produce effective “hotter” noise or persistent driving forces. Extending our coupled FLE framework to incorporate such FDT-violating contributions would yield a more realistic description of chromatin and other complex materials. Additionally, our theoretical framework opens promising avenues to explore entropy production, energy dissipation, and thermodynamic efficiency in stochastic, non-equilibrium viscoelastic systems. 
{Furthermore, extending our framework to include pinning potentials and periodic driving would expand the scope of the model to understand the stability and synchronism phenomena of coupled many viscoelastic systems~\cite{mengTemperedAnomalousDynamics2024}.
A comprehensive analysis of this many-body, heterogeneous case is technically challenging but represents an exciting direction for future work. }

\section*{Acknowledgment}
We thank E. Barkai for stimulating discussions. This work was supported by the National Research Foundation (NRF) of Korea, Grant No.~2021R1A6A1A10042944, RS-2023-00218927, \& RS-2024-00343900.

\appendix
\counterwithin{figure}{section}

\section{Derivation of the coupled FLE \eqref{eq:coupled_FLE} from polymer examples \label{sec:FLE_derivation}}
In this appendix, we present the derivation of the coupled FLE~\eqref{eq:coupled_FLE} that is conceptually explained in Sec.~\ref{sec:Model}. 
We work out the system of a flexible polymer bound to a Brownian macromolecule [Fig.~\ref{fig1}(a)], and subsequently discuss the generalization to other physical models.
As described in the main text, the equation of motion for individual beads comprising the given polymer system is governed by the Langevin equation~\eqref{eq:poly_EOM}, along with the total potential energy Eq.~\eqref{eq:poly_E_tot}.  
We derive the effective one-particle coupled FLE for the tracers $\mathbf r^{A}_{a}$ and $\mathbf r^{B}_{b}$ in sub-systems $A$ and $B$ by integrating out all bead coordinates except for the two tracer variables. Our coarse-graining integration is based on the techniques employed in  Refs.~\cite{panjaGeneralizedLangevinEquation2010,panjaAnomalousPolymerDynamics2010,taloniGeneralizedElasticModel2010,lizanaFoundationFractionalLangevin2010,maesLangevinGeneralizedLangevin2013,durangGeneralizedLangevinEquation2024,hanNonequilibriumDiffusionActive2023,vandebroekGeneralizedLangevinEquation2017,sakaueMemoryEffectFluctuating2013,shinkaiGeneralizedLangevinDynamics2024}.

\subsection{Derivation for a flexible polymer bound to a Brownian particle}
As an illustrative example, let system $A$ be a flexible Rouse chain with $M_A=2N_A+1$ beads, whose coordinate is represented by $\{\mathbf r^{A}_{n}\}$ ($n=1,\ldots,M_A$) and the central bead of index $a=N_A+1$ is the tracer.  The polymer bead is connected by harmonic springs of stiffness $k^{A}$, and each bead experiences an isotropic Stokes friction $\gamma^{A}$. 
System $B$ is a single Brownian particle with a friction $\gamma^{B}$, and its coordinate is represented by $\{\mathbf r^{B}_{n}\}$ where the number of beads is $M_B=1$ and the bead index is $b=1$. 
The potential energy of this Rouse (Gaussian) chain is
\[
U^{A}=\frac{k^{A}}{2}\sum_{n=1}^{2N_A}\bigl(\mathbf r^{A}_{n+1}-\mathbf r^{A}_{n}\bigr)^2,
\]
with the free-end (phantom) boundary conditions $\mathbf r^{A}_0=\mathbf r^{A}_1$ and $\mathbf r^{A}_{2N_A+2}=\mathbf r^{A}_{2N_A+1}$.
The Langevin equation of a bead in $A$ reads (where $u=1,\dots,d$ is the Cartesian component)
\begin{multline}
\label{eq:rouse-A-bead}
\gamma^A\dot r^{A}_{n,u}(t)
=-\,k^{A}\!\left(2r^{A}_{n,u}-r^{A}_{n-1,u}-r^{A}_{n+1,u}\right)\\
-k\!\left(r^{A}_{a,u}-r^{B}_{b,u}\right)\delta_{n,a}
+\eta^{A}_{n,u}(t),
\end{multline}
while the Brownian particle in $B$ obeys
\begin{equation}\label{eq:Brownian-B-bead}
    \gamma^B \dot r^{B}_{b,u}(t)=-k\!\left(r^{B}_{b,u}-r^{A}_{a,u}\right) + \eta^B_{b,u}(t).
\end{equation}
The thermal noises satisfy $\langle \eta^{Z}_{n,u}(t)\eta^{Z}_{m,v}(t')\rangle
=2\gamma^{Z} k_BT\delta_{nm}\delta_{uv}\delta(t-t')$ for $Z\in\{A,B\}$.

\paragraph{Mode decomposition.}
We split the polymer $A$ into the left arm $n=1,\dots,N_A$ and the right arm $n=N_A+2,\dots,2N_A+1$, both attached to the center tracer $\mathbf r^{A}_{a}$.
Introduce the arm's normal modes (for $p=1,\dots,N_A$) \cite{vandebroekGeneralizedLangevinEquation2017,durangGeneralizedLangevinEquation2024}
\begin{equation}\label{eq:arm-mode}
\begin{aligned}
X^{L}_{p,u}(t)&=\frac{2}{\sqrt{2N_A+1}}\sum_{n=1}^{N_A} c^{(p)}_{n}\, r^{A}_{n,u}(t),\\
X^{R}_{p,u}(t)&=\frac{2}{\sqrt{2N_A+1}}\sum_{n=N_A+2}^{2N_A+1} c^{(p)}_{n}\, r^{A}_{n,u}(t),
\end{aligned}
\end{equation}
with the orthogonal basis
\[
c^{(p)}_{n}=\cos\!\left(\frac{2n-1}{2}\,\theta_p\right),\qquad
\theta_p=\frac{2p-1}{2N_A+1}\,\pi .
\]

Projecting \eqref{eq:rouse-A-bead} onto $X^{\sigma}_{p,u}$ with $\sigma\in\{L,R\}$ and using
$2c^{(p)}_n-c^{(p)}_{n-1}-c^{(p)}_{n+1}=4\sin^2(\tfrac{\theta_p}{2})\,c^{(p)}_n$,
$c^{(p)}_{N_A+1}=0$, $c^{(p)}_{0}=c^{(p)}_{1}$, we obtain
\begin{equation}\label{eq:arm-mode-eom}
\gamma^{A}\,\dot X^{\sigma}_{p,u}(t)
= -\,\lambda_p\,k^{A}\,X^{\sigma}_{p,u}(t)
+ k^{A}\,s_p\, r^{A}_{a,u}(t) + \xi^{\sigma}_{p,u}(t),
\end{equation}
where $s_p \equiv \tfrac{2}{\sqrt{2N_A+1}}\,c^{(p)}_{N_A}$ and
\begin{equation}
\begin{aligned}
\xi^{L}_{p,u}(t) &\equiv \tfrac{2}{\sqrt{2N_A+1}}\sum_{n=1}^{N_A} c^{(p)}_{n}\,\eta^{A}_{n,u}(t),\\
\xi^{R}_{p,u}(t) &\equiv \tfrac{2}{\sqrt{2N_A+1}}\sum_{n=N_A+2}^{2N_A+1} c^{(p)}_{n}\,\eta^{A}_{n,u}(t),    
\end{aligned}
\end{equation}
for which $\langle \xi^{\sigma}_{p,u}(t)\,\xi^{\sigma'}_{p',v}(t')\rangle
= 2\gamma^{A} k_BT\delta_{\sigma\sigma'}\delta_{pp'}\delta_{uv}\delta(t-t')$.
The mode factor is
\begin{equation}\label{eq:lambda-def}
\lambda_p \;=\; 4\sin^2\!\Bigl(\tfrac{\theta_p}{2}\Bigr),
\end{equation}
and the relaxation time associated with the mode $p$ is
\begin{equation}\label{eq:tau-def}
\tau_p \;=\; \frac{\gamma^{A}}{k^{A}\lambda_p}.
\end{equation}
{
Note that the normal-mode equation~\eqref{eq:arm-mode-eom} contains no explicit dependence on system $B$ or on the coupling constant $k$. The mode decomposition diagonalizes the internal elastic interactions of polymer $A$, while the external coupling to system $B$ acts only on bead $a$ (the central bead). Because the basis functions satisfy $c_a^{(p)}=0$, the external force $-k(r_{a,u}^A-r_{b,u}^B)$ does not project onto any polymer mode {in Eq.~\eqref{eq:arm-mode}}. Hence, the relaxation spectrum $\{\lambda_p, \tau_p\}$ reflects only the intrinsic viscoelastic response of polymer $A$ and is unaffected by the coupling. 
}

The formal solution of Eq.~\eqref{eq:arm-mode-eom} is
\begin{multline}
\label{eq:arm-mode-sol}
X^\sigma_{p,u}(t)=X^\sigma_{p,u}(0)e^{-t/\tau_p}
\\+\frac{1}{\gamma^{A}}\!\int_0^t\! \mathrm dt'\, e^{-(t-t')/\tau_p}\!\left[k^{A} s_p\, r^{A}_{a,u}(t')+\xi^\sigma_{p,u}(t')\right].
\end{multline}

\paragraph{Tracer FLEs.}
The equation for the center tracer (the bead index $a$) in $A$ follows from Eq.~\eqref{eq:rouse-A-bead}:
\begin{equation}\label{eq:center-bead-raw}
\begin{aligned}
\gamma^{A}\dot r^{A}_{a,u}(t)
= &-\,k\!\left(r^{A}_{a,u}(t)-r^{B}_{b,u}(t)\right)\\
&+ k^{A}\sum_{p=1}^{N_A}(-1)^{p-1}s_p\!\left[X^{L}_{p,u}(t)+X^{R}_{p,u}(t)\right]\\
&+\eta^{A}_{a,u}(t) - 2k^{A} r^{A}_{a,u}(t).
\end{aligned}
\end{equation}
Substituting Eq.~\eqref{eq:arm-mode-sol} into Eq.~\eqref{eq:center-bead-raw} yields the generalized Langevin equation
\begin{equation}\label{eq:GLE-A}
\int_0^t\! K_A(t-t')\,\dot r^{A}_{a,u}(t')\,\mathrm dt'
= -\,k\!\left(r^{A}_{a,u}(t)-r^{B}_{b,u}(t)\right) + \Xi^{A}_{\mu}(t),
\end{equation}
with the memory kernel (exact for finite $N_A$)
\begin{equation}\label{eq:KA-sum}
K_A(t)=\gamma^{A}\delta(t)
+\frac{2(k^{A})^{2}}{\gamma^{A}}\sum_{p=1}^{N_A} s_p^{2}\,\tau_p\,e^{-t/\tau_p}
\end{equation}
and the effective Gaussian noise obeying the FDT
\begin{equation}\label{eq:FDT-A}
\big\langle \Xi^{A}_{\mu}(t)\,\Xi^{A}_{\nu}(t')\big\rangle
\simeq k_B T\,K_A(|t-t'|)\,\delta_{\mu\nu}
\end{equation}
at a long time.

For $N_A\gg1$, the spectrum $\{\tau_p\}$ becomes quasi-continuous between
$\tau_{0}^A\sim \tau_{N_A}\sim \gamma^A/(\pi^2 k^A)$ and $\tau_{R}^A\sim \tau_{1}\sim \gamma^{A}(2N_A+1)^2/\bigl(\pi^2 k^{A}\bigr)$.
In the broad intermediate window $\tau_{0}^A\ll t\ll \tau_{R}^A$, the sum in \eqref{eq:KA-sum} is well approximated by an integral, leading to the Rouse long-time tail
\begin{equation}\label{eq:KA-asym}
K_A^\mathrm{flex}(t)\simeq\sqrt{\tfrac{4k^A\gamma^A}{\pi( k_B T)^2} } t^{-1/2}.
\end{equation}
Hence, the generalized Langevin equation~\eqref{eq:GLE-A} reduces to a FLE with the exponent $\alpha=\frac{1}{2}$ for system $A$.

Collecting the tracer equations for $\mathbf r^{A}_{a}$ and $\mathbf r^{B}_{b}$ [Eqs.~\eqref{eq:GLE-A} and \eqref{eq:Brownian-B-bead}], finally,  results in the coupled FLE:
\begin{equation}\label{eq:coupled_FLE_flex_bead}
\begin{aligned}
    \int_0^t\! K_{A}^\mathrm{flex}(t-t')\,\dot r^{A}_{a,u}(t')\,\mathrm dt'
= -\,k\!\left(r^{A}_{a,u}-r^{B}_{b,u}\right)+\Xi^{A}_{u}(t),\\
\gamma^B \dot r^{B}_{b,u}(t)=-k\!\left(r^{B}_{b,u}-r^{A}_{a,u}\right) + \eta^B_{b,u}(t).
\end{aligned}
\end{equation}
Identifying $x_{\alpha}\equiv r^{A}_{a,u}$ and $x_{\beta}\equiv r^{B}_{b,u}$, we obtain the coupled FLE~\eqref{eq:coupled_FLE} for a flexible polymer bound to a Brownian particle [Fig.~\ref{fig1}(a)].

\subsection{Generalization}
The above coarse-graining for the Rouse chain extends to other polymer models or systems by modifying the mode spectrum $\{\lambda_p,\tau_p\}$ according to the underlying elastic and hydrodynamic physics~\cite{panjaGeneralizedLangevinEquation2010,sakaueMemoryEffectFluctuating2013}. Examples include self-avoiding or fractally packed chains (altered long-wavelength dispersion and density of modes)~\cite{panjaGeneralizedLangevinEquation2010,tammAnomalousDiffusionFractal2015,sakaueMemoryEffectFluctuating2013} and hydrodynamic interactions (Zimm-type relaxation spectrum)~\cite{panjaGeneralizedLangevinEquation2010,sakaueMemoryEffectFluctuating2013}. In each case, the kernel retains the form \eqref{eq:KA-sum} with a model-specific set of $\{\tau_p,s_p\}$, leading to a power-law tail $K_A(t)\propto t^{-\alpha}$ with an exponent $\alpha$ characteristic of the chosen polymer model.

In addition, the coarse-graining procedure for the polymer ($A$) bound to a Brownian particle ($B$) can be readily extended to polymer--polymer junction. Notably, the derivation of Eq.~\eqref{eq:GLE-A} for system $A$ is independent of the detailed structure of system $B$. Hence, Eq.~\eqref{eq:GLE-A} remains valid even when system $B$ itself represents another polymer, as illustrated in Fig.~\ref{fig1}(b). Applying the same procedure to system $B$, characterized by its own parameters $(k^{B}, \gamma^{B}, M_B)$ and the coupling site $b$, yields an analogous generalized Langevin equation:
\begin{align}
\int_0^t\! K_{A}(t-t')\,\dot r^{A}_{a,u}(t')\,\mathrm dt'
&= -\,k\!\left(r^{A}_{a,u}-r^{B}_{b,u}\right)+\Xi^{A}_{u}(t),\\
\int_0^t\! K_{B}(t-t')\,\dot r^{B}_{b,u}(t')\,\mathrm dt'
&= -\,k\!\left(r^{B}_{b,u}-r^{A}_{a,u}\right)+\Xi^{B}_{u}(t),
\end{align}
where the effective noises satisfy the fluctuation–dissipation relation
\begin{equation}
\big\langle \Xi^{Z}_{u}(t)\,\Xi^{Z}_{v}(t')\big\rangle
= k_B T\,K^{Z}(|t-t'|)\,\delta_{uv}, \qquad Z\in\{A,B\}.    
\end{equation}
With the identifications $x_{\alpha}\equiv r^{A}_{a,u}$ and $x_{\beta}\equiv r^{B}_{b,u}$, this reproduces the coupled FLEs \eqref{eq:coupled_FLE} in full generality.

\section{The correlation functions of the effective noises\label{sec:noise_correlation}}
In this section, we derive the correlation functions of $\eta_{\alpha\beta}(t)$ [Eq.~\eqref{eq:noise_rab}] and $\xi_{\alpha}^\mathrm{(eff)}(t)$ [Eq.~\eqref{eq:noise_eff_laplace}]. For a stationary process $\phi(t)$ with the correlation function $\langle\phi(t)\phi(t')\rangle = \psi(|t-t'|)$, its Laplace transform satisfies
\begin{equation}\label{eq:correlation_laplace}
    \langle \tilde\phi(s)\tilde\phi(s')\rangle = \frac{\tilde \psi(s)+\tilde \psi(s')}{s+s'}.
\end{equation}
Employing this result to fractional Gaussian noise $\xi_p$, which obeys the FDT
\begin{equation}\label{eq:FDT_laplace}
    \langle \tilde \xi_\mu(s)\tilde \xi_\mu(s')\rangle = k_B T \frac{\tilde K_\mu(s)+\tilde K_\mu(s')}{s+s'},
\end{equation}
where $\tilde K_\mu(s)=\gamma_\mu s^{-1+\mu}$, we proceed to compute the correlations.

\subsection{The correlation function of \texorpdfstring{$\eta_{\alpha\beta}(t)$}{eta ab}}
Using Eq.~\eqref{eq:FDT_laplace}, the correlation of $\tilde \eta_{\alpha\beta}(s)$ yields
\begin{equation}\label{eq:noise_correlation_derivation}
\begin{aligned}
    \langle \tilde\eta_{\alpha\beta}(s) \tilde\eta_{\alpha\beta}(s')\rangle = &
    \tfrac{\tilde K_\beta(s)}{\tilde K_\alpha(s)+\tilde K_\beta(s)}
  \tfrac{\tilde K_\beta(s')}{\tilde K_\alpha(s')+\tilde K_\beta(s')} \langle \tilde{\xi}_{\alpha} (s) \tilde{\xi}_{\alpha} (s')\rangle\\
  &+  \tfrac{\tilde K_\alpha(s)}{\tilde K_\alpha(s)+\tilde K_\beta(s)}
  \tfrac{\tilde K_\alpha(s')}{\tilde K_\alpha(s')+\tilde K_\beta(s')} \langle \tilde{\xi}_{\beta} (s) \tilde{\xi}_{\beta} (s')\rangle\\[6pt]
  =& \tfrac{k_B T}{s+s'} \tfrac{  K_\beta(s)K_{\beta}(s')\left(\tilde K_\alpha(s)+\tilde K_\alpha(s')\right) }{  \left(\tilde K_\alpha(s)+\tilde K_\beta(s)\right)\left(\tilde K_\alpha(s')+\tilde K_\beta(s')\right)  }\\
&+  \tfrac{k_B T}{s+s'} \tfrac{  K_\alpha(s)K_{\alpha}(s')\left(\tilde K_\beta(s)+\tilde K_\beta(s')\right) }{  \left(\tilde K_\alpha(s)+\tilde K_\beta(s)\right)\left(\tilde K_\alpha(s')+\tilde K_\beta(s')\right)  }.
\end{aligned}
\end{equation}
After rearranging terms, we arrive at the expression:
\begin{equation}
    \langle \tilde\eta_{\alpha\beta}(s) \tilde\eta_{\alpha\beta}(s')\rangle = 
  k_{B}T \frac{\frac{\tilde K_\alpha(s)\tilde K_\beta(s)}{\tilde K_\alpha(s)+\tilde K_{\beta}(s)} + \frac{\tilde K_\alpha(s')\tilde K_\beta(s')}{\tilde K_\alpha(s')+\tilde K_\beta(s')}}{s+s'}.
\end{equation}
Since the obtained expression follows the form defined in Eq.~\eqref{eq:correlation_laplace}, we can identify the correlation function in the time domain, which satisfies the FDT in the following:
\begin{equation}
    \langle \eta_{\alpha\beta}(t)\eta_{\alpha\beta}(t')\rangle = k_B T K^{\mathrm{(eff)}}_{\alpha\beta}(|t-t'|).
\end{equation}
This is Eq.~\eqref{eq:FDT_rab} in the main text. Because no specific assumption on $\tilde K_\mu(s)$ was made, this result holds for any memory kernels other than the power-law.

\subsection{The correlation function of \texorpdfstring{$\xi_{\alpha}^\mathrm{(eff)}(t)$}{eta b}}
Similarly, using Eq.~\eqref{eq:FDT_laplace}, we calculate the correlation function of $\tilde \eta_{\beta}(s)$ as:
\begin{equation}
    \begin{aligned}
        \frac{\langle \tilde\eta_{\beta}(s) \tilde\eta_{\beta}(s')\rangle}{k_BT}
        =&  \tfrac{k }{s\tilde K_\beta(s) + k}\tfrac{k }{s'\tilde K_\beta(s') + k} \tfrac{\tilde K_\beta(s)+\tilde K_\beta(s')}{s+s'}\\[8pt]
        =&  \tfrac{k\tilde K_\beta(s)\left(s'\tilde K_\beta(s') + k\right) + k\tilde K_\beta(s')\left(s\tilde K_\beta(s) + k\right) }{\left(s\tilde K_\beta(s)+k\right)\left(s'\tilde K_\beta(s') + k\right)(s+s')}\\
        &- \tfrac{k\tilde K_\beta(s)\tilde K_\beta(s')(s+s')}{\left(s\tilde K_\beta(s)+k\right)\left(s'\tilde K_\beta(s') + k\right)(s+s')}\\[8pt]
        =& \tfrac{\frac{k\tilde K_\beta(s)}{s\tilde K_\beta(s') + k} + \frac{k\tilde K_\beta(s)}{s'\tilde K_\beta(s') + k}}{s+s'} \\
        &-k\tfrac{\tilde K_\beta(s)}{s\tilde K_\beta(s) + k}\tfrac{\tilde K_\beta(s')}{s'\tilde K_\beta(s') + k}.
    \end{aligned}
\end{equation}
Through the inverse Laplace transform we obtain
\begin{multline}
\label{eq:eta_corr_exact}
    \langle\eta_\beta(t)\eta_\beta(t')\rangle = k_B T \Phi_\beta(|t-t'|) \\- k_B Tk E_{\beta,1} \left(-(t/\tau_\beta)^\beta\right) \; E_{\beta,1} \left(-(t'/\tau_\beta)^\beta\right).
\end{multline}
Here, the Mittag-Leffler function asymptotically decays as $E_{\beta,1} \left(-z^\beta\right)\simeq z^{-\beta}$ for $z\gg1$, so the non-stationary terms, $E_{\beta,1} \left(-(t/\tau_\beta)^\beta\right)$ and $E_{\beta,1} \left(-(t'/\tau_\beta)^\beta\right)$ become negligible as $t/\tau_\beta\to\infty$ and $t'/\tau_\beta\to\infty$. Finally, neglecting non-stationary contributions and using the independency between the noises ($\langle \xi_\alpha(t)\eta_\beta(t')\rangle = 0$), we obtain the FDT for the effective noise $\xi_\alpha^\mathrm{(eff)}(t) = \xi_\alpha(t) + \eta_\beta(t)$ as follows:
\begin{equation}
     \langle\xi_\alpha^\mathrm{(eff)}(t)\xi_\alpha^\mathrm{(eff)}(t')\rangle \simeq k_B T \big[K_\alpha(|t-t'|)+\Phi_\beta(|t-t'|)\big].
\end{equation}

\section{Derivation of MSDs~\eqref{eq:MSD_general_confined} and \eqref{eq:MSD_general} \label{sec:GLE_MSD}}
In this appendix, we derive the MSDs for the generalized Langevin equations quoted in Eqs.~\eqref{eq:MSD_general_confined} and \eqref{eq:MSD_general}.

\subsection{The GLE subject to a harmonic potential}

We begin with a GLE described with a memory kernel \(K(t)\) and subject to a harmonic potential of stiffness \(k\):
\begin{equation}
\label{eq:b5}
    \int_0^t K(t-t') \dot x(t') dt' = -kx(t) + \xi(t).
\end{equation}
The free-space GLE is recovered in the limit of $k\!\to\!0$.

In the Laplace domain, the solution of the GLE above can be written as
\begin{equation}\label{eq:sol_laplace}
    \tilde x(s) = \frac{\tilde K(s)}{s\tilde K(s)+k} x(0) + \frac{\tilde \xi(s)}{s\tilde K(s) + k}.
\end{equation}
Using this, we obtain the expression for the two-point correlation function as follows:
\begin{equation}
\begin{aligned}
\langle &\tilde x(s)\tilde x(s')\rangle 
=\langle x^2(0)\rangle\tilde \Psi(s) \tilde \Psi(s') +\tfrac{\langle\tilde \xi_\alpha(s)\tilde \xi_\alpha(s')\rangle}{[s\tilde K(s) + k][s'\tilde K(s') + k]}
\end{aligned}
\end{equation}
where $\tilde\Psi(s)$ is defined by
\begin{equation}
    \tilde\Psi(s) \equiv \frac{\tilde K(s)}{s\tilde K(s)+k},
\end{equation}
and $\langle x(0) \tilde \xi(s) \rangle = 0$ is used.

Employing the Laplace-transformed FDT \eqref{eq:FDT_laplace}, we have
\begin{equation}\label{eq:corr_laplace}
\begin{aligned}
\langle\tilde x(s)\tilde x(s')\rangle 
&=\langle x^2(0)\rangle\tilde \Psi(s) \tilde \Psi(s')\\
&+ \frac{k_B T}{k}\left[\frac{\tilde \Psi(s)+\tilde \Psi(s')}{s+s'} -\tilde \Psi(s) \tilde \Psi(s')\right].
\end{aligned}
\end{equation}
The inverse Laplace transformation of the above results is performed to obtain the autocovariance of $x(t)$:
\begin{equation}
\begin{aligned}
    \langle x(t)x(t') \rangle = & \left[\langle  x^2(0)\rangle - \frac{k_B T}{k}\right] \Psi(t)\Psi(t') \\
    &+\frac{k_B T}{k} \Psi(|t-t'|).
\end{aligned}    
\end{equation}
If we set the equilibrium initial condition, $\langle x^2(0)\rangle = \frac{k_B T}{k}$ and the first term in R.H.S. vanishes. Assuming $\Psi(0)=1$ (i.e., $\lim_{s\to\infty}s\tilde K(s)\!=\!\infty$), the MSD
\begin{multline}\label{eq:MSD_derivation}
    \langle [x(t+t_0)-x(t_0)]^2\rangle = \langle x^2(t+t_0)\rangle + \langle x^2(t_0)\rangle\\ - 2 \langle x(t+t_0)x(t_0) \rangle,
\end{multline}
is simplified to
\begin{equation}
    \langle [x(t+t_0)-x(t_0)]^2\rangle =\frac{2k_B T}{k}\left[1-\Psi(t)\right].
\end{equation}
Equivalently, in the Laplace domain, this can be written as
\begin{equation}\label{eq:MSD_confined_app}
    \langle [x(t+t_0)-x(t_0)]^2\rangle = 2k_B T \mathcal{L}^{-1} \left\{ \frac{1}{s(s\tilde K(s) + k)} \right\} (t),
\end{equation}
which recovers Eq.~\eqref{eq:MSD_general_confined}.

\subsection{GLE in free space}
As explained above, the GLE~\eqref{eq:b5} converges to that in free space in the limit of $k\to 0$. Using this property, we obtain the expression for MSD in free space by taking $k\to0$ in Eq.~\eqref{eq:corr_laplace}. In this limit, the autocovariance is
\begin{equation}
\begin{aligned}
\langle\tilde x(s)\tilde x(s')\rangle 
&=\frac{\langle x^2(0)\rangle}{ss'}
+ \frac{k_B T [\tilde K(s)+\tilde K(s')]}{ss'(s+s')\tilde K(s)\tilde K(s')}.
\end{aligned}
\end{equation}
Now we define $\tilde G(s) = \frac{1}{s^2 \tilde K(s)}$ and rewrite the autocovariance in terms of $G$ as such:
\begin{multline}
\langle\tilde x(s)\tilde x(s')\rangle 
=\frac{\langle x^2(0)\rangle}{ss'}\\
+ k_B T \left[\frac{\tilde G(s)}{s'}+\frac{\tilde G(s')}{s}-\frac{\tilde G(s)+\tilde G(s')}{s+s'}\right].
\end{multline}
Transforming this expression back to the time domain yields
\begin{multline}
\label{eq:x(t)x(t')_fin}
    \langle x(t)x(t')\rangle = \langle x_0^2\rangle \\
    + k_B T\left[ G(t) + G(t') - G(|t-t'|)\right].
\end{multline}
Inserting Eq.~\eqref{eq:x(t)x(t')_fin} into the definition of MSD, Eq.~\eqref{eq:MSD_derivation}, we finally obtain the expression for the MSD
\begin{equation}\label{eq:MSD_free_app}
    \langle [x(t+t_0)-x(t_0)]^2\rangle = 2k_B T G(t),
\end{equation}
which recovers Eq.~\eqref{eq:MSD_general}. We can also confirm that taking $k\to0$ in Eq.~\eqref{eq:MSD_confined_app} yields Eq.~\eqref{eq:MSD_free_app}, which ensures a smooth connection between the MSDs at the conditions of confined and free space.

\section{The exact solution of Eq.~\texorpdfstring{\eqref{eq:CFLE_dev3}}{22}}\label{sec:MSD_exact}
% \reply{REVISED}
In Sec.~\ref{sec:GLE_a_b}, we neglected the initial-position term in Eq.~\eqref{eq:CFLE_dev3} and assumed a stationary noise correlation. Under this condition, we obtained the effective GLE~\eqref{eq:GLE_eff} with the MSD~\eqref{eq:MSD_general}. Here, we retain every initial-condition contribution and demonstrate that the full solution still reproduces the same MSD and two-point correlation.

We take the Laplace transform of Eq.~\eqref{eq:CFLE_dev3} to find the formal solution
\begin{multline}
    \tilde x_\alpha(s)= \tfrac{1}{s}x_\alpha(0)
    - \tfrac{\tilde \Phi_\beta(s) }{s\tilde K_\alpha^\mathrm{(eff)}(s)}r_{\alpha\beta}(0) 
    +\tfrac{\tilde \xi_\alpha(s) + \tilde \eta_\beta(s)}{s\tilde K_\alpha^\mathrm{(eff)}(s)}.
\end{multline}
Using this, we obtain the autocovariance as
\begin{equation}
\begin{aligned}
\langle &\tilde x_\alpha(s)\tilde x_\alpha(s')\rangle 
=\tfrac{\langle x_\alpha^2(0)\rangle}{ss'}\\&
+\langle x_\alpha(0) r_{\alpha\beta}(0)\rangle 
\left[\tfrac{\tilde\Phi_\beta(s')}{ss'\tilde K^\mathrm{(eff)}_\alpha(s')}+\tfrac{\tilde\Phi_\beta(s)}{s's\tilde K^\mathrm{(eff)}_\alpha(s)}\right]\\
&+\langle r_{\alpha\beta}^2(0) \rangle \tfrac{\tilde\Phi_\beta(s)}{s\tilde K^\mathrm{(eff)}_\alpha(s)}\tfrac{\tilde\Phi_\beta(s')}{s'\tilde K^\mathrm{(eff)}_\alpha(s')}\\
&+\tfrac{1}{s\tilde K_\alpha^\mathrm{(eff)}(s)s'\tilde K_\alpha^\mathrm{(eff)}(s')}\left[\langle\tilde \xi_\alpha(s)\tilde \xi_\alpha(s')\rangle + \langle \tilde \eta_\beta(s)\tilde \eta_\beta(s')\rangle\right].
\end{aligned}
\end{equation}
Using the relations of Eqs.~\eqref{eq:memory_eff} and \eqref{eq:noise_eff_laplace}, we can rearrange the above expression as follows:
\begin{equation}
\begin{aligned}
\langle &\tilde x_\alpha(s)\tilde x_\alpha(s')\rangle 
=\tfrac{\langle x_\alpha^2(0)\rangle}{ss'}\\&
+\langle x_\alpha(0) r_{\alpha\beta}(0)\rangle 
\left[\tfrac{\tilde\Phi_\beta(s')}{ss'\tilde K^\mathrm{(eff)}_\alpha(s')}+\tfrac{\tilde\Phi_\beta(s)}{s's\tilde K^\mathrm{(eff)}_\alpha(s)}\right]\\
&+\left[\langle r_{\alpha\beta}^2(0)\rangle -\tfrac{k_B T}{k}\right] \tfrac{\tilde\Phi_\beta(s)}{s\tilde K^\mathrm{(eff)}_\alpha(s)}\tfrac{\tilde\Phi_\beta(s')}{s'\tilde K^\mathrm{(eff)}_\alpha(s')}\\
&+ \tfrac{k_B T}{k}\left[\tfrac{\tilde G^\mathrm{(eff)}(s)}{s'}+\tfrac{\tilde G^\mathrm{(eff)}(s')}{s}-\tfrac{\tilde G^\mathrm{(eff)}(s)+\tilde G^\mathrm{(eff)}(s')}{s+s'}\right],
\end{aligned}
\end{equation}
where $ \tilde G^\mathrm{(eff)}(s) = \frac{1}{s^2\tilde K^\mathrm{(eff)}_\alpha(s)}$. Performing the inverse Laplace transformation, we obtain the autocovariance in the time domain
\begin{equation}\label{eq:corr_exact}
\begin{aligned}
\langle & x_\alpha(t) x_\alpha(t')\rangle 
= \langle x_\alpha^2(0)\rangle\\&
+\langle x_\alpha(0) r_{\alpha\beta}(0) \rangle [\Psi_{\alpha\beta}(t)+\Psi_{\alpha\beta}(t')]\\
&+\left[\langle r_{\alpha\beta}^2 (0)\rangle - \tfrac{k_B T}{k}\right] \Psi_{\alpha\beta}(t)\Psi_{\alpha\beta}(t')\\
&+ k_B T\left[G^\mathrm{(eff)}(t) + G^\mathrm{(eff)}(t') - G^\mathrm{(eff)}(|t-t'|)\right],
\end{aligned}
\end{equation}
with 
\begin{equation}
\begin{aligned}
    \Psi_{\alpha\beta}(t) &= \mathcal{L}^{-1} \left\{\frac{\tilde \Phi_\beta(s)}{s \tilde K^\mathrm{(eff)}_\alpha(s)}\right\}(t)\\
    &\simeq (t/\tau_\beta)^{-\beta} ,& t\gg \tau_\alpha,\tau_\beta,\tau_c.
\end{aligned}
\end{equation}
The above Eq.~\eqref{eq:corr_exact} is the most general expression obtained under the non-equilibrium initial condition. Even in such a case, the second and third terms in R.H.S. decay out in the infinite-time limit as $\Psi_{\alpha\beta}(t)\sim (t/\tau_\beta)^{-\beta}$. These contributions are absent either in the long-time limit $t\gg\tau_\beta$ (or if the process starts at $t=-\infty$) or under the equilibrium initial conditions: $\langle r_{\alpha\beta}^{2}(0)\rangle=k_B T/k$ and $\langle r_{\alpha\beta}(0)x_\alpha(0)\rangle=0$. In such cases, Eq.~\eqref{eq:corr_exact} reduces to the form, Eq.~\eqref{eq:x(t)x(t')_fin}, presented in the main text.  Finally, Eq.~\eqref{eq:MSD_derivation} is used to estimate MSD, yielding
\begin{equation}
    \langle [x(t+t_0)-x(t_0)]^2\rangle = 2k_B T G^\mathrm{(eff)}(t).
\end{equation}
Thus, the MSD is stationary and identical to the expression shown in Eq.~\eqref{eq:MSD_general}.

\section{Polymer simulations}\label{sec:simulation_detail}
In Sec.~\ref{sec:simulation} and Fig.~\ref{fig7}, we present the Langevin dynamics simulation results of the bipartite polymer systems. Here, we describe the simulation details.

We have performed the overdamped Langevin dynamics simulations of the coupled polymer systems introduced in Figs.~\ref{fig1}(a) \& \ref{fig1}(b). The setup is identical to the two-component model discussed in Sec.~\ref{sec:Model}, which consists of two polymeric components, systems $A$ and $B$, each represented by a chain of beads. The total potential energy is given by Eq.~\eqref{eq:poly_E_tot}, and the equation of motion follows the overdamped Langevin equation~\eqref{eq:poly_EOM}. The initial positions of all beads are sampled from the Boltzmann distribution, $P(\{\mathbf{r}_i^A\},\{\mathbf{r}_i^B\})\propto e^{-U_\mathrm{tot}}$. We numerically integrate the equations of motion using a second-order Langevin integrator~\cite{vandenSecondIntegratorsLangevin}, with a time step of $dt=0.001$. Each simulation runs for $10^7$ time steps. The MSDs are computed as both time- and ensemble-averaged quantities over $10^3$ independent trajectories.

Below, we specify the system parameters for the two cases studied in our simulation.

\subsection{Flexible polymer bound to a Brownian particle}
In System \emph{A} [Fig.~\ref{fig1}(a)], we model the polymer with a Rouse chain connected via the potential
\begin{equation}\label{eq:U_Rouse2}
U^{A}=\frac{k^{A}}{2}\sum_{i=1}^{M_A-1}
\bigl | \mathbf r^{A}_{i+1}-\mathbf r^{A}_{i}\bigr |^{2},
\end{equation}
with $M_A=2001$. We set $k^{A}=10$ and $\gamma^{A}=0.1$, which makes the characteristic time of a monomer and the Rouse relaxation time
\begin{equation}
\tau_0^A \simeq \frac{\gamma^{A}}{k^{A}} \simeq 10^{-2}, \quad
\tau_R^A \simeq (M_A/2)^2 \tau_0^A \simeq 10^4,
\end{equation}
respectively. As shown in Appendix~\ref{sec:FLE_derivation}, a Rouse monomer in the intermediate timescale $\tau_0^A \ll t \ll \tau_R^A$ obeys a fractional Langevin equation (FLE) with the memory kernel given by Eq.~\eqref{eq:KA-asym}.
The central monomer ($a=1001$) is connected to a single Brownian particle (System \emph{B}): $U^{B}=0$, $M_B=1$, and $b=1$. The central monomer interacts with the Brownian particle via a harmonic spring of stiffness $k=0.3$. Combining $K_\alpha(t)\simeq K_A^\mathrm{flex}(t)$  explained in Eq.~\eqref{eq:KA-asym} and $K_\beta(t)=2\gamma^B \delta(t)$, the critical coupling strength $k_c$ is obtained from Eq.~\eqref{eq:k_c}. The corresponding dimensionless interaction strength is
\begin{equation}
\frac{k}{k_c} \simeq 7.5 > 1,
\end{equation}
placing the system in the strong-interaction regime.

\subsection{Crosslinked flexible and semiflexible polymer}
The system $A$ is modeled as a Rouse chain with the same potential of Eq.~\eqref{eq:U_Rouse2}. We set $M_A=2001$, $a=1001$, $k^A=40$, and $\gamma^A=0.4$, which results in
\begin{equation}
\tau_0^A \simeq \frac{\gamma^{A}}{k^{A}} \simeq 10^{-2}, \quad
\tau_R^A \simeq (M_A/2)^2 \tau_0^A \simeq 10^4.
\end{equation}
%so that $\mathbf{r}_a^A$ is well described by the FLE for $\tau_0^A \ll t \ll \tau_R^A$.
The system $B$ is a Gaussian semiflexible chain~\cite{winklerModelsEquilibriumProperties1994,durangGeneralizedLangevinEquation2024} with $M_B=2N_B+1$ beads. In this polymer model, the potential energy is given by
\begin{equation}
U^{B}=\frac{k^{B}}{2}\left[
\sum_{i=1}^{M_B-1}\lambda^{B}_i\,| \mathbf q_i^{B}|^2
+\sum_{i=1}^{M_B-2}\mu^{B}\,\mathbf q_i^{B}\cdot\mathbf q_{i+1}^{B}\right]
\end{equation}
where $\mathbf q_i^{B}=\mathbf r^{B}_{i+1}-\mathbf r^{B}_{i}$ is the bond vector. In R.H.S., the first term represents bond stretching and the second term accounts for bending energy. The spring constant is given by $k^{B}=d\,k_B T/\ell$, where $\ell$ is the bond length, yielding the contour length $L_c = 2N_B \ell$. The Lagrange multipliers $\lambda_i^B$ and $\mu^B$ are:
\begin{equation}
    \begin{aligned}
        \lambda_i^B&=
        \begin{cases}
            \frac{1}{2}\frac{1+\kappa^2}{1-\kappa^2}, &i=2,3,...2N_B-1\\
            \frac{1}{2}\frac{1}{1-\kappa^2}, &i=1,2N_B
        \end{cases},\\
        \mu^B&= \frac{\kappa}{1-\kappa^2},
    \end{aligned}
\end{equation}
which enforce the constraints
\begin{equation}
\begin{aligned}
    \langle |\mathbf{q}_{i}|^2\rangle  &= \ell^2,\\
\langle \mathbf{q}_{i+1} \cdot \mathbf{q}_{i}\rangle &=\kappa \ell^2.
\end{aligned}
\end{equation}
The parameter $0\leq\kappa<1$ determines the persistence length
\begin{equation}
    L_p = \frac{\ell}{1-\kappa}.
\end{equation}
For $L_c \lesssim L_p$, the chain behaves as a stiff polymer whose local fluctuations are described by an FLE with kernel~\cite{hanNonequilibriumDiffusionActive2023}
\begin{equation}\label{eq:kernel_stiff}
    K_B^\mathrm{stiff}(t) = c_\mathrm{stiff}\left((\gamma^B)^3  k^B\tfrac{ L_p }{ \ell}\right)^{1/4}t^{-3/4},
\end{equation}
where $c_\mathrm{stiff}$  is a numerical prefactor that originates from the limitation of the scaling theory.

We use $M_B=101$, $b=51$, $k^B=1$, $d=3$, $\gamma^B=2$, and $\kappa=0.995$, so that the persistence length exceeds the contour length ($L_p/L_c\simeq 2$). The characteristic timescales are
\begin{equation}
    \tau_0^B \simeq \frac{\gamma^{B}}{k^B}\frac{\ell}{L_p} \simeq 10^{-2}, \quad \tau^B_R \simeq (M_B/2)^2 \tau_0^A\simeq6.25\times10^4.
\end{equation}
In between these two timescales, the dynamics of the central monomer ($b$) is governed by the FLE with the kernel~\eqref{eq:kernel_stiff}. We have simulated the free Gaussian semiflexible chain ($k=0$) and obtained $c_\mathrm{stiff}\approx0.66$ by fitting the MSD.

The spring constant is set to $k=0.5$. Combining $K_\alpha(t)\simeq K_A^\mathrm{flex}(t)$ and $K_\beta(t)=K_B^\mathrm{stiff}(t)$, we find from Eq.~\eqref{eq:k_c} that
\begin{equation}
\frac{k}{k_c} \simeq 0.22 \lesssim 1,
\end{equation}
which places the system in the weak-interaction regime.

% \bibliography{references}
% \bibliography{references}

\begin{thebibliography}{85}%
	\makeatletter
	\providecommand \@ifxundefined [1]{%
		\@ifx{#1\undefined}
	}%
	\providecommand \@ifnum [1]{%
		\ifnum #1\expandafter \@firstoftwo
		\else \expandafter \@secondoftwo
		\fi
	}%
	\providecommand \@ifx [1]{%
		\ifx #1\expandafter \@firstoftwo
		\else \expandafter \@secondoftwo
		\fi
	}%
	\providecommand \natexlab [1]{#1}%
	\providecommand \enquote  [1]{``#1''}%
	\providecommand \bibnamefont  [1]{#1}%
	\providecommand \bibfnamefont [1]{#1}%
	\providecommand \citenamefont [1]{#1}%
	\providecommand \href@noop [0]{\@secondoftwo}%
	\providecommand \href [0]{\begingroup \@sanitize@url \@href}%
	\providecommand \@href[1]{\@@startlink{#1}\@@href}%
	\providecommand \@@href[1]{\endgroup#1\@@endlink}%
	\providecommand \@sanitize@url [0]{\catcode `\\12\catcode `\$12\catcode
		`\&12\catcode `\#12\catcode `\^12\catcode `\_12\catcode `\%12\relax}%
	\providecommand \@@startlink[1]{}%
	\providecommand \@@endlink[0]{}%
	\providecommand \url  [0]{\begingroup\@sanitize@url \@url }%
	\providecommand \@url [1]{\endgroup\@href {#1}{\urlprefix }}%
	\providecommand \urlprefix  [0]{URL }%
	\providecommand \Eprint [0]{\href }%
	\providecommand \doibase [0]{https://doi.org/}%
	\providecommand \selectlanguage [0]{\@gobble}%
	\providecommand \bibinfo  [0]{\@secondoftwo}%
	\providecommand \bibfield  [0]{\@secondoftwo}%
	\providecommand \translation [1]{[#1]}%
	\providecommand \BibitemOpen [0]{}%
	\providecommand \bibitemStop [0]{}%
	\providecommand \bibitemNoStop [0]{.\EOS\space}%
	\providecommand \EOS [0]{\spacefactor3000\relax}%
	\providecommand \BibitemShut  [1]{\csname bibitem#1\endcsname}%
	\let\auto@bib@innerbib\@empty
	%</preamble>
	\bibitem [{\citenamefont {Metzler}\ \emph {et~al.}(2014)\citenamefont
		{Metzler}, \citenamefont {Jeon}, \citenamefont {Cherstvy},\ and\
		\citenamefont {Barkai}}]{metzlerAnomalousDiffusionModels2014}%
	\BibitemOpen
	\bibfield  {author} {\bibinfo {author} {\bibfnamefont {R.}~\bibnamefont
			{Metzler}}, \bibinfo {author} {\bibfnamefont {J.-H.}\ \bibnamefont {Jeon}},
		\bibinfo {author} {\bibfnamefont {A.~G.}\ \bibnamefont {Cherstvy}},\ and\
		\bibinfo {author} {\bibfnamefont {E.}~\bibnamefont {Barkai}},\ }\bibfield
	{title} {\bibinfo {title} {Anomalous diffusion models and their properties:
			Non-stationarity, non-ergodicity, and ageing at the centenary of single
			particle tracking},\ }\href {https://doi.org/10.1039/C4CP03465A} {\bibfield
		{journal} {\bibinfo  {journal} {Physical Chemistry Chemical Physics}\
		}\textbf {\bibinfo {volume} {16}},\ \bibinfo {pages} {24128} (\bibinfo {year}
		{2014})}\BibitemShut {NoStop}%
	\bibitem [{\citenamefont {Baggioli}\ \emph {et~al.}(2021)\citenamefont
		{Baggioli}, \citenamefont {La~Nave},\ and\ \citenamefont
		{Phillips}}]{baggioliAnomalousDiffusionNoethers2021}%
	\BibitemOpen
	\bibfield  {author} {\bibinfo {author} {\bibfnamefont {M.}~\bibnamefont
			{Baggioli}}, \bibinfo {author} {\bibfnamefont {G.}~\bibnamefont {La~Nave}},\
		and\ \bibinfo {author} {\bibfnamefont {P.~W.}\ \bibnamefont {Phillips}},\
	}\bibfield  {title} {\bibinfo {title} {Anomalous diffusion and {{Noether}}'s
			second theorem},\ }\href {https://doi.org/10.1103/PhysRevE.103.032115}
	{\bibfield  {journal} {\bibinfo  {journal} {Physical Review E}\ }\textbf
		{\bibinfo {volume} {103}},\ \bibinfo {pages} {032115} (\bibinfo {year}
		{2021})}\BibitemShut {NoStop}%
	\bibitem [{\citenamefont {Ford}\ \emph {et~al.}(1965)\citenamefont {Ford},
		\citenamefont {Kac},\ and\ \citenamefont
		{Mazur}}]{fordStatisticalMechanicsAssemblies1965}%
	\BibitemOpen
	\bibfield  {author} {\bibinfo {author} {\bibfnamefont {G.~W.}\ \bibnamefont
			{Ford}}, \bibinfo {author} {\bibfnamefont {M.}~\bibnamefont {Kac}},\ and\
		\bibinfo {author} {\bibfnamefont {P.}~\bibnamefont {Mazur}},\ }\bibfield
	{title} {\bibinfo {title} {Statistical {{Mechanics}} of {{Assemblies}} of
			{{Coupled Oscillators}}},\ }\href {https://doi.org/10.1063/1.1704304}
	{\bibfield  {journal} {\bibinfo  {journal} {Journal of Mathematical Physics}\
		}\textbf {\bibinfo {volume} {6}},\ \bibinfo {pages} {504} (\bibinfo {year}
		{1965})}\BibitemShut {NoStop}%
	\bibitem [{\citenamefont {Mori}(1965)}]{moriTransportCollectiveMotion1965}%
	\BibitemOpen
	\bibfield  {author} {\bibinfo {author} {\bibfnamefont {H.}~\bibnamefont
			{Mori}},\ }\bibfield  {title} {\bibinfo {title} {Transport, {{Collective
					Motion}}, and {{Brownian Motion}}},\ }\href
	{https://doi.org/10.1143/PTP.33.423} {\bibfield  {journal} {\bibinfo
			{journal} {Progress of Theoretical Physics}\ }\textbf {\bibinfo {volume}
			{33}},\ \bibinfo {pages} {423} (\bibinfo {year} {1965})}\BibitemShut
	{NoStop}%
	\bibitem [{\citenamefont
		{Zwanzig}(1973)}]{zwanzigNonlinearGeneralizedLangevin1973}%
	\BibitemOpen
	\bibfield  {author} {\bibinfo {author} {\bibfnamefont {R.}~\bibnamefont
			{Zwanzig}},\ }\bibfield  {title} {\bibinfo {title} {Nonlinear generalized
			{{Langevin}} equations},\ }\href {https://doi.org/10.1007/BF01008729}
	{\bibfield  {journal} {\bibinfo  {journal} {Journal of Statistical Physics}\
		}\textbf {\bibinfo {volume} {9}},\ \bibinfo {pages} {215} (\bibinfo {year}
		{1973})}\BibitemShut {NoStop}%
	\bibitem [{\citenamefont {Kubo}(1966)}]{kuboFluctuationdissipationTheorem1966}%
	\BibitemOpen
	\bibfield  {author} {\bibinfo {author} {\bibfnamefont {R.}~\bibnamefont
			{Kubo}},\ }\bibfield  {title} {\bibinfo {title} {The fluctuation-dissipation
			theorem},\ }\href {https://doi.org/10.1088/0034-4885/29/1/306} {\bibfield
		{journal} {\bibinfo  {journal} {Reports on Progress in Physics}\ }\textbf
		{\bibinfo {volume} {29}},\ \bibinfo {pages} {255} (\bibinfo {year}
		{1966})}\BibitemShut {NoStop}%
	\bibitem [{\citenamefont {Lutz}(2001)}]{lutzFractionalLangevinEquation2001}%
	\BibitemOpen
	\bibfield  {author} {\bibinfo {author} {\bibfnamefont {E.}~\bibnamefont
			{Lutz}},\ }\bibfield  {title} {\bibinfo {title} {Fractional {{Langevin}}
			equation},\ }\href {https://doi.org/10.1103/PhysRevE.64.051106} {\bibfield
		{journal} {\bibinfo  {journal} {Physical Review E}\ }\textbf {\bibinfo
			{volume} {64}},\ \bibinfo {pages} {051106} (\bibinfo {year}
		{2001})}\BibitemShut {NoStop}%
	\bibitem [{\citenamefont
		{Goychuk}(2009)}]{goychukViscoelasticSubdiffusionAnomalous2009}%
	\BibitemOpen
	\bibfield  {author} {\bibinfo {author} {\bibfnamefont {I.}~\bibnamefont
			{Goychuk}},\ }\bibfield  {title} {\bibinfo {title} {Viscoelastic
			subdiffusion: {{From}} anomalous to normal},\ }\href
	{https://doi.org/10.1103/PhysRevE.80.046125} {\bibfield  {journal} {\bibinfo
			{journal} {Physical Review E}\ }\textbf {\bibinfo {volume} {80}},\ \bibinfo
		{pages} {046125} (\bibinfo {year} {2009})}\BibitemShut {NoStop}%
	\bibitem [{\citenamefont
		{Goychuk}(2012)}]{goychukViscoelasticSubdiffusionGeneralized2012}%
	\BibitemOpen
	\bibfield  {author} {\bibinfo {author} {\bibfnamefont {I.}~\bibnamefont
			{Goychuk}},\ }\bibfield  {title} {\bibinfo {title} {Viscoelastic
			{{Subdiffusion}}: {{Generalized Langevin Equation Approach}}},\ }in\ \href
	{https://doi.org/10.1002/9781118197714.ch5} {\emph {\bibinfo {booktitle}
			{Advances in {{Chemical Physics}}}}}\ (\bibinfo  {publisher} {John Wiley \&
		Sons, Ltd},\ \bibinfo {year} {2012})\ pp.\ \bibinfo {pages}
	{187--253}\BibitemShut {NoStop}%
	\bibitem [{\citenamefont {Kou}\ and\ \citenamefont
		{Xie}(2004)}]{kouGeneralizedLangevinEquation2004}%
	\BibitemOpen
	\bibfield  {author} {\bibinfo {author} {\bibfnamefont {S.~C.}\ \bibnamefont
			{Kou}}\ and\ \bibinfo {author} {\bibfnamefont {X.~S.}\ \bibnamefont {Xie}},\
	}\bibfield  {title} {\bibinfo {title} {Generalized {{Langevin Equation}} with
			{{Fractional Gaussian Noise}}: {{Subdiffusion}} within a {{Single Protein
					Molecule}}},\ }\href {https://doi.org/10.1103/PhysRevLett.93.180603}
	{\bibfield  {journal} {\bibinfo  {journal} {Physical Review Letters}\
		}\textbf {\bibinfo {volume} {93}},\ \bibinfo {pages} {180603} (\bibinfo
		{year} {2004})}\BibitemShut {NoStop}%
	\bibitem [{\citenamefont {Min}\ \emph {et~al.}(2005)\citenamefont {Min},
		\citenamefont {Luo}, \citenamefont {Cherayil}, \citenamefont {Kou},\ and\
		\citenamefont {Xie}}]{minObservationPowerLawMemory2005}%
	\BibitemOpen
	\bibfield  {author} {\bibinfo {author} {\bibfnamefont {W.}~\bibnamefont
			{Min}}, \bibinfo {author} {\bibfnamefont {G.}~\bibnamefont {Luo}}, \bibinfo
		{author} {\bibfnamefont {B.~J.}\ \bibnamefont {Cherayil}}, \bibinfo {author}
		{\bibfnamefont {S.~C.}\ \bibnamefont {Kou}},\ and\ \bibinfo {author}
		{\bibfnamefont {X.~S.}\ \bibnamefont {Xie}},\ }\bibfield  {title} {\bibinfo
		{title} {Observation of a {{Power-Law Memory Kernel}} for {{Fluctuations}}
			within a {{Single Protein Molecule}}},\ }\href
	{https://doi.org/10.1103/PhysRevLett.94.198302} {\bibfield  {journal}
		{\bibinfo  {journal} {Physical Review Letters}\ }\textbf {\bibinfo {volume}
			{94}},\ \bibinfo {pages} {198302} (\bibinfo {year} {2005})}\BibitemShut
	{NoStop}%
	\bibitem [{\citenamefont {Fabry}\ \emph {et~al.}(2001)\citenamefont {Fabry},
		\citenamefont {Maksym}, \citenamefont {Butler}, \citenamefont {Glogauer},
		\citenamefont {Navajas},\ and\ \citenamefont
		{Fredberg}}]{fabryScalingMicrorheologyLiving2001}%
	\BibitemOpen
	\bibfield  {author} {\bibinfo {author} {\bibfnamefont {B.}~\bibnamefont
			{Fabry}}, \bibinfo {author} {\bibfnamefont {G.~N.}\ \bibnamefont {Maksym}},
		\bibinfo {author} {\bibfnamefont {J.~P.}\ \bibnamefont {Butler}}, \bibinfo
		{author} {\bibfnamefont {M.}~\bibnamefont {Glogauer}}, \bibinfo {author}
		{\bibfnamefont {D.}~\bibnamefont {Navajas}},\ and\ \bibinfo {author}
		{\bibfnamefont {J.~J.}\ \bibnamefont {Fredberg}},\ }\bibfield  {title}
	{\bibinfo {title} {Scaling the {{Microrheology}} of {{Living Cells}}},\
	}\href {https://doi.org/10.1103/PhysRevLett.87.148102} {\bibfield  {journal}
		{\bibinfo  {journal} {Physical Review Letters}\ }\textbf {\bibinfo {volume}
			{87}},\ \bibinfo {pages} {148102} (\bibinfo {year} {2001})}\BibitemShut
	{NoStop}%
	\bibitem [{\citenamefont
		{Wilhelm}(2008)}]{wilhelmOutofEquilibriumMicrorheologyLiving2008}%
	\BibitemOpen
	\bibfield  {author} {\bibinfo {author} {\bibfnamefont {C.}~\bibnamefont
			{Wilhelm}},\ }\bibfield  {title} {\bibinfo {title} {Out-of-{{Equilibrium
					Microrheology}} inside {{Living Cells}}},\ }\href
	{https://doi.org/10.1103/PhysRevLett.101.028101} {\bibfield  {journal}
		{\bibinfo  {journal} {Physical Review Letters}\ }\textbf {\bibinfo {volume}
			{101}},\ \bibinfo {pages} {028101} (\bibinfo {year} {2008})}\BibitemShut
	{NoStop}%
	\bibitem [{\citenamefont {Guo}\ \emph {et~al.}(2014)\citenamefont {Guo},
		\citenamefont {Ehrlicher}, \citenamefont {Jensen}, \citenamefont {Renz},
		\citenamefont {Moore}, \citenamefont {Goldman}, \citenamefont
		{{Lippincott-Schwartz}}, \citenamefont {Mackintosh},\ and\ \citenamefont
		{Weitz}}]{guoProbingStochasticMotorDriven2014}%
	\BibitemOpen
	\bibfield  {author} {\bibinfo {author} {\bibfnamefont {M.}~\bibnamefont
			{Guo}}, \bibinfo {author} {\bibfnamefont {A.~J.}\ \bibnamefont {Ehrlicher}},
		\bibinfo {author} {\bibfnamefont {M.~H.}\ \bibnamefont {Jensen}}, \bibinfo
		{author} {\bibfnamefont {M.}~\bibnamefont {Renz}}, \bibinfo {author}
		{\bibfnamefont {J.~R.}\ \bibnamefont {Moore}}, \bibinfo {author}
		{\bibfnamefont {R.~D.}\ \bibnamefont {Goldman}}, \bibinfo {author}
		{\bibfnamefont {J.}~\bibnamefont {{Lippincott-Schwartz}}}, \bibinfo {author}
		{\bibfnamefont {F.~C.}\ \bibnamefont {Mackintosh}},\ and\ \bibinfo {author}
		{\bibfnamefont {D.~A.}\ \bibnamefont {Weitz}},\ }\bibfield  {title} {\bibinfo
		{title} {Probing the {{Stochastic}}, {{Motor-Driven Properties}} of the
			{{Cytoplasm Using Force Spectrum Microscopy}}},\ }\href
	{https://doi.org/10.1016/j.cell.2014.06.051} {\bibfield  {journal} {\bibinfo
			{journal} {Cell}\ }\textbf {\bibinfo {volume} {158}},\ \bibinfo {pages} {822}
		(\bibinfo {year} {2014})}\BibitemShut {NoStop}%
	\bibitem [{\citenamefont {Weber}\ \emph {et~al.}(2010)\citenamefont {Weber},
		\citenamefont {Spakowitz},\ and\ \citenamefont
		{Theriot}}]{weberBacterialChromosomalLoci2010}%
	\BibitemOpen
	\bibfield  {author} {\bibinfo {author} {\bibfnamefont {S.~C.}\ \bibnamefont
			{Weber}}, \bibinfo {author} {\bibfnamefont {A.~J.}\ \bibnamefont
			{Spakowitz}},\ and\ \bibinfo {author} {\bibfnamefont {J.~A.}\ \bibnamefont
			{Theriot}},\ }\bibfield  {title} {\bibinfo {title} {Bacterial {{Chromosomal
					Loci Move Subdiffusively}} through a {{Viscoelastic Cytoplasm}}},\ }\href
	{https://doi.org/10.1103/PhysRevLett.104.238102} {\bibfield  {journal}
		{\bibinfo  {journal} {Physical Review Letters}\ }\textbf {\bibinfo {volume}
			{104}},\ \bibinfo {pages} {238102} (\bibinfo {year} {2010})}\BibitemShut
	{NoStop}%
	\bibitem [{\citenamefont {Lampo}\ \emph {et~al.}(2015)\citenamefont {Lampo},
		\citenamefont {Kuwada}, \citenamefont {Wiggins},\ and\ \citenamefont
		{Spakowitz}}]{lampoPhysicalModelingChromosome2015}%
	\BibitemOpen
	\bibfield  {author} {\bibinfo {author} {\bibfnamefont {T.~J.}\ \bibnamefont
			{Lampo}}, \bibinfo {author} {\bibfnamefont {N.~J.}\ \bibnamefont {Kuwada}},
		\bibinfo {author} {\bibfnamefont {P.~A.}\ \bibnamefont {Wiggins}},\ and\
		\bibinfo {author} {\bibfnamefont {A.~J.}\ \bibnamefont {Spakowitz}},\
	}\bibfield  {title} {\bibinfo {title} {Physical {{Modeling}} of {{Chromosome
					Segregation}} in {{{\emph{Escherichia}}}}{\emph{ coli}} {{Reveals Impact}} of
			{{Force}} and {{DNA Relaxation}}},\ }\href
	{https://doi.org/10.1016/j.bpj.2014.10.074} {\bibfield  {journal} {\bibinfo
			{journal} {Biophysical Journal}\ }\textbf {\bibinfo {volume} {108}},\
		\bibinfo {pages} {146} (\bibinfo {year} {2015})}\BibitemShut {NoStop}%
	\bibitem [{\citenamefont {Mason}\ \emph {et~al.}(2000)\citenamefont {Mason},
		\citenamefont {Gisler}, \citenamefont {Kroy}, \citenamefont {Frey},\ and\
		\citenamefont {Weitz}}]{masonRheologyFactinSolutions2000}%
	\BibitemOpen
	\bibfield  {author} {\bibinfo {author} {\bibfnamefont {T.~G.}\ \bibnamefont
			{Mason}}, \bibinfo {author} {\bibfnamefont {T.}~\bibnamefont {Gisler}},
		\bibinfo {author} {\bibfnamefont {K.}~\bibnamefont {Kroy}}, \bibinfo {author}
		{\bibfnamefont {E.}~\bibnamefont {Frey}},\ and\ \bibinfo {author}
		{\bibfnamefont {D.~A.}\ \bibnamefont {Weitz}},\ }\bibfield  {title} {\bibinfo
		{title} {Rheology of {{F-actin}} solutions determined from thermally driven
			tracer motion},\ }\href {https://doi.org/10.1122/1.551113} {\bibfield
		{journal} {\bibinfo  {journal} {Journal of Rheology}\ }\textbf {\bibinfo
			{volume} {44}},\ \bibinfo {pages} {917} (\bibinfo {year} {2000})}\BibitemShut
	{NoStop}%
	\bibitem [{\citenamefont {Gisler}\ and\ \citenamefont
		{Weitz}(1999)}]{gislerScalingMicrorheologySemidilute1999}%
	\BibitemOpen
	\bibfield  {author} {\bibinfo {author} {\bibfnamefont {T.}~\bibnamefont
			{Gisler}}\ and\ \bibinfo {author} {\bibfnamefont {D.~A.}\ \bibnamefont
			{Weitz}},\ }\bibfield  {title} {\bibinfo {title} {Scaling of the
			{{Microrheology}} of {{Semidilute F-Actin Solutions}}},\ }\href
	{https://doi.org/10.1103/PhysRevLett.82.1606} {\bibfield  {journal} {\bibinfo
			{journal} {Physical Review Letters}\ }\textbf {\bibinfo {volume} {82}},\
		\bibinfo {pages} {1606} (\bibinfo {year} {1999})}\BibitemShut {NoStop}%
	\bibitem [{\citenamefont {Schnurr}\ \emph {et~al.}(1997)\citenamefont
		{Schnurr}, \citenamefont {Gittes}, \citenamefont {MacKintosh},\ and\
		\citenamefont {Schmidt}}]{schnurrDeterminingMicroscopicViscoelasticity1997}%
	\BibitemOpen
	\bibfield  {author} {\bibinfo {author} {\bibfnamefont {B.}~\bibnamefont
			{Schnurr}}, \bibinfo {author} {\bibfnamefont {F.}~\bibnamefont {Gittes}},
		\bibinfo {author} {\bibfnamefont {F.~C.}\ \bibnamefont {MacKintosh}},\ and\
		\bibinfo {author} {\bibfnamefont {C.~F.}\ \bibnamefont {Schmidt}},\
	}\bibfield  {title} {\bibinfo {title} {Determining {{Microscopic
					Viscoelasticity}} in {{Flexible}} and {{Semiflexible Polymer Networks}} from
			{{Thermal Fluctuations}}},\ }\href {https://doi.org/10.1021/ma970555n}
	{\bibfield  {journal} {\bibinfo  {journal} {Macromolecules}\ }\textbf
		{\bibinfo {volume} {30}},\ \bibinfo {pages} {7781} (\bibinfo {year}
		{1997})}\BibitemShut {NoStop}%
	\bibitem [{\citenamefont {Dasgupta}\ \emph {et~al.}(2002)\citenamefont
		{Dasgupta}, \citenamefont {Tee}, \citenamefont {Crocker}, \citenamefont
		{Frisken},\ and\ \citenamefont
		{Weitz}}]{dasguptaMicrorheologyPolyethyleneOxide2002}%
	\BibitemOpen
	\bibfield  {author} {\bibinfo {author} {\bibfnamefont {B.~R.}\ \bibnamefont
			{Dasgupta}}, \bibinfo {author} {\bibfnamefont {S.-Y.}\ \bibnamefont {Tee}},
		\bibinfo {author} {\bibfnamefont {J.~C.}\ \bibnamefont {Crocker}}, \bibinfo
		{author} {\bibfnamefont {B.~J.}\ \bibnamefont {Frisken}},\ and\ \bibinfo
		{author} {\bibfnamefont {D.~A.}\ \bibnamefont {Weitz}},\ }\bibfield  {title}
	{\bibinfo {title} {Microrheology of polyethylene oxide using diffusing wave
			spectroscopy and single scattering},\ }\href
	{https://doi.org/10.1103/PhysRevE.65.051505} {\bibfield  {journal} {\bibinfo
			{journal} {Physical Review E}\ }\textbf {\bibinfo {volume} {65}},\ \bibinfo
		{pages} {051505} (\bibinfo {year} {2002})}\BibitemShut {NoStop}%
	\bibitem [{\citenamefont {Caputo}(1967)}]{caputoLinearModelsDissipation1967}%
	\BibitemOpen
	\bibfield  {author} {\bibinfo {author} {\bibfnamefont {M.}~\bibnamefont
			{Caputo}},\ }\bibfield  {title} {\bibinfo {title} {Linear {{Models}} of
			{{Dissipation}} whose {{Q}} is almost {{Frequency Independent}}---{{II}}},\
	}\href {https://doi.org/10.1111/j.1365-246X.1967.tb02303.x} {\bibfield
		{journal} {\bibinfo  {journal} {Geophysical Journal International}\ }\textbf
		{\bibinfo {volume} {13}},\ \bibinfo {pages} {529} (\bibinfo {year}
		{1967})}\BibitemShut {NoStop}%
	\bibitem [{\citenamefont {{de Oliveira}}\ and\ \citenamefont
		{Tenreiro~Machado}(2014)}]{deoliveiraReviewDefinitionsFractional2014}%
	\BibitemOpen
	\bibfield  {author} {\bibinfo {author} {\bibfnamefont {E.~C.}\ \bibnamefont
			{{de Oliveira}}}\ and\ \bibinfo {author} {\bibfnamefont {J.~A.}\ \bibnamefont
			{Tenreiro~Machado}},\ }\bibfield  {title} {\bibinfo {title} {A {{Review}} of
			{{Definitions}} for {{Fractional Derivatives}} and {{Integral}}},\ }\href
	{https://doi.org/10.1155/2014/238459} {\bibfield  {journal} {\bibinfo
			{journal} {Mathematical Problems in Engineering}\ }\textbf {\bibinfo {volume}
			{2014}},\ \bibinfo {pages} {238459} (\bibinfo {year} {2014})}\BibitemShut
	{NoStop}%
	\bibitem [{\citenamefont {Mason}\ and\ \citenamefont
		{Weitz}(1995)}]{masonOpticalMeasurementsFrequencyDependent1995}%
	\BibitemOpen
	\bibfield  {author} {\bibinfo {author} {\bibfnamefont {T.~G.}\ \bibnamefont
			{Mason}}\ and\ \bibinfo {author} {\bibfnamefont {D.~A.}\ \bibnamefont
			{Weitz}},\ }\bibfield  {title} {\bibinfo {title} {Optical {{Measurements}} of
			{{Frequency-Dependent Linear Viscoelastic Moduli}} of {{Complex Fluids}}},\
	}\href {https://doi.org/10.1103/PhysRevLett.74.1250} {\bibfield  {journal}
		{\bibinfo  {journal} {Physical Review Letters}\ }\textbf {\bibinfo {volume}
			{74}},\ \bibinfo {pages} {1250} (\bibinfo {year} {1995})}\BibitemShut
	{NoStop}%
	\bibitem [{\citenamefont {Goychuk}\ and\ \citenamefont
		{H{\"a}nggi}(2007)}]{goychukAnomalousEscapeGoverned2007}%
	\BibitemOpen
	\bibfield  {author} {\bibinfo {author} {\bibfnamefont {I.}~\bibnamefont
			{Goychuk}}\ and\ \bibinfo {author} {\bibfnamefont {P.}~\bibnamefont
			{H{\"a}nggi}},\ }\bibfield  {title} {\bibinfo {title} {Anomalous {{Escape
					Governed}} by {{Thermal}} 1/f {{Noise}}},\ }\href
	{https://doi.org/10.1103/PhysRevLett.99.200601} {\bibfield  {journal}
		{\bibinfo  {journal} {Physical Review Letters}\ }\textbf {\bibinfo {volume}
			{99}},\ \bibinfo {pages} {200601} (\bibinfo {year} {2007})}\BibitemShut
	{NoStop}%
	\bibitem [{\citenamefont {Taloni}\ \emph {et~al.}(2010)\citenamefont {Taloni},
		\citenamefont {Chechkin},\ and\ \citenamefont
		{Klafter}}]{taloniGeneralizedElasticModel2010}%
	\BibitemOpen
	\bibfield  {author} {\bibinfo {author} {\bibfnamefont {A.}~\bibnamefont
			{Taloni}}, \bibinfo {author} {\bibfnamefont {A.}~\bibnamefont {Chechkin}},\
		and\ \bibinfo {author} {\bibfnamefont {J.}~\bibnamefont {Klafter}},\
	}\bibfield  {title} {\bibinfo {title} {Generalized {{Elastic Model Yields}} a
			{{Fractional Langevin Equation Description}}},\ }\href
	{https://doi.org/10.1103/PhysRevLett.104.160602} {\bibfield  {journal}
		{\bibinfo  {journal} {Physical Review Letters}\ }\textbf {\bibinfo {volume}
			{104}},\ \bibinfo {pages} {160602} (\bibinfo {year} {2010})}\BibitemShut
	{NoStop}%
	\bibitem [{\citenamefont
		{Panja}(2010{\natexlab{a}})}]{panjaAnomalousPolymerDynamics2010}%
	\BibitemOpen
	\bibfield  {author} {\bibinfo {author} {\bibfnamefont {D.}~\bibnamefont
			{Panja}},\ }\bibfield  {title} {\bibinfo {title} {Anomalous polymer dynamics
			is non-{{Markovian}}: Memory effects and the generalized {{Langevin}}
			equation formulation},\ }\href
	{https://doi.org/10.1088/1742-5468/2010/06/P06011} {\bibfield  {journal}
		{\bibinfo  {journal} {Journal of Statistical Mechanics: Theory and
				Experiment}\ }\textbf {\bibinfo {volume} {2010}},\ \bibinfo {pages} {P06011}
		(\bibinfo {year} {2010}{\natexlab{a}})}\BibitemShut {NoStop}%
	\bibitem [{\citenamefont {Lizana}\ \emph {et~al.}(2010)\citenamefont {Lizana},
		\citenamefont {Ambj{\"o}rnsson}, \citenamefont {Taloni}, \citenamefont
		{Barkai},\ and\ \citenamefont
		{Lomholt}}]{lizanaFoundationFractionalLangevin2010}%
	\BibitemOpen
	\bibfield  {author} {\bibinfo {author} {\bibfnamefont {L.}~\bibnamefont
			{Lizana}}, \bibinfo {author} {\bibfnamefont {T.}~\bibnamefont
			{Ambj{\"o}rnsson}}, \bibinfo {author} {\bibfnamefont {A.}~\bibnamefont
			{Taloni}}, \bibinfo {author} {\bibfnamefont {E.}~\bibnamefont {Barkai}},\
		and\ \bibinfo {author} {\bibfnamefont {M.~A.}\ \bibnamefont {Lomholt}},\
	}\bibfield  {title} {\bibinfo {title} {Foundation of fractional {{Langevin}}
			equation: {{Harmonization}} of a many-body problem},\ }\href
	{https://doi.org/10.1103/PhysRevE.81.051118} {\bibfield  {journal} {\bibinfo
			{journal} {Physical Review E}\ }\textbf {\bibinfo {volume} {81}},\ \bibinfo
		{pages} {051118} (\bibinfo {year} {2010})}\BibitemShut {NoStop}%
	\bibitem [{\citenamefont {Vandebroek}\ and\ \citenamefont
		{Vanderzande}(2017)}]{vandebroekGeneralizedLangevinEquation2017}%
	\BibitemOpen
	\bibfield  {author} {\bibinfo {author} {\bibfnamefont {H.}~\bibnamefont
			{Vandebroek}}\ and\ \bibinfo {author} {\bibfnamefont {C.}~\bibnamefont
			{Vanderzande}},\ }\bibfield  {title} {\bibinfo {title} {On the {{Generalized
					Langevin Equation}} for a {{Rouse Bead}} in a {{Nonequilibrium Bath}}},\
	}\href {https://doi.org/10.1007/s10955-017-1734-x} {\bibfield  {journal}
		{\bibinfo  {journal} {Journal of Statistical Physics}\ }\textbf {\bibinfo
			{volume} {167}},\ \bibinfo {pages} {14} (\bibinfo {year} {2017})}\BibitemShut
	{NoStop}%
	\bibitem [{\citenamefont {Maes}\ and\ \citenamefont
		{Thomas}(2013)}]{maesLangevinGeneralizedLangevin2013}%
	\BibitemOpen
	\bibfield  {author} {\bibinfo {author} {\bibfnamefont {C.}~\bibnamefont
			{Maes}}\ and\ \bibinfo {author} {\bibfnamefont {S.~R.}\ \bibnamefont
			{Thomas}},\ }\bibfield  {title} {\bibinfo {title} {From {{Langevin}} to
			generalized {{Langevin}} equations for the nonequilibrium {{Rouse}} model},\
	}\href {https://doi.org/10.1103/PhysRevE.87.022145} {\bibfield  {journal}
		{\bibinfo  {journal} {Physical Review E}\ }\textbf {\bibinfo {volume} {87}},\
		\bibinfo {pages} {022145} (\bibinfo {year} {2013})}\BibitemShut {NoStop}%
	\bibitem [{\citenamefont {Shinkai}\ \emph {et~al.}(2024)\citenamefont
		{Shinkai}, \citenamefont {Onami},\ and\ \citenamefont
		{Miyaguchi}}]{shinkaiGeneralizedLangevinDynamics2024}%
	\BibitemOpen
	\bibfield  {author} {\bibinfo {author} {\bibfnamefont {S.}~\bibnamefont
			{Shinkai}}, \bibinfo {author} {\bibfnamefont {S.}~\bibnamefont {Onami}},\
		and\ \bibinfo {author} {\bibfnamefont {T.}~\bibnamefont {Miyaguchi}},\
	}\bibfield  {title} {\bibinfo {title} {Generalized {{Langevin}} dynamics for
			single beads in linear elastic networks},\ }\href
	{https://doi.org/10.1103/PhysRevE.110.044136} {\bibfield  {journal} {\bibinfo
			{journal} {Physical Review E}\ }\textbf {\bibinfo {volume} {110}},\ \bibinfo
		{pages} {044136} (\bibinfo {year} {2024})}\BibitemShut {NoStop}%
	\bibitem [{\citenamefont {Han}\ \emph {et~al.}(2023)\citenamefont {Han},
		\citenamefont {Joo}, \citenamefont {Sakaue},\ and\ \citenamefont
		{Jeon}}]{hanNonequilibriumDiffusionActive2023}%
	\BibitemOpen
	\bibfield  {author} {\bibinfo {author} {\bibfnamefont {H.-T.}\ \bibnamefont
			{Han}}, \bibinfo {author} {\bibfnamefont {S.}~\bibnamefont {Joo}}, \bibinfo
		{author} {\bibfnamefont {T.}~\bibnamefont {Sakaue}},\ and\ \bibinfo {author}
		{\bibfnamefont {J.-H.}\ \bibnamefont {Jeon}},\ }\bibfield  {title} {\bibinfo
		{title} {Nonequilibrium diffusion of active particles bound to a semiflexible
			polymer network: {{Simulations}} and fractional {{Langevin}} equation},\
	}\href {https://doi.org/10.1063/5.0150224} {\bibfield  {journal} {\bibinfo
			{journal} {The Journal of Chemical Physics}\ }\textbf {\bibinfo {volume}
			{159}},\ \bibinfo {pages} {024901} (\bibinfo {year} {2023})}\BibitemShut
	{NoStop}%
	\bibitem [{\citenamefont {Durang}\ \emph {et~al.}(2024)\citenamefont {Durang},
		\citenamefont {Lim},\ and\ \citenamefont
		{Jeon}}]{durangGeneralizedLangevinEquation2024}%
	\BibitemOpen
	\bibfield  {author} {\bibinfo {author} {\bibfnamefont {X.}~\bibnamefont
			{Durang}}, \bibinfo {author} {\bibfnamefont {C.}~\bibnamefont {Lim}},\ and\
		\bibinfo {author} {\bibfnamefont {J.-H.}\ \bibnamefont {Jeon}},\ }\bibfield
	{title} {\bibinfo {title} {Generalized {{Langevin}} equation for a tagged
			monomer in a {{Gaussian}} semiflexible polymer},\ }\href
	{https://doi.org/10.1063/5.0229919} {\bibfield  {journal} {\bibinfo
			{journal} {The Journal of Chemical Physics}\ }\textbf {\bibinfo {volume}
			{161}},\ \bibinfo {pages} {244906} (\bibinfo {year} {2024})}\BibitemShut
	{NoStop}%
	\bibitem [{\citenamefont {Sakaue}(2013)}]{sakaueMemoryEffectFluctuating2013}%
	\BibitemOpen
	\bibfield  {author} {\bibinfo {author} {\bibfnamefont {T.}~\bibnamefont
			{Sakaue}},\ }\bibfield  {title} {\bibinfo {title} {Memory effect and
			fluctuating anomalous dynamics of a tagged monomer},\ }\href
	{https://doi.org/10.1103/PhysRevE.87.040601} {\bibfield  {journal} {\bibinfo
			{journal} {Physical Review E}\ }\textbf {\bibinfo {volume} {87}},\ \bibinfo
		{pages} {040601} (\bibinfo {year} {2013})}\BibitemShut {NoStop}%
	\bibitem [{\citenamefont {Kappler}\ \emph {et~al.}(2019)\citenamefont
		{Kappler}, \citenamefont {No{\'e}},\ and\ \citenamefont
		{Netz}}]{kapplerCyclizationRelaxationDynamics2019}%
	\BibitemOpen
	\bibfield  {author} {\bibinfo {author} {\bibfnamefont {J.}~\bibnamefont
			{Kappler}}, \bibinfo {author} {\bibfnamefont {F.}~\bibnamefont {No{\'e}}},\
		and\ \bibinfo {author} {\bibfnamefont {R.~R.}\ \bibnamefont {Netz}},\
	}\bibfield  {title} {\bibinfo {title} {Cyclization and {{Relaxation
					Dynamics}} of {{Finite-Length Collapsed Self-Avoiding Polymers}}},\ }\href
	{https://doi.org/10.1103/PhysRevLett.122.067801} {\bibfield  {journal}
		{\bibinfo  {journal} {Physical Review Letters}\ }\textbf {\bibinfo {volume}
			{122}},\ \bibinfo {pages} {067801} (\bibinfo {year} {2019})}\BibitemShut
	{NoStop}%
	\bibitem [{\citenamefont
		{Panja}(2010{\natexlab{b}})}]{panjaGeneralizedLangevinEquation2010}%
	\BibitemOpen
	\bibfield  {author} {\bibinfo {author} {\bibfnamefont {D.}~\bibnamefont
			{Panja}},\ }\bibfield  {title} {\bibinfo {title} {Generalized {{Langevin}}
			equation formulation for anomalous polymer dynamics},\ }\href
	{https://doi.org/10.1088/1742-5468/2010/02/L02001} {\bibfield  {journal}
		{\bibinfo  {journal} {Journal of Statistical Mechanics: Theory and
				Experiment}\ }\textbf {\bibinfo {volume} {2010}},\ \bibinfo {pages} {L02001}
		(\bibinfo {year} {2010}{\natexlab{b}})}\BibitemShut {NoStop}%
	\bibitem [{\citenamefont {Rigato}\ \emph {et~al.}(2017)\citenamefont {Rigato},
		\citenamefont {Miyagi}, \citenamefont {Scheuring},\ and\ \citenamefont
		{Rico}}]{rigatoHighfrequencyMicrorheologyReveals2017}%
	\BibitemOpen
	\bibfield  {author} {\bibinfo {author} {\bibfnamefont {A.}~\bibnamefont
			{Rigato}}, \bibinfo {author} {\bibfnamefont {A.}~\bibnamefont {Miyagi}},
		\bibinfo {author} {\bibfnamefont {S.}~\bibnamefont {Scheuring}},\ and\
		\bibinfo {author} {\bibfnamefont {F.}~\bibnamefont {Rico}},\ }\bibfield
	{title} {\bibinfo {title} {High-frequency microrheology reveals cytoskeleton
			dynamics in living cells},\ }\href {https://doi.org/10.1038/nphys4104}
	{\bibfield  {journal} {\bibinfo  {journal} {Nature Physics}\ }\textbf
		{\bibinfo {volume} {13}},\ \bibinfo {pages} {771} (\bibinfo {year}
		{2017})}\BibitemShut {NoStop}%
	\bibitem [{\citenamefont {Jeon}\ \emph {et~al.}(2012)\citenamefont {Jeon},
		\citenamefont {Monne}, \citenamefont {Javanainen},\ and\ \citenamefont
		{Metzler}}]{jeonAnomalousDiffusionPhospholipids2012}%
	\BibitemOpen
	\bibfield  {author} {\bibinfo {author} {\bibfnamefont {J.-H.}\ \bibnamefont
			{Jeon}}, \bibinfo {author} {\bibfnamefont {H.~M.-S.}\ \bibnamefont {Monne}},
		\bibinfo {author} {\bibfnamefont {M.}~\bibnamefont {Javanainen}},\ and\
		\bibinfo {author} {\bibfnamefont {R.}~\bibnamefont {Metzler}},\ }\bibfield
	{title} {\bibinfo {title} {Anomalous {{Diffusion}} of {{Phospholipids}} and
			{{Cholesterols}} in a {{Lipid Bilayer}} and its {{Origins}}},\ }\href
	{https://doi.org/10.1103/PhysRevLett.109.188103} {\bibfield  {journal}
		{\bibinfo  {journal} {Physical Review Letters}\ }\textbf {\bibinfo {volume}
			{109}},\ \bibinfo {pages} {188103} (\bibinfo {year} {2012})}\BibitemShut
	{NoStop}%
	\bibitem [{\citenamefont {Yamamoto}\ \emph {et~al.}(2017)\citenamefont
		{Yamamoto}, \citenamefont {Akimoto}, \citenamefont {Kalli}, \citenamefont
		{Yasuoka},\ and\ \citenamefont
		{Sansom}}]{yamamotoDynamicInteractionsMembrane2017}%
	\BibitemOpen
	\bibfield  {author} {\bibinfo {author} {\bibfnamefont {E.}~\bibnamefont
			{Yamamoto}}, \bibinfo {author} {\bibfnamefont {T.}~\bibnamefont {Akimoto}},
		\bibinfo {author} {\bibfnamefont {A.~C.}\ \bibnamefont {Kalli}}, \bibinfo
		{author} {\bibfnamefont {K.}~\bibnamefont {Yasuoka}},\ and\ \bibinfo {author}
		{\bibfnamefont {M.~S.~P.}\ \bibnamefont {Sansom}},\ }\bibfield  {title}
	{\bibinfo {title} {Dynamic interactions between a membrane binding protein
			and lipids induce fluctuating diffusivity},\ }\href
	{https://doi.org/10.1126/sciadv.1601871} {\bibfield  {journal} {\bibinfo
			{journal} {Science Advances}\ }\textbf {\bibinfo {volume} {3}},\ \bibinfo
		{pages} {e1601871} (\bibinfo {year} {2017})}\BibitemShut {NoStop}%
	\bibitem [{\citenamefont {Watanabe}\ \emph {et~al.}(2024)\citenamefont
		{Watanabe}, \citenamefont {Akimoto}, \citenamefont {Best}, \citenamefont
		{{Lindorff-Larsen}}, \citenamefont {Metzler},\ and\ \citenamefont
		{Yamamoto}}]{watanabeDiffusionIntrinsicallyDisordered2024}%
	\BibitemOpen
	\bibfield  {author} {\bibinfo {author} {\bibfnamefont {F.}~\bibnamefont
			{Watanabe}}, \bibinfo {author} {\bibfnamefont {T.}~\bibnamefont {Akimoto}},
		\bibinfo {author} {\bibfnamefont {R.~B.}\ \bibnamefont {Best}}, \bibinfo
		{author} {\bibfnamefont {K.}~\bibnamefont {{Lindorff-Larsen}}}, \bibinfo
		{author} {\bibfnamefont {R.}~\bibnamefont {Metzler}},\ and\ \bibinfo {author}
		{\bibfnamefont {E.}~\bibnamefont {Yamamoto}},\ }\href
	{https://doi.org/10.48550/arXiv.2401.10438} {\bibinfo {title} {Diffusion of
			intrinsically disordered proteins within viscoelastic membraneless droplets}}
	(\bibinfo {year} {2024}),\ \Eprint {https://arxiv.org/abs/2401.10438}
	{arXiv:2401.10438 [cond-mat]} \BibitemShut {NoStop}%
	\bibitem [{\citenamefont {Germier}\ \emph {et~al.}(2017)\citenamefont
		{Germier}, \citenamefont {Kocanova}, \citenamefont {Walther}, \citenamefont
		{Bancaud}, \citenamefont {Shaban}, \citenamefont {Sellou}, \citenamefont
		{Politi}, \citenamefont {Ellenberg}, \citenamefont {Gallardo},\ and\
		\citenamefont {Bystricky}}]{germierRealTimeImagingSingle2017}%
	\BibitemOpen
	\bibfield  {author} {\bibinfo {author} {\bibfnamefont {T.}~\bibnamefont
			{Germier}}, \bibinfo {author} {\bibfnamefont {S.}~\bibnamefont {Kocanova}},
		\bibinfo {author} {\bibfnamefont {N.}~\bibnamefont {Walther}}, \bibinfo
		{author} {\bibfnamefont {A.}~\bibnamefont {Bancaud}}, \bibinfo {author}
		{\bibfnamefont {H.~A.}\ \bibnamefont {Shaban}}, \bibinfo {author}
		{\bibfnamefont {H.}~\bibnamefont {Sellou}}, \bibinfo {author} {\bibfnamefont
			{A.~Z.}\ \bibnamefont {Politi}}, \bibinfo {author} {\bibfnamefont
			{J.}~\bibnamefont {Ellenberg}}, \bibinfo {author} {\bibfnamefont
			{F.}~\bibnamefont {Gallardo}},\ and\ \bibinfo {author} {\bibfnamefont
			{K.}~\bibnamefont {Bystricky}},\ }\bibfield  {title} {\bibinfo {title}
		{Real-{{Time Imaging}} of a {{Single Gene Reveals Transcription-Initiated
					Local Confinement}}},\ }\href {https://doi.org/10.1016/j.bpj.2017.08.014}
	{\bibfield  {journal} {\bibinfo  {journal} {Biophysical Journal}\ }\textbf
		{\bibinfo {volume} {113}},\ \bibinfo {pages} {1383} (\bibinfo {year}
		{2017})}\BibitemShut {NoStop}%
	\bibitem [{\citenamefont {Zidovska}\ \emph {et~al.}(2013)\citenamefont
		{Zidovska}, \citenamefont {Weitz},\ and\ \citenamefont
		{Mitchison}}]{zidovskaMicronscaleCoherenceInterphase2013}%
	\BibitemOpen
	\bibfield  {author} {\bibinfo {author} {\bibfnamefont {A.}~\bibnamefont
			{Zidovska}}, \bibinfo {author} {\bibfnamefont {D.~A.}\ \bibnamefont
			{Weitz}},\ and\ \bibinfo {author} {\bibfnamefont {T.~J.}\ \bibnamefont
			{Mitchison}},\ }\bibfield  {title} {\bibinfo {title} {Micron-scale coherence
			in interphase chromatin dynamics},\ }\href
	{https://doi.org/10.1073/pnas.1220313110} {\bibfield  {journal} {\bibinfo
			{journal} {Proceedings of the National Academy of Sciences}\ }\textbf
		{\bibinfo {volume} {110}},\ \bibinfo {pages} {15555} (\bibinfo {year}
		{2013})}\BibitemShut {NoStop}%
	\bibitem [{\citenamefont {Bronstein}\ \emph {et~al.}(2009)\citenamefont
		{Bronstein}, \citenamefont {Israel}, \citenamefont {Kepten}, \citenamefont
		{Mai}, \citenamefont {{Shav-Tal}}, \citenamefont {Barkai},\ and\
		\citenamefont {Garini}}]{bronsteinTransientAnomalousDiffusion2009}%
	\BibitemOpen
	\bibfield  {author} {\bibinfo {author} {\bibfnamefont {I.}~\bibnamefont
			{Bronstein}}, \bibinfo {author} {\bibfnamefont {Y.}~\bibnamefont {Israel}},
		\bibinfo {author} {\bibfnamefont {E.}~\bibnamefont {Kepten}}, \bibinfo
		{author} {\bibfnamefont {S.}~\bibnamefont {Mai}}, \bibinfo {author}
		{\bibfnamefont {Y.}~\bibnamefont {{Shav-Tal}}}, \bibinfo {author}
		{\bibfnamefont {E.}~\bibnamefont {Barkai}},\ and\ \bibinfo {author}
		{\bibfnamefont {Y.}~\bibnamefont {Garini}},\ }\bibfield  {title} {\bibinfo
		{title} {Transient {{Anomalous Diffusion}} of {{Telomeres}} in the
			{{Nucleus}} of {{Mammalian Cells}}},\ }\href
	{https://doi.org/10.1103/PhysRevLett.103.018102} {\bibfield  {journal}
		{\bibinfo  {journal} {Physical Review Letters}\ }\textbf {\bibinfo {volume}
			{103}},\ \bibinfo {pages} {018102} (\bibinfo {year} {2009})}\BibitemShut
	{NoStop}%
	\bibitem [{\citenamefont {Bronshtein}\ \emph {et~al.}(2015)\citenamefont
		{Bronshtein}, \citenamefont {Kepten}, \citenamefont {Kanter}, \citenamefont
		{Berezin}, \citenamefont {Lindner}, \citenamefont {Redwood}, \citenamefont
		{Mai}, \citenamefont {Gonzalo}, \citenamefont {Foisner}, \citenamefont
		{{Shav-Tal}},\ and\ \citenamefont
		{Garini}}]{bronshteinLossLaminFunction2015}%
	\BibitemOpen
	\bibfield  {author} {\bibinfo {author} {\bibfnamefont {I.}~\bibnamefont
			{Bronshtein}}, \bibinfo {author} {\bibfnamefont {E.}~\bibnamefont {Kepten}},
		\bibinfo {author} {\bibfnamefont {I.}~\bibnamefont {Kanter}}, \bibinfo
		{author} {\bibfnamefont {S.}~\bibnamefont {Berezin}}, \bibinfo {author}
		{\bibfnamefont {M.}~\bibnamefont {Lindner}}, \bibinfo {author} {\bibfnamefont
			{A.~B.}\ \bibnamefont {Redwood}}, \bibinfo {author} {\bibfnamefont
			{S.}~\bibnamefont {Mai}}, \bibinfo {author} {\bibfnamefont {S.}~\bibnamefont
			{Gonzalo}}, \bibinfo {author} {\bibfnamefont {R.}~\bibnamefont {Foisner}},
		\bibinfo {author} {\bibfnamefont {Y.}~\bibnamefont {{Shav-Tal}}},\ and\
		\bibinfo {author} {\bibfnamefont {Y.}~\bibnamefont {Garini}},\ }\bibfield
	{title} {\bibinfo {title} {Loss of lamin {{A}} function increases chromatin
			dynamics in the nuclear interior},\ }\href
	{https://doi.org/10.1038/ncomms9044} {\bibfield  {journal} {\bibinfo
			{journal} {Nature Communications}\ }\textbf {\bibinfo {volume} {6}},\
		\bibinfo {pages} {8044} (\bibinfo {year} {2015})}\BibitemShut {NoStop}%
	\bibitem [{\citenamefont {Ashwin}\ \emph {et~al.}(2019)\citenamefont {Ashwin},
		\citenamefont {Nozaki}, \citenamefont {Maeshima},\ and\ \citenamefont
		{Sasai}}]{ashwinOrganizationFastSlow2019}%
	\BibitemOpen
	\bibfield  {author} {\bibinfo {author} {\bibfnamefont {S.~S.}\ \bibnamefont
			{Ashwin}}, \bibinfo {author} {\bibfnamefont {T.}~\bibnamefont {Nozaki}},
		\bibinfo {author} {\bibfnamefont {K.}~\bibnamefont {Maeshima}},\ and\
		\bibinfo {author} {\bibfnamefont {M.}~\bibnamefont {Sasai}},\ }\bibfield
	{title} {\bibinfo {title} {Organization of fast and slow chromatin revealed
			by single-nucleosome dynamics},\ }\href
	{https://doi.org/10.1073/pnas.1907342116} {\bibfield  {journal} {\bibinfo
			{journal} {Proceedings of the National Academy of Sciences}\ }\textbf
		{\bibinfo {volume} {116}},\ \bibinfo {pages} {19939} (\bibinfo {year}
		{2019})}\BibitemShut {NoStop}%
	\bibitem [{\citenamefont {Khanna}\ \emph {et~al.}(2019)\citenamefont {Khanna},
		\citenamefont {Zhang}, \citenamefont {Lucas}, \citenamefont {Dudko},\ and\
		\citenamefont {Murre}}]{khannaChromosomeDynamicsSolgel2019}%
	\BibitemOpen
	\bibfield  {author} {\bibinfo {author} {\bibfnamefont {N.}~\bibnamefont
			{Khanna}}, \bibinfo {author} {\bibfnamefont {Y.}~\bibnamefont {Zhang}},
		\bibinfo {author} {\bibfnamefont {J.~S.}\ \bibnamefont {Lucas}}, \bibinfo
		{author} {\bibfnamefont {O.~K.}\ \bibnamefont {Dudko}},\ and\ \bibinfo
		{author} {\bibfnamefont {C.}~\bibnamefont {Murre}},\ }\bibfield  {title}
	{\bibinfo {title} {Chromosome dynamics near the sol-gel phase transition
			dictate the timing of remote genomic interactions},\ }\href
	{https://doi.org/10.1038/s41467-019-10628-9} {\bibfield  {journal} {\bibinfo
			{journal} {Nature Communications}\ }\textbf {\bibinfo {volume} {10}},\
		\bibinfo {pages} {2771} (\bibinfo {year} {2019})}\BibitemShut {NoStop}%
	\bibitem [{\citenamefont {Levi}\ \emph {et~al.}(2005)\citenamefont {Levi},
		\citenamefont {Ruan}, \citenamefont {Plutz}, \citenamefont {Belmont},\ and\
		\citenamefont {Gratton}}]{leviChromatinDynamicsInterphase2005}%
	\BibitemOpen
	\bibfield  {author} {\bibinfo {author} {\bibfnamefont {V.}~\bibnamefont
			{Levi}}, \bibinfo {author} {\bibfnamefont {Q.}~\bibnamefont {Ruan}}, \bibinfo
		{author} {\bibfnamefont {M.}~\bibnamefont {Plutz}}, \bibinfo {author}
		{\bibfnamefont {A.~S.}\ \bibnamefont {Belmont}},\ and\ \bibinfo {author}
		{\bibfnamefont {E.}~\bibnamefont {Gratton}},\ }\bibfield  {title} {\bibinfo
		{title} {Chromatin {{Dynamics}} in {{Interphase Cells Revealed}} by
			{{Tracking}} in a {{Two-Photon Excitation Microscope}}},\ }\href
	{https://doi.org/10.1529/biophysj.105.066670} {\bibfield  {journal} {\bibinfo
			{journal} {Biophysical Journal}\ }\textbf {\bibinfo {volume} {89}},\
		\bibinfo {pages} {4275} (\bibinfo {year} {2005})}\BibitemShut {NoStop}%
	\bibitem [{\citenamefont {Hoffman}\ \emph {et~al.}(2006)\citenamefont
		{Hoffman}, \citenamefont {Massiera}, \citenamefont {Van~Citters},\ and\
		\citenamefont {Crocker}}]{hoffmanConsensusMechanicsCultured2006}%
	\BibitemOpen
	\bibfield  {author} {\bibinfo {author} {\bibfnamefont {B.~D.}\ \bibnamefont
			{Hoffman}}, \bibinfo {author} {\bibfnamefont {G.}~\bibnamefont {Massiera}},
		\bibinfo {author} {\bibfnamefont {K.~M.}\ \bibnamefont {Van~Citters}},\ and\
		\bibinfo {author} {\bibfnamefont {J.~C.}\ \bibnamefont {Crocker}},\
	}\bibfield  {title} {\bibinfo {title} {The consensus mechanics of cultured
			mammalian cells},\ }\href {https://doi.org/10.1073/pnas.0510348103}
	{\bibfield  {journal} {\bibinfo  {journal} {Proceedings of the National
				Academy of Sciences}\ }\textbf {\bibinfo {volume} {103}},\ \bibinfo {pages}
		{10259} (\bibinfo {year} {2006})}\BibitemShut {NoStop}%
	\bibitem [{\citenamefont {Otten}\ \emph {et~al.}(2012)\citenamefont {Otten},
		\citenamefont {Nandi}, \citenamefont {Arcizet}, \citenamefont {Gorelashvili},
		\citenamefont {Lindner},\ and\ \citenamefont
		{Heinrich}}]{ottenLocalMotionAnalysis2012}%
	\BibitemOpen
	\bibfield  {author} {\bibinfo {author} {\bibfnamefont {M.}~\bibnamefont
			{Otten}}, \bibinfo {author} {\bibfnamefont {A.}~\bibnamefont {Nandi}},
		\bibinfo {author} {\bibfnamefont {D.}~\bibnamefont {Arcizet}}, \bibinfo
		{author} {\bibfnamefont {M.}~\bibnamefont {Gorelashvili}}, \bibinfo {author}
		{\bibfnamefont {B.}~\bibnamefont {Lindner}},\ and\ \bibinfo {author}
		{\bibfnamefont {D.}~\bibnamefont {Heinrich}},\ }\bibfield  {title} {\bibinfo
		{title} {Local {{Motion Analysis Reveals Impact}} of the {{Dynamic
					Cytoskeleton}} on {{Intracellular Subdiffusion}}},\ }\href
	{https://doi.org/10.1016/j.bpj.2011.12.057} {\bibfield  {journal} {\bibinfo
			{journal} {Biophysical Journal}\ }\textbf {\bibinfo {volume} {102}},\
		\bibinfo {pages} {758} (\bibinfo {year} {2012})}\BibitemShut {NoStop}%
	\bibitem [{\citenamefont {Annibale}\ and\ \citenamefont
		{Gratton}(2015)}]{annibaleSingleCellVisualization2015}%
	\BibitemOpen
	\bibfield  {author} {\bibinfo {author} {\bibfnamefont {P.}~\bibnamefont
			{Annibale}}\ and\ \bibinfo {author} {\bibfnamefont {E.}~\bibnamefont
			{Gratton}},\ }\bibfield  {title} {\bibinfo {title} {Single cell visualization
			of transcription kinetics variance of highly mobile identical genes using
			{{3D}} nanoimaging},\ }\href {https://doi.org/10.1038/srep09258} {\bibfield
		{journal} {\bibinfo  {journal} {Scientific Reports}\ }\textbf {\bibinfo
			{volume} {5}},\ \bibinfo {pages} {9258} (\bibinfo {year} {2015})}\BibitemShut
	{NoStop}%
	\bibitem [{\citenamefont {Mak}\ \emph {et~al.}(2017)\citenamefont {Mak},
		\citenamefont {Anderson}, \citenamefont {McDonough}, \citenamefont {Spill},
		\citenamefont {Kim}, \citenamefont {{Boussommier-Calleja}}, \citenamefont
		{Zaman},\ and\ \citenamefont
		{Kamm}}]{makIntegratedAnalysisIntracellular2017}%
	\BibitemOpen
	\bibfield  {author} {\bibinfo {author} {\bibfnamefont {M.}~\bibnamefont
			{Mak}}, \bibinfo {author} {\bibfnamefont {S.}~\bibnamefont {Anderson}},
		\bibinfo {author} {\bibfnamefont {M.~C.}\ \bibnamefont {McDonough}}, \bibinfo
		{author} {\bibfnamefont {F.}~\bibnamefont {Spill}}, \bibinfo {author}
		{\bibfnamefont {J.~E.}\ \bibnamefont {Kim}}, \bibinfo {author} {\bibfnamefont
			{A.}~\bibnamefont {{Boussommier-Calleja}}}, \bibinfo {author} {\bibfnamefont
			{M.~H.}\ \bibnamefont {Zaman}},\ and\ \bibinfo {author} {\bibfnamefont
			{R.~D.}\ \bibnamefont {Kamm}},\ }\bibfield  {title} {\bibinfo {title}
		{Integrated {{Analysis}} of {{Intracellular Dynamics}} of {{MenaINV Cancer
					Cells}} in a {{3D Matrix}}},\ }\href
	{https://doi.org/10.1016/j.bpj.2017.03.030} {\bibfield  {journal} {\bibinfo
			{journal} {Biophysical Journal}\ }\textbf {\bibinfo {volume} {112}},\
		\bibinfo {pages} {1874} (\bibinfo {year} {2017})}\BibitemShut {NoStop}%
	\bibitem [{\citenamefont {Rogers}\ \emph {et~al.}(2008)\citenamefont {Rogers},
		\citenamefont {Waigh},\ and\ \citenamefont
		{Lu}}]{rogersIntracellularMicrorheologyMotile2008}%
	\BibitemOpen
	\bibfield  {author} {\bibinfo {author} {\bibfnamefont {S.~S.}\ \bibnamefont
			{Rogers}}, \bibinfo {author} {\bibfnamefont {T.~A.}\ \bibnamefont {Waigh}},\
		and\ \bibinfo {author} {\bibfnamefont {J.~R.}\ \bibnamefont {Lu}},\
	}\bibfield  {title} {\bibinfo {title} {Intracellular {{Microrheology}} of
			{{Motile}} {{{\emph{Amoeba}}}}{\emph{ proteus}}},\ }\href
	{https://doi.org/10.1529/biophysj.107.123851} {\bibfield  {journal} {\bibinfo
			{journal} {Biophysical Journal}\ }\textbf {\bibinfo {volume} {94}},\
		\bibinfo {pages} {3313} (\bibinfo {year} {2008})}\BibitemShut {NoStop}%
	\bibitem [{\citenamefont {Lee}\ \emph {et~al.}(2024)\citenamefont {Lee},
		\citenamefont {Kim}, \citenamefont {Song}, \citenamefont {Shin},
		\citenamefont {Song}, \citenamefont {Lee}, \citenamefont {Goh}, \citenamefont
		{Lim}, \citenamefont {Kim}, \citenamefont {Sung},\ and\ \citenamefont
		{Lee}}]{leeRealTimeTrackingVesicles2024}%
	\BibitemOpen
	\bibfield  {author} {\bibinfo {author} {\bibfnamefont {E.}~\bibnamefont
			{Lee}}, \bibinfo {author} {\bibfnamefont {D.}~\bibnamefont {Kim}}, \bibinfo
		{author} {\bibfnamefont {Y.~H.}\ \bibnamefont {Song}}, \bibinfo {author}
		{\bibfnamefont {K.}~\bibnamefont {Shin}}, \bibinfo {author} {\bibfnamefont
			{S.}~\bibnamefont {Song}}, \bibinfo {author} {\bibfnamefont {M.}~\bibnamefont
			{Lee}}, \bibinfo {author} {\bibfnamefont {Y.}~\bibnamefont {Goh}}, \bibinfo
		{author} {\bibfnamefont {M.~H.}\ \bibnamefont {Lim}}, \bibinfo {author}
		{\bibfnamefont {J.-H.}\ \bibnamefont {Kim}}, \bibinfo {author} {\bibfnamefont
			{J.}~\bibnamefont {Sung}},\ and\ \bibinfo {author} {\bibfnamefont {K.~T.}\
			\bibnamefont {Lee}},\ }\bibfield  {title} {\bibinfo {title} {Real-{{Time
					Tracking}} of {{Vesicles}} in {{Living Cells Reveals That
					Tau-Hyperphosphorylation Suppresses Unidirectional Transport}} by {{Motor
					Proteins}}},\ }\href {https://doi.org/10.1021/cbmi.4c00016} {\bibfield
		{journal} {\bibinfo  {journal} {Chemical \& Biomedical Imaging}\ }\textbf
		{\bibinfo {volume} {2}},\ \bibinfo {pages} {362} (\bibinfo {year}
		{2024})}\BibitemShut {NoStop}%
	\bibitem [{\citenamefont {Murase}\ \emph {et~al.}(2004)\citenamefont {Murase},
		\citenamefont {Fujiwara}, \citenamefont {Umemura}, \citenamefont {Suzuki},
		\citenamefont {Iino}, \citenamefont {Yamashita}, \citenamefont {Saito},
		\citenamefont {Murakoshi}, \citenamefont {Ritchie},\ and\ \citenamefont
		{Kusumi}}]{muraseUltrafineMembraneCompartments2004}%
	\BibitemOpen
	\bibfield  {author} {\bibinfo {author} {\bibfnamefont {K.}~\bibnamefont
			{Murase}}, \bibinfo {author} {\bibfnamefont {T.}~\bibnamefont {Fujiwara}},
		\bibinfo {author} {\bibfnamefont {Y.}~\bibnamefont {Umemura}}, \bibinfo
		{author} {\bibfnamefont {K.}~\bibnamefont {Suzuki}}, \bibinfo {author}
		{\bibfnamefont {R.}~\bibnamefont {Iino}}, \bibinfo {author} {\bibfnamefont
			{H.}~\bibnamefont {Yamashita}}, \bibinfo {author} {\bibfnamefont
			{M.}~\bibnamefont {Saito}}, \bibinfo {author} {\bibfnamefont
			{H.}~\bibnamefont {Murakoshi}}, \bibinfo {author} {\bibfnamefont
			{K.}~\bibnamefont {Ritchie}},\ and\ \bibinfo {author} {\bibfnamefont
			{A.}~\bibnamefont {Kusumi}},\ }\bibfield  {title} {\bibinfo {title}
		{Ultrafine {{Membrane Compartments}} for {{Molecular Diffusion}} as
			{{Revealed}} by {{Single Molecule Techniques}}},\ }\href
	{https://doi.org/10.1529/biophysj.103.035717} {\bibfield  {journal} {\bibinfo
			{journal} {Biophysical Journal}\ }\textbf {\bibinfo {volume} {86}},\
		\bibinfo {pages} {4075} (\bibinfo {year} {2004})}\BibitemShut {NoStop}%
	\bibitem [{\citenamefont {Suzuki}\ \emph {et~al.}(2005)\citenamefont {Suzuki},
		\citenamefont {Ritchie}, \citenamefont {Kajikawa}, \citenamefont {Fujiwara},\
		and\ \citenamefont {Kusumi}}]{suzukiRapidHopDiffusion2005}%
	\BibitemOpen
	\bibfield  {author} {\bibinfo {author} {\bibfnamefont {K.}~\bibnamefont
			{Suzuki}}, \bibinfo {author} {\bibfnamefont {K.}~\bibnamefont {Ritchie}},
		\bibinfo {author} {\bibfnamefont {E.}~\bibnamefont {Kajikawa}}, \bibinfo
		{author} {\bibfnamefont {T.}~\bibnamefont {Fujiwara}},\ and\ \bibinfo
		{author} {\bibfnamefont {A.}~\bibnamefont {Kusumi}},\ }\bibfield  {title}
	{\bibinfo {title} {Rapid {{Hop Diffusion}} of a {{G-Protein-Coupled
					Receptor}} in the {{Plasma Membrane}} as {{Revealed}} by {{Single-Molecule
					Techniques}}},\ }\href {https://doi.org/10.1529/biophysj.104.048538}
	{\bibfield  {journal} {\bibinfo  {journal} {Biophysical Journal}\ }\textbf
		{\bibinfo {volume} {88}},\ \bibinfo {pages} {3659} (\bibinfo {year}
		{2005})}\BibitemShut {NoStop}%
	\bibitem [{\citenamefont {Wulstein}\ \emph {et~al.}(2019)\citenamefont
		{Wulstein}, \citenamefont {Regan}, \citenamefont {Garamella}, \citenamefont
		{McGorty},\ and\ \citenamefont
		{{Robertson-Anderson}}}]{wulsteinTopologydependentAnomalousDynamics2019}%
	\BibitemOpen
	\bibfield  {author} {\bibinfo {author} {\bibfnamefont {D.~M.}\ \bibnamefont
			{Wulstein}}, \bibinfo {author} {\bibfnamefont {K.~E.}\ \bibnamefont {Regan}},
		\bibinfo {author} {\bibfnamefont {J.}~\bibnamefont {Garamella}}, \bibinfo
		{author} {\bibfnamefont {R.~J.}\ \bibnamefont {McGorty}},\ and\ \bibinfo
		{author} {\bibfnamefont {R.~M.}\ \bibnamefont {{Robertson-Anderson}}},\
	}\bibfield  {title} {\bibinfo {title} {Topology-dependent anomalous dynamics
			of ring and linear {{DNA}} are sensitive to cytoskeleton crosslinking},\
	}\href {https://doi.org/10.1126/sciadv.aay5912} {\bibfield  {journal}
		{\bibinfo  {journal} {Science Advances}\ }\textbf {\bibinfo {volume} {5}},\
		\bibinfo {pages} {eaay5912} (\bibinfo {year} {2019})}\BibitemShut {NoStop}%
	\bibitem [{\citenamefont {Garamella}\ \emph {et~al.}(2020)\citenamefont
		{Garamella}, \citenamefont {Regan}, \citenamefont {Aguirre}, \citenamefont
		{McGorty},\ and\ \citenamefont
		{{Robertson-Anderson}}}]{garamellaAnomalousHeterogeneousDNA2020}%
	\BibitemOpen
	\bibfield  {author} {\bibinfo {author} {\bibfnamefont {J.}~\bibnamefont
			{Garamella}}, \bibinfo {author} {\bibfnamefont {K.}~\bibnamefont {Regan}},
		\bibinfo {author} {\bibfnamefont {G.}~\bibnamefont {Aguirre}}, \bibinfo
		{author} {\bibfnamefont {R.~J.}\ \bibnamefont {McGorty}},\ and\ \bibinfo
		{author} {\bibfnamefont {R.~M.}\ \bibnamefont {{Robertson-Anderson}}},\
	}\bibfield  {title} {\bibinfo {title} {Anomalous and heterogeneous {{DNA}}
			transport in biomimetic cytoskeleton networks},\ }\href
	{https://doi.org/10.1039/D0SM00544D} {\bibfield  {journal} {\bibinfo
			{journal} {Soft Matter}\ }\textbf {\bibinfo {volume} {16}},\ \bibinfo {pages}
		{6344} (\bibinfo {year} {2020})}\BibitemShut {NoStop}%
	\bibitem [{\citenamefont {Anderson}\ \emph {et~al.}(2021)\citenamefont
		{Anderson}, \citenamefont {Garamella}, \citenamefont {Adalbert},
		\citenamefont {McGorty},\ and\ \citenamefont
		{{Robertson-Anderson}}}]{andersonSubtleChangesCrosslinking2021}%
	\BibitemOpen
	\bibfield  {author} {\bibinfo {author} {\bibfnamefont {S.~J.}\ \bibnamefont
			{Anderson}}, \bibinfo {author} {\bibfnamefont {J.}~\bibnamefont {Garamella}},
		\bibinfo {author} {\bibfnamefont {S.}~\bibnamefont {Adalbert}}, \bibinfo
		{author} {\bibfnamefont {R.~J.}\ \bibnamefont {McGorty}},\ and\ \bibinfo
		{author} {\bibfnamefont {R.~M.}\ \bibnamefont {{Robertson-Anderson}}},\
	}\bibfield  {title} {\bibinfo {title} {Subtle changes in crosslinking drive
			diverse anomalous transport characteristics in actin--microtubule networks},\
	}\href {https://doi.org/10.1039/D1SM00093D} {\bibfield  {journal} {\bibinfo
			{journal} {Soft Matter}\ }\textbf {\bibinfo {volume} {17}},\ \bibinfo {pages}
		{4375} (\bibinfo {year} {2021})}\BibitemShut {NoStop}%
	\bibitem [{\citenamefont {Lee}\ \emph {et~al.}(2021)\citenamefont {Lee},
		\citenamefont {Leech}, \citenamefont {Rust}, \citenamefont {Das},
		\citenamefont {McGorty}, \citenamefont {Ross},\ and\ \citenamefont
		{{Robertson-Anderson}}}]{leeMyosindrivenActinmicrotubuleNetworks2021}%
	\BibitemOpen
	\bibfield  {author} {\bibinfo {author} {\bibfnamefont {G.}~\bibnamefont
			{Lee}}, \bibinfo {author} {\bibfnamefont {G.}~\bibnamefont {Leech}}, \bibinfo
		{author} {\bibfnamefont {M.~J.}\ \bibnamefont {Rust}}, \bibinfo {author}
		{\bibfnamefont {M.}~\bibnamefont {Das}}, \bibinfo {author} {\bibfnamefont
			{R.~J.}\ \bibnamefont {McGorty}}, \bibinfo {author} {\bibfnamefont {J.~L.}\
			\bibnamefont {Ross}},\ and\ \bibinfo {author} {\bibfnamefont {R.~M.}\
			\bibnamefont {{Robertson-Anderson}}},\ }\bibfield  {title} {\bibinfo {title}
		{Myosin-driven actin-microtubule networks exhibit self-organized contractile
			dynamics},\ }\href {https://doi.org/10.1126/sciadv.abe4334} {\bibfield
		{journal} {\bibinfo  {journal} {Science Advances}\ }\textbf {\bibinfo
			{volume} {7}},\ \bibinfo {pages} {eabe4334} (\bibinfo {year}
		{2021})}\BibitemShut {NoStop}%
	\bibitem [{\citenamefont {Sheung}\ \emph {et~al.}(2022)\citenamefont {Sheung},
		\citenamefont {Garamella}, \citenamefont {Kahl}, \citenamefont {Lee},
		\citenamefont {McGorty},\ and\ \citenamefont
		{{Robertson-Anderson}}}]{sheungMotordrivenAdvectionCompetes2022}%
	\BibitemOpen
	\bibfield  {author} {\bibinfo {author} {\bibfnamefont {J.~Y.}\ \bibnamefont
			{Sheung}}, \bibinfo {author} {\bibfnamefont {J.}~\bibnamefont {Garamella}},
		\bibinfo {author} {\bibfnamefont {S.~K.}\ \bibnamefont {Kahl}}, \bibinfo
		{author} {\bibfnamefont {B.~Y.}\ \bibnamefont {Lee}}, \bibinfo {author}
		{\bibfnamefont {R.~J.}\ \bibnamefont {McGorty}},\ and\ \bibinfo {author}
		{\bibfnamefont {R.~M.}\ \bibnamefont {{Robertson-Anderson}}},\ }\bibfield
	{title} {\bibinfo {title} {Motor-driven advection competes with crowding to
			drive spatiotemporally heterogeneous transport in cytoskeleton composites},\
	}\href {https://doi.org/10.3389/fphy.2022.1055441} {\bibfield  {journal}
		{\bibinfo  {journal} {Frontiers in Physics}\ }\textbf {\bibinfo {volume}
			{10}},\ \bibinfo {pages} {1055441} (\bibinfo {year} {2022})}\BibitemShut
	{NoStop}%
	\bibitem [{\citenamefont {Krajina}\ \emph {et~al.}(2021)\citenamefont
		{Krajina}, \citenamefont {LeSavage}, \citenamefont {Roth}, \citenamefont
		{Zhu}, \citenamefont {Cai}, \citenamefont {Spakowitz},\ and\ \citenamefont
		{Heilshorn}}]{krajinaMicrorheologyRevealsSimultaneous2021}%
	\BibitemOpen
	\bibfield  {author} {\bibinfo {author} {\bibfnamefont {B.~A.}\ \bibnamefont
			{Krajina}}, \bibinfo {author} {\bibfnamefont {B.~L.}\ \bibnamefont
			{LeSavage}}, \bibinfo {author} {\bibfnamefont {J.~G.}\ \bibnamefont {Roth}},
		\bibinfo {author} {\bibfnamefont {A.~W.}\ \bibnamefont {Zhu}}, \bibinfo
		{author} {\bibfnamefont {P.~C.}\ \bibnamefont {Cai}}, \bibinfo {author}
		{\bibfnamefont {A.~J.}\ \bibnamefont {Spakowitz}},\ and\ \bibinfo {author}
		{\bibfnamefont {S.~C.}\ \bibnamefont {Heilshorn}},\ }\bibfield  {title}
	{\bibinfo {title} {Microrheology reveals simultaneous cell-mediated matrix
			stiffening and fluidization that underlie breast cancer invasion},\ }\href
	{https://doi.org/10.1126/sciadv.abe1969} {\bibfield  {journal} {\bibinfo
			{journal} {Science Advances}\ }\textbf {\bibinfo {volume} {7}},\ \bibinfo
		{pages} {eabe1969} (\bibinfo {year} {2021})}\BibitemShut {NoStop}%
	\bibitem [{\citenamefont {Baker}\ \emph {et~al.}(2009)\citenamefont {Baker},
		\citenamefont {Bonnecaze},\ and\ \citenamefont
		{Zaman}}]{bakerExtracellularMatrixStiffness2009}%
	\BibitemOpen
	\bibfield  {author} {\bibinfo {author} {\bibfnamefont {E.~L.}\ \bibnamefont
			{Baker}}, \bibinfo {author} {\bibfnamefont {R.~T.}\ \bibnamefont
			{Bonnecaze}},\ and\ \bibinfo {author} {\bibfnamefont {M.~H.}\ \bibnamefont
			{Zaman}},\ }\bibfield  {title} {\bibinfo {title} {Extracellular {{Matrix
					Stiffness}} and {{Architecture Govern Intracellular Rheology}} in
			{{Cancer}}},\ }\href {https://doi.org/10.1016/j.bpj.2009.05.054} {\bibfield
		{journal} {\bibinfo  {journal} {Biophysical Journal}\ }\textbf {\bibinfo
			{volume} {97}},\ \bibinfo {pages} {1013} (\bibinfo {year}
		{2009})}\BibitemShut {NoStop}%
	\bibitem [{\citenamefont {Higgins}\ \emph {et~al.}(2023)\citenamefont
		{Higgins}, \citenamefont {Higgins}, \citenamefont {Peres}, \citenamefont
		{Lang}, \citenamefont {Abdalrahman}, \citenamefont {Zaman}, \citenamefont
		{Prince},\ and\ \citenamefont
		{Franz}}]{higginsIntracellularMechanicsTBX32023}%
	\BibitemOpen
	\bibfield  {author} {\bibinfo {author} {\bibfnamefont {G.}~\bibnamefont
			{Higgins}}, \bibinfo {author} {\bibfnamefont {F.}~\bibnamefont {Higgins}},
		\bibinfo {author} {\bibfnamefont {J.}~\bibnamefont {Peres}}, \bibinfo
		{author} {\bibfnamefont {D.~M.}\ \bibnamefont {Lang}}, \bibinfo {author}
		{\bibfnamefont {T.}~\bibnamefont {Abdalrahman}}, \bibinfo {author}
		{\bibfnamefont {M.~H.}\ \bibnamefont {Zaman}}, \bibinfo {author}
		{\bibfnamefont {S.}~\bibnamefont {Prince}},\ and\ \bibinfo {author}
		{\bibfnamefont {T.}~\bibnamefont {Franz}},\ }\bibfield  {title} {\bibinfo
		{title} {Intracellular mechanics and {{TBX3}} expression jointly dictate the
			spreading mode of melanoma cells in {{3D}} environments},\ }\href
	{https://doi.org/10.1016/j.yexcr.2023.113633} {\bibfield  {journal} {\bibinfo
			{journal} {Experimental Cell Research}\ }\textbf {\bibinfo {volume} {428}},\
		\bibinfo {pages} {113633} (\bibinfo {year} {2023})}\BibitemShut {NoStop}%
	\bibitem [{\citenamefont {Spitz}\ and\ \citenamefont
		{Furlong}(2012)}]{spitzTranscriptionFactorsEnhancer2012}%
	\BibitemOpen
	\bibfield  {author} {\bibinfo {author} {\bibfnamefont {F.}~\bibnamefont
			{Spitz}}\ and\ \bibinfo {author} {\bibfnamefont {E.~E.~M.}\ \bibnamefont
			{Furlong}},\ }\bibfield  {title} {\bibinfo {title} {Transcription factors:
			From enhancer binding to developmental control},\ }\href
	{https://doi.org/10.1038/nrg3207} {\bibfield  {journal} {\bibinfo  {journal}
			{Nature Reviews Genetics}\ }\textbf {\bibinfo {volume} {13}},\ \bibinfo
		{pages} {613} (\bibinfo {year} {2012})}\BibitemShut {NoStop}%
	\bibitem [{\citenamefont {Clapier}\ \emph {et~al.}(2017)\citenamefont
		{Clapier}, \citenamefont {Iwasa}, \citenamefont {Cairns},\ and\ \citenamefont
		{Peterson}}]{clapierMechanismsActionRegulation2017}%
	\BibitemOpen
	\bibfield  {author} {\bibinfo {author} {\bibfnamefont {C.~R.}\ \bibnamefont
			{Clapier}}, \bibinfo {author} {\bibfnamefont {J.}~\bibnamefont {Iwasa}},
		\bibinfo {author} {\bibfnamefont {B.~R.}\ \bibnamefont {Cairns}},\ and\
		\bibinfo {author} {\bibfnamefont {C.~L.}\ \bibnamefont {Peterson}},\
	}\bibfield  {title} {\bibinfo {title} {Mechanisms of action and regulation of
			{{ATP-dependent}} chromatin-remodelling complexes},\ }\href
	{https://doi.org/10.1038/nrm.2017.26} {\bibfield  {journal} {\bibinfo
			{journal} {Nature Reviews Molecular Cell Biology}\ }\textbf {\bibinfo
			{volume} {18}},\ \bibinfo {pages} {407} (\bibinfo {year} {2017})}\BibitemShut
	{NoStop}%
	\bibitem [{\citenamefont {Su}\ \emph {et~al.}(2021)\citenamefont {Su},
		\citenamefont {Mehta},\ and\ \citenamefont
		{Zhang}}]{suLiquidliquidPhaseSeparation2021}%
	\BibitemOpen
	\bibfield  {author} {\bibinfo {author} {\bibfnamefont {Q.}~\bibnamefont
			{Su}}, \bibinfo {author} {\bibfnamefont {S.}~\bibnamefont {Mehta}},\ and\
		\bibinfo {author} {\bibfnamefont {J.}~\bibnamefont {Zhang}},\ }\bibfield
	{title} {\bibinfo {title} {Liquid-liquid phase separation: {{Orchestrating}}
			cell signaling through time and space},\ }\href
	{https://doi.org/10.1016/j.molcel.2021.09.010} {\bibfield  {journal}
		{\bibinfo  {journal} {Molecular Cell}\ }\textbf {\bibinfo {volume} {81}},\
		\bibinfo {pages} {4137} (\bibinfo {year} {2021})}\BibitemShut {NoStop}%
	\bibitem [{\citenamefont {Hirose}\ \emph {et~al.}(2023)\citenamefont {Hirose},
		\citenamefont {Ninomiya}, \citenamefont {Nakagawa},\ and\ \citenamefont
		{Yamazaki}}]{hiroseGuideMembranelessOrganelles2023}%
	\BibitemOpen
	\bibfield  {author} {\bibinfo {author} {\bibfnamefont {T.}~\bibnamefont
			{Hirose}}, \bibinfo {author} {\bibfnamefont {K.}~\bibnamefont {Ninomiya}},
		\bibinfo {author} {\bibfnamefont {S.}~\bibnamefont {Nakagawa}},\ and\
		\bibinfo {author} {\bibfnamefont {T.}~\bibnamefont {Yamazaki}},\ }\bibfield
	{title} {\bibinfo {title} {A guide to membraneless organelles and their
			various roles in gene regulation},\ }\href
	{https://doi.org/10.1038/s41580-022-00558-8} {\bibfield  {journal} {\bibinfo
			{journal} {Nature Reviews Molecular Cell Biology}\ }\textbf {\bibinfo
			{volume} {24}},\ \bibinfo {pages} {288} (\bibinfo {year} {2023})}\BibitemShut
	{NoStop}%
	\bibitem [{\citenamefont {Ricketts}\ \emph {et~al.}(2018)\citenamefont
		{Ricketts}, \citenamefont {Ross},\ and\ \citenamefont
		{{Robertson-Anderson}}}]{rickettsCoEntangledActinMicrotubuleComposites2018}%
	\BibitemOpen
	\bibfield  {author} {\bibinfo {author} {\bibfnamefont {S.~N.}\ \bibnamefont
			{Ricketts}}, \bibinfo {author} {\bibfnamefont {J.~L.}\ \bibnamefont {Ross}},\
		and\ \bibinfo {author} {\bibfnamefont {R.~M.}\ \bibnamefont
			{{Robertson-Anderson}}},\ }\bibfield  {title} {\bibinfo {title}
		{Co-{{Entangled Actin-Microtubule Composites Exhibit Tunable Stiffness}} and
			{{Power-Law Stress Relaxation}}},\ }\href
	{https://doi.org/10.1016/j.bpj.2018.08.010} {\bibfield  {journal} {\bibinfo
			{journal} {Biophysical Journal}\ }\textbf {\bibinfo {volume} {115}},\
		\bibinfo {pages} {1055} (\bibinfo {year} {2018})}\BibitemShut {NoStop}%
	\bibitem [{\citenamefont {Gittes}\ \emph {et~al.}(1993)\citenamefont {Gittes},
		\citenamefont {Mickey}, \citenamefont {Nettleton},\ and\ \citenamefont
		{Howard}}]{gittesFlexuralRigidityMicrotubules1993}%
	\BibitemOpen
	\bibfield  {author} {\bibinfo {author} {\bibfnamefont {F.}~\bibnamefont
			{Gittes}}, \bibinfo {author} {\bibfnamefont {B.}~\bibnamefont {Mickey}},
		\bibinfo {author} {\bibfnamefont {J.}~\bibnamefont {Nettleton}},\ and\
		\bibinfo {author} {\bibfnamefont {J.}~\bibnamefont {Howard}},\ }\bibfield
	{title} {\bibinfo {title} {Flexural rigidity of microtubules and actin
			filaments measured from thermal fluctuations in shape.},\ }\href
	{https://doi.org/10.1083/jcb.120.4.923} {\bibfield  {journal} {\bibinfo
			{journal} {Journal of Cell Biology}\ }\textbf {\bibinfo {volume} {120}},\
		\bibinfo {pages} {923} (\bibinfo {year} {1993})}\BibitemShut {NoStop}%
	\bibitem [{\citenamefont {Kikumoto}\ \emph {et~al.}(2006)\citenamefont
		{Kikumoto}, \citenamefont {Kurachi}, \citenamefont {Tosa},\ and\
		\citenamefont {Tashiro}}]{kikumotoFlexuralRigidityIndividual2006}%
	\BibitemOpen
	\bibfield  {author} {\bibinfo {author} {\bibfnamefont {M.}~\bibnamefont
			{Kikumoto}}, \bibinfo {author} {\bibfnamefont {M.}~\bibnamefont {Kurachi}},
		\bibinfo {author} {\bibfnamefont {V.}~\bibnamefont {Tosa}},\ and\ \bibinfo
		{author} {\bibfnamefont {H.}~\bibnamefont {Tashiro}},\ }\bibfield  {title}
	{\bibinfo {title} {Flexural {{Rigidity}} of {{Individual Microtubules
					Measured}} by a {{Buckling Force}} with {{Optical Traps}}},\ }\href
	{https://doi.org/10.1529/biophysj.104.055483} {\bibfield  {journal} {\bibinfo
			{journal} {Biophysical Journal}\ }\textbf {\bibinfo {volume} {90}},\
		\bibinfo {pages} {1687} (\bibinfo {year} {2006})}\BibitemShut {NoStop}%
	\bibitem [{\citenamefont {Huber}\ \emph {et~al.}(2015)\citenamefont {Huber},
		\citenamefont {Boire}, \citenamefont {L{\'o}pez},\ and\ \citenamefont
		{Koenderink}}]{huberCytoskeletalCrosstalkWhen2015}%
	\BibitemOpen
	\bibfield  {author} {\bibinfo {author} {\bibfnamefont {F.}~\bibnamefont
			{Huber}}, \bibinfo {author} {\bibfnamefont {A.}~\bibnamefont {Boire}},
		\bibinfo {author} {\bibfnamefont {M.~P.}\ \bibnamefont {L{\'o}pez}},\ and\
		\bibinfo {author} {\bibfnamefont {G.~H.}\ \bibnamefont {Koenderink}},\
	}\bibfield  {title} {\bibinfo {title} {Cytoskeletal crosstalk: When three
			different personalities team up},\ }\href
	{https://doi.org/10.1016/j.ceb.2014.10.005} {\bibfield  {journal} {\bibinfo
			{journal} {Current Opinion in Cell Biology}\ }\textbf {\bibinfo {volume}
			{32}},\ \bibinfo {pages} {39} (\bibinfo {year} {2015})}\BibitemShut {NoStop}%
	\bibitem [{\citenamefont {Schulz}\ \emph {et~al.}(2001)\citenamefont {Schulz},
		\citenamefont {Stepanow},\ and\ \citenamefont
		{Trimper}}]{schulzTwoHarmonicallyCoupled2001}%
	\BibitemOpen
	\bibfield  {author} {\bibinfo {author} {\bibfnamefont {M.}~\bibnamefont
			{Schulz}}, \bibinfo {author} {\bibfnamefont {S.}~\bibnamefont {Stepanow}},\
		and\ \bibinfo {author} {\bibfnamefont {S.}~\bibnamefont {Trimper}},\
	}\bibfield  {title} {\bibinfo {title} {Two harmonically coupled {{Brownian}}
			particles in random media},\ }\href
	{https://doi.org/10.1209/epl/i2001-00258-6} {\bibfield  {journal} {\bibinfo
			{journal} {Europhysics Letters}\ }\textbf {\bibinfo {volume} {54}},\ \bibinfo
		{pages} {424} (\bibinfo {year} {2001})}\BibitemShut {NoStop}%
	\bibitem [{\citenamefont {Haubold}\ \emph {et~al.}(2011)\citenamefont
		{Haubold}, \citenamefont {Mathai},\ and\ \citenamefont
		{Saxena}}]{hauboldMittagLefflerFunctionsTheir2011}%
	\BibitemOpen
	\bibfield  {author} {\bibinfo {author} {\bibfnamefont {H.~J.}\ \bibnamefont
			{Haubold}}, \bibinfo {author} {\bibfnamefont {A.~M.}\ \bibnamefont
			{Mathai}},\ and\ \bibinfo {author} {\bibfnamefont {R.~K.}\ \bibnamefont
			{Saxena}},\ }\bibfield  {title} {\bibinfo {title} {Mittag-{{Leffler
					Functions}} and {{Their Applications}}},\ }\href
	{https://doi.org/10.1155/2011/298628} {\bibfield  {journal} {\bibinfo
			{journal} {Journal of Applied Mathematics}\ }\textbf {\bibinfo {volume}
			{2011}},\ \bibinfo {pages} {298628} (\bibinfo {year} {2011})}\BibitemShut
	{NoStop}%
	\bibitem [{\citenamefont
		{Mainardi}(2020)}]{mainardiWhyMittagLefflerFunction2020}%
	\BibitemOpen
	\bibfield  {author} {\bibinfo {author} {\bibfnamefont {F.}~\bibnamefont
			{Mainardi}},\ }\bibfield  {title} {\bibinfo {title} {Why the {{Mittag-Leffler
					Function Can Be Considered}} the {{Queen Function}} of the {{Fractional
					Calculus}}?},\ }\href {https://doi.org/10.3390/e22121359} {\bibfield
		{journal} {\bibinfo  {journal} {Entropy}\ }\textbf {\bibinfo {volume} {22}},\
		\bibinfo {pages} {1359} (\bibinfo {year} {2020})}\BibitemShut {NoStop}%
	\bibitem [{\citenamefont {Stehfest}(1970)}]{stehfestAlgorithm368Numerical1970}%
	\BibitemOpen
	\bibfield  {author} {\bibinfo {author} {\bibfnamefont {H.}~\bibnamefont
			{Stehfest}},\ }\bibfield  {title} {\bibinfo {title} {Algorithm 368:
			{{Numerical}} inversion of {{Laplace}} transforms [{{D5}}]},\ }\href
	{https://doi.org/10.1145/361953.361969} {\bibfield  {journal} {\bibinfo
			{journal} {Commun. ACM}\ }\textbf {\bibinfo {volume} {13}},\ \bibinfo {pages}
		{47} (\bibinfo {year} {1970})}\BibitemShut {NoStop}%
	\bibitem [{\citenamefont {Jeon}\ and\ \citenamefont
		{Metzler}(2012)}]{jeonInequivalenceTimeEnsemble2012}%
	\BibitemOpen
	\bibfield  {author} {\bibinfo {author} {\bibfnamefont {J.-H.}\ \bibnamefont
			{Jeon}}\ and\ \bibinfo {author} {\bibfnamefont {R.}~\bibnamefont {Metzler}},\
	}\bibfield  {title} {\bibinfo {title} {Inequivalence of time and ensemble
			averages in ergodic systems: {{Exponential}} versus power-law relaxation in
			confinement},\ }\href {https://doi.org/10.1103/PhysRevE.85.021147} {\bibfield
		{journal} {\bibinfo  {journal} {Physical Review E}\ }\textbf {\bibinfo
			{volume} {85}},\ \bibinfo {pages} {021147} (\bibinfo {year}
		{2012})}\BibitemShut {NoStop}%
	\bibitem [{\citenamefont {Rippe}\ and\ \citenamefont
		{Papantonis}(2025)}]{rippeRNAPolymeraseII2025}%
	\BibitemOpen
	\bibfield  {author} {\bibinfo {author} {\bibfnamefont {K.}~\bibnamefont
			{Rippe}}\ and\ \bibinfo {author} {\bibfnamefont {A.}~\bibnamefont
			{Papantonis}},\ }\bibfield  {title} {\bibinfo {title} {{{RNA}} polymerase
			{{II}} transcription compartments --- from factories to condensates},\ }\href
	{https://doi.org/10.1038/s41576-025-00859-6} {\bibfield  {journal} {\bibinfo
			{journal} {Nature Reviews Genetics}\ ,\ \bibinfo {pages} {1}} (\bibinfo
		{year} {2025})}\BibitemShut {NoStop}%
	\bibitem [{\citenamefont {Alghoul}\ \emph {et~al.}(2023)\citenamefont
		{Alghoul}, \citenamefont {Basbous},\ and\ \citenamefont
		{Constantinou}}]{alghoulCompartmentalizationDNADamage2023}%
	\BibitemOpen
	\bibfield  {author} {\bibinfo {author} {\bibfnamefont {E.}~\bibnamefont
			{Alghoul}}, \bibinfo {author} {\bibfnamefont {J.}~\bibnamefont {Basbous}},\
		and\ \bibinfo {author} {\bibfnamefont {A.}~\bibnamefont {Constantinou}},\
	}\bibfield  {title} {\bibinfo {title} {Compartmentalization of the {{DNA}}
			damage response: {{Mechanisms}} and functions},\ }\href
	{https://doi.org/10.1016/j.dnarep.2023.103524} {\bibfield  {journal}
		{\bibinfo  {journal} {DNA Repair}\ }\textbf {\bibinfo {volume} {128}},\
		\bibinfo {pages} {103524} (\bibinfo {year} {2023})}\BibitemShut {NoStop}%
	\bibitem [{\citenamefont {Larson}\ \emph {et~al.}(2017)\citenamefont {Larson},
		\citenamefont {Elnatan}, \citenamefont {Keenen}, \citenamefont {Trnka},
		\citenamefont {Johnston}, \citenamefont {Burlingame}, \citenamefont {Agard},
		\citenamefont {Redding},\ and\ \citenamefont
		{Narlikar}}]{larsonLiquidDropletFormation2017}%
	\BibitemOpen
	\bibfield  {author} {\bibinfo {author} {\bibfnamefont {A.~G.}\ \bibnamefont
			{Larson}}, \bibinfo {author} {\bibfnamefont {D.}~\bibnamefont {Elnatan}},
		\bibinfo {author} {\bibfnamefont {M.~M.}\ \bibnamefont {Keenen}}, \bibinfo
		{author} {\bibfnamefont {M.~J.}\ \bibnamefont {Trnka}}, \bibinfo {author}
		{\bibfnamefont {J.~B.}\ \bibnamefont {Johnston}}, \bibinfo {author}
		{\bibfnamefont {A.~L.}\ \bibnamefont {Burlingame}}, \bibinfo {author}
		{\bibfnamefont {D.~A.}\ \bibnamefont {Agard}}, \bibinfo {author}
		{\bibfnamefont {S.}~\bibnamefont {Redding}},\ and\ \bibinfo {author}
		{\bibfnamefont {G.~J.}\ \bibnamefont {Narlikar}},\ }\bibfield  {title}
	{\bibinfo {title} {Liquid droplet formation by {{HP1$\alpha$}} suggests a
			role for phase separation in heterochromatin},\ }\href
	{https://doi.org/10.1038/nature22822} {\bibfield  {journal} {\bibinfo
			{journal} {Nature}\ }\textbf {\bibinfo {volume} {547}},\ \bibinfo {pages}
		{236} (\bibinfo {year} {2017})}\BibitemShut {NoStop}%
	\bibitem [{\citenamefont {Wachsmuth}\ \emph {et~al.}(2000)\citenamefont
		{Wachsmuth}, \citenamefont {Waldeck},\ and\ \citenamefont
		{Langowski}}]{wachsmuthAnomalousDiffusionFluorescent2000}%
	\BibitemOpen
	\bibfield  {author} {\bibinfo {author} {\bibfnamefont {M.}~\bibnamefont
			{Wachsmuth}}, \bibinfo {author} {\bibfnamefont {W.}~\bibnamefont {Waldeck}},\
		and\ \bibinfo {author} {\bibfnamefont {J.}~\bibnamefont {Langowski}},\
	}\bibfield  {title} {\bibinfo {title} {Anomalous diffusion of fluorescent
			probes inside living cell nuclei investigated by spatially-resolved
			fluorescence correlation spectroscopy1},\ }\href
	{https://doi.org/10.1006/jmbi.2000.3692} {\bibfield  {journal} {\bibinfo
			{journal} {Journal of Molecular Biology}\ }\textbf {\bibinfo {volume}
			{298}},\ \bibinfo {pages} {677} (\bibinfo {year} {2000})}\BibitemShut
	{NoStop}%
	\bibitem [{\citenamefont {Bancaud}\ \emph {et~al.}(2009)\citenamefont
		{Bancaud}, \citenamefont {Huet}, \citenamefont {Daigle}, \citenamefont
		{Mozziconacci}, \citenamefont {Beaudouin},\ and\ \citenamefont
		{Ellenberg}}]{bancaudMolecularCrowdingAffects2009}%
	\BibitemOpen
	\bibfield  {author} {\bibinfo {author} {\bibfnamefont {A.}~\bibnamefont
			{Bancaud}}, \bibinfo {author} {\bibfnamefont {S.}~\bibnamefont {Huet}},
		\bibinfo {author} {\bibfnamefont {N.}~\bibnamefont {Daigle}}, \bibinfo
		{author} {\bibfnamefont {J.}~\bibnamefont {Mozziconacci}}, \bibinfo {author}
		{\bibfnamefont {J.}~\bibnamefont {Beaudouin}},\ and\ \bibinfo {author}
		{\bibfnamefont {J.}~\bibnamefont {Ellenberg}},\ }\bibfield  {title} {\bibinfo
		{title} {Molecular crowding affects diffusion and binding of nuclear proteins
			in heterochromatin and reveals the fractal organization of chromatin},\
	}\href {https://doi.org/10.1038/emboj.2009.340} {\bibfield  {journal}
		{\bibinfo  {journal} {The EMBO Journal}\ }\textbf {\bibinfo {volume} {28}},\
		\bibinfo {pages} {3785} (\bibinfo {year} {2009})}\BibitemShut {NoStop}%
	\bibitem [{\citenamefont {Daddysman}\ and\ \citenamefont
		{Fecko}(2013)}]{daddysmanRevisitingPointFRAP2013}%
	\BibitemOpen
	\bibfield  {author} {\bibinfo {author} {\bibfnamefont {M.~K.}\ \bibnamefont
			{Daddysman}}\ and\ \bibinfo {author} {\bibfnamefont {C.~J.}\ \bibnamefont
			{Fecko}},\ }\bibfield  {title} {\bibinfo {title} {Revisiting {{Point FRAP}}
			to {{Quantitatively Characterize Anomalous Diffusion}} in {{Live Cells}}},\
	}\href {https://doi.org/10.1021/jp310348s} {\bibfield  {journal} {\bibinfo
			{journal} {The Journal of Physical Chemistry B}\ }\textbf {\bibinfo {volume}
			{117}},\ \bibinfo {pages} {1241} (\bibinfo {year} {2013})}\BibitemShut
	{NoStop}%
	\bibitem [{\citenamefont {Meng}\ \emph {et~al.}(2024)\citenamefont {Meng},
		\citenamefont {Zhang}, \citenamefont {Yu},\ and\ \citenamefont
		{Wang}}]{mengTemperedAnomalousDynamics2024}%
	\BibitemOpen
	\bibfield  {author} {\bibinfo {author} {\bibfnamefont {L.}~\bibnamefont
			{Meng}}, \bibinfo {author} {\bibfnamefont {R.}~\bibnamefont {Zhang}},
		\bibinfo {author} {\bibfnamefont {L.}~\bibnamefont {Yu}},\ and\ \bibinfo
		{author} {\bibfnamefont {H.}~\bibnamefont {Wang}},\ }\bibfield  {title}
	{\bibinfo {title} {Tempered anomalous dynamics of globally coupled harmonic
			oscillators in the fluctuating potential field: Stability, synchronism, and
			collective behaviors},\ }\href
	{https://doi.org/10.1140/epjp/s13360-024-04865-1} {\bibfield  {journal}
		{\bibinfo  {journal} {The European Physical Journal Plus}\ }\textbf {\bibinfo
			{volume} {139}},\ \bibinfo {pages} {63} (\bibinfo {year} {2024})}\BibitemShut
	{NoStop}%
	\bibitem [{\citenamefont {Tamm}\ \emph {et~al.}(2015)\citenamefont {Tamm},
		\citenamefont {Nazarov}, \citenamefont {Gavrilov},\ and\ \citenamefont
		{Chertovich}}]{tammAnomalousDiffusionFractal2015}%
	\BibitemOpen
	\bibfield  {author} {\bibinfo {author} {\bibfnamefont {M.~V.}\ \bibnamefont
			{Tamm}}, \bibinfo {author} {\bibfnamefont {L.~I.}\ \bibnamefont {Nazarov}},
		\bibinfo {author} {\bibfnamefont {A.~A.}\ \bibnamefont {Gavrilov}},\ and\
		\bibinfo {author} {\bibfnamefont {A.~V.}\ \bibnamefont {Chertovich}},\
	}\bibfield  {title} {\bibinfo {title} {Anomalous {{Diffusion}} in {{Fractal
					Globules}}},\ }\href {https://doi.org/10.1103/PhysRevLett.114.178102}
	{\bibfield  {journal} {\bibinfo  {journal} {Physical Review Letters}\
		}\textbf {\bibinfo {volume} {114}},\ \bibinfo {pages} {178102} (\bibinfo
		{year} {2015})}\BibitemShut {NoStop}%
	\bibitem [{\citenamefont {Vanden-Eijnden}\ and\ \citenamefont
		{Ciccotti}(2006)}]{vandenSecondIntegratorsLangevin}%
	\BibitemOpen
	\bibfield  {author} {\bibinfo {author} {\bibfnamefont {E.}~\bibnamefont
			{Vanden-Eijnden}}\ and\ \bibinfo {author} {\bibfnamefont {G.}~\bibnamefont
			{Ciccotti}},\ }\bibfield  {title} {\bibinfo {title} {Second-order integrators
			for langevin equations with holonomic constraints},\ }\href
	{https://doi.org/https://doi.org/10.1016/j.cplett.2006.07.086} {\bibfield
		{journal} {\bibinfo  {journal} {Chemical Physics Letters}\ }\textbf {\bibinfo
			{volume} {429}},\ \bibinfo {pages} {310} (\bibinfo {year}
		{2006})}\BibitemShut {NoStop}%
	\bibitem [{\citenamefont {Winkler}\ \emph {et~al.}(1994)\citenamefont
		{Winkler}, \citenamefont {Reineker},\ and\ \citenamefont
		{Harnau}}]{winklerModelsEquilibriumProperties1994}%
	\BibitemOpen
	\bibfield  {author} {\bibinfo {author} {\bibfnamefont {R.~G.}\ \bibnamefont
			{Winkler}}, \bibinfo {author} {\bibfnamefont {P.}~\bibnamefont {Reineker}},\
		and\ \bibinfo {author} {\bibfnamefont {L.}~\bibnamefont {Harnau}},\
	}\bibfield  {title} {\bibinfo {title} {Models and equilibrium properties of
			stiff molecular chains},\ }\href {https://doi.org/10.1063/1.468239}
	{\bibfield  {journal} {\bibinfo  {journal} {The Journal of Chemical Physics}\
		}\textbf {\bibinfo {volume} {101}},\ \bibinfo {pages} {8119} (\bibinfo {year}
		{1994})}\BibitemShut {NoStop}%
\end{thebibliography}
% \bibliographystyle{unsrt}


\end{document}